\documentclass[11pt, oneside]{article}

\usepackage[paper=a4paper, total={16cm,23cm}]{geometry}
\usepackage[T1]{fontenc}
\usepackage{setspace} % line-spacing
\usepackage[hang]{footmisc} % footnotes
\usepackage{titling} % title page
\usepackage{tocloft} % table of contents
\usepackage{parskip} % paragraph indentation
\usepackage{comment}

\usepackage{amsfonts}
\usepackage{amsmath}
\usepackage{amssymb}
\usepackage{mathrsfs, dsfont, amsthm}
\usepackage{physics}
\usepackage{mathtools}
\usepackage{upgreek}
\numberwithin{equation}{section} % numbering equations
\usepackage{wasysym}
\usepackage{marvosym}

\usepackage{xcolor}
\usepackage{colortbl}
\definecolor{dark-red}{rgb}{0.50,0.12,0.12} % links
\definecolor{mblue}{rgb}{0.30, 0.45, 0.70}
\definecolor{mred}{rgb}{0.70, 0.20, 0.20}
\definecolor{mgray}{rgb}{0.63, 0.63, 0.63}

\usepackage{hyperref} % hyperlink refs
\hypersetup{colorlinks=true, allcolors=dark-red, linktoc=all}

\usepackage{cite}

\usepackage{tikz}
\usepackage{tikzlings}
\usetikzlibrary{calc, patterns, decorations, decorations.markings, shapes, positioning, intersections, quotes, angles}
\usepackage[subrefformat=parens]{subcaption}
\usepackage{pgfplots}
\pgfplotsset{compat=newest}
\newcommand{\mathdefault}[1][]{}

\def \cA {\mathcal{A}}
\def \cM {\mathcal{M}}

\newcommand{\beq}{\begin{equation}}
\newcommand{\eeq}{\end{equation}} 

\newcommand{\lc}{\left(}
\newcommand{\rc}{\right)}

\newcommand{\ic}{{\rm i}}
\newcommand{\ee}{\mathrm{e}}
\newcommand{\cG}{\mathcal G}

\newcommand{\diff}{\mathrm{d}}

\newcommand{\zh}{z_\mathrm{h}}

\newcommand{\tO}{\widetilde{\mathcal O}}
\newcommand{\by}{\bar y}

\begin{document}
% -------------------------
% title page
% -------------------------
\begin{titlingpage}
    \vspace*{3em}
    \onehalfspacing
    \begin{center}
    {\LARGE Imaging the black hole interior \\[0.25em]
    with two-sided four-point correlators}
    \end{center}
    \singlespacing
    \vspace*{2em}
      \begin{center}
        % authors
        \textbf{
       Joydeep Chakravarty
        }
    \end{center}
    \vspace*{1em}
    \begin{center}
        \textsl{
        % affiliations
      Department of Physics, Harvard University, Cambridge, MA, USA \\
       \vspace*{1em}
    Department of Physics, McGill University,
        Montr\'eal, QC, Canada \\[\baselineskip]
        }
        % emails
 \href{mailto:joydeepchakravarty@fas.harvard.edu}{\small joydeepchakravarty@fas.harvard.edu}
    \end{center}
    \vspace*{3em}
    \begin{abstract}
We propose a boundary four-point function smeared with wavepackets, which probes local physics behind the horizon of a two-sided AdS black hole. At high frequencies, boundary wavepackets aimed at an interior bulk point travel along geodesics and scatter locally about the bulk point, factorizing the correlator into a flat-space like scattering amplitude. To obtain this configuration, we utilize a non-trivial time ordering of operators, and also use one state-dependent operator insertion moved halfway across the thermal circle. An interesting property of this observable is that it also probes features of nontrivial states: null shocks and matter shells behind the horizon show up as jumps in the observed signal as the bulk point crosses them, thereby imaging the black hole interior.
    \end{abstract}
\end{titlingpage}
\tableofcontents

% -------------------------
% introduction
% -------------------------

\section{Introduction}

The AdS/CFT correspondence gives a nonperturbative definition of quantum gravity in asymptotically anti-de Sitter spacetimes in terms of an ordinary quantum field theory
without gravity \cite{Maldacena:1997re,Gubser:1998bc,Witten:1998qj}. The description of black holes as strongly coupled gauge theory at a temperature has strong implications for both thermal systems as well as for puzzles associated with the black hole interior.  

Generally it is difficult to utilize simple correlators to extract reliable statements about the interior dual to a thermal density matrix (or very close to it) in holography. In our present work, we extract implications of bulk locality in the two-sided black hole interior by suitably smearing the four-point correlator. We do this by evaluating the correlator on a cousin of the thermal density matrix, i.e. the thermofield-doubled state (TFD) and its deformations. The TFD is a state of two entangled CFTs at a finite temperature and provides a particularly clean arena to explore questions about bulk locality in the actual black hole interior. Note that the TFD state is a purification of the thermal density matrix and prepares a two-sided bulk geometry with a shared interior \cite{Maldacena:2001kr, Maldacena:2013xja}. 

On a related note, concrete ways to \textit{prepare} the single-side interior geometry by studying analytic properties of simple observables have also been proposed, which works by continuing exterior data to prepare the classical interior geometry. This follows a line of work which has shown that certain boundary two-point  correlators, under special circumstances, can carry  signatures of the black hole singularity  \cite{Kraus:2002iv,Fidkowski:2003nf,Festuccia:2005pi,Hubeny:2006yu,Festuccia:2008zx,Hartman:2013qma,Ceplak:2024bja}, and also in the one-point function \cite{Grinberg:2020fdj}. The procedure to prepare the classical interior geometry via analytic continuation  works for light operator insertions over the thermal state, and can be shown to work for light operator insertions over the thermal state, and can be shown to work using the two-point function \cite{Afkhami-Jeddi:2025wra,  Jia:2025jbi, Giombi:2026kdz, Jia:2026ryl, Chakravarty:2026fdm, Jia:2026qvp, Oh:2026qcm}, and for the four-point function \cite{Chakravarty:2025ncy}. While this technique of analytic continuation does not address  whether there is a \textit{dragon} sitting in the interior waiting to burn a curious explorer, i.e. questions about the experience of infalling observers, it shows how  concrete operations upon the thermal state can be used to prepare an analytically continued classical geometry behind the horizon.  See also \cite{Ceplak:2025dds, Dodelson:2025jff, Arnaudo:2026tcy, Grozdanov:2026cut, Hartnoll:2026vhu, Coviello:2026hdc, Xiao:2026pir, David:2022nfn,  Singhi:2024sdr, Dey:2026ckh, Goldar:2026bij, Singhi:2026xbc, Grassi:2026ekz, Jia:2026pmv, Ceplak:2026ard, Arnaudo:2026der, Arnaudo:2026axe, Buric:2026qsp, Huang:2026jbv, Huang:2026uif, Sorokhaibam:2025nur, DeClerck:2025mem, Blacker:2026poc, AliAhmad:2026wem, Aalsma:2026kqj, Grozdanov:2026ktq, Grozdanov:2026lnc, Chen:2025jbf, Chen:2025ywj, Chen:2026efa, Shi:2026cfq, Saha:2026ows, Dodelson:2026gak, Gao:2026ieh, Oh:2026dky, Lyu:2026xec} for recent work on thermal correlators, on geodesics corresponding to lightcone singularities, quasinormal modes and the geometry near the singularity, and \cite{Simon_unpub, Komargodski:2026oob, Batra:2026gbq, Batra:2026otu, Buric:2026sit, Liu:2026luf, Almheiri:2026cpv, Dodelson:2026hjm} for the fate of the bouncing geodesic beyond the leading semiclassical approximation. 

In our present work we ask whether a correlator with support on both boundaries can do better, i.e. whether it can be tuned to record a local scattering event in the actual shared interior, as opposed to preparing an analytically continued geometry from one side. We find that using one left boundary operator, two right boundary operators and one right-boundary representation of a left operator moved by half of the thermal circle, gives an observable which sees the interior event. Since the legs reach the interior by different routes, it also provides a way to diagnose the geometry. 

The underlying reason as to why our local experiments can extract flat-space like physics about an interior bulk event builds upon the fact that bulk locality arises in the large-$N$ regime with a large mass-gap \cite{Heemskerk:2009pn}. This is essential to understand local flat space-like features arising from boundary observables \cite{Susskind:1998vk, Polchinski:1999ry, Giddings:1999jq, Gary:2009ae, Gary:2009mi, Okuda:2010ym, Penedones:2010ue, Fitzpatrick:2011ia, Fitzpatrick:2011hu, Fitzpatrick:2011jn, Maldacena:2015iua, Paulos:2016fap, Hijano:2019qmi, Penedones:2019tng,  Komatsu:2020sag, Hijano:2020szl, Chandorkar:2021viw, Li:2021snj, Gadde:2022ghy,Banerjee:2022oll, Jain:2023fxc, vanRees:2023fcf, Alday:2024yyj, deGioia:2024yne, Duary:2024kxl, Navarro:2025xln, Alday:2025pmg,  Navarro:2026rna}. Understanding how bulk locality is encoded in
boundary observables is especially subtle behind the black hole horizon, where no single asymptotic
observer can both send and receive ordinary (future-directed) causal signals from the interior.

 \begin{figure}[t!]
\centering
\begin{subfigure}{0.45\textwidth}
\begin{tikzpicture}[scale=0.5]
\node (I)    at ( 4,0)   {};
\node (II)   at (-4,0)   {};
\path  (II) +(90:4)  coordinate (IItop) +(-90:4) coordinate (IIbot) +(0:4) coordinate (IIright);
\draw (IItop) -- (IIright) -- (IIbot)--cycle;
\draw[dashed, mred, thick]  (-4,4) ..controls (-1.0,3) and (1.0,3).. (4,4);
\path (I) +(90:4)  coordinate (Itop) +(-90:4) coordinate (Ibot) +(180:4) coordinate (Ileft);
\draw  (Ileft) -- (Itop) -- (Ibot) -- (Ileft) -- cycle;
\draw[dashed, mred, thick] (IIbot) to[out=+20,in=-200] (Ibot) ;
% fold of O_R(x),O_R(x') from the right, fold of O_L(y),tO_R(y') from the left, meeting at X in the future interior
\draw[mblue, thick] (4,-1.5) -- (0.05,2.6) to[out=200,in=-150] (0.05, 2.55) -- (4,-1.8);
\draw[mblue, thick] (-4,-1.5) -- (-0.05,2.6) to[out=-350,in=50] (-0.05, 2.55);
\draw[mblue, thick, dashed] (-0.2, 2.35) -- (-4,-1.8); 
\node[below] at (0,2.2) {$X$};
\filldraw (4,-1.5) circle (2pt); \filldraw (4,-1.8) circle (2pt); \filldraw (-4,-1.5) circle (2pt); \filldraw (-4,-1.8) circle (2pt); 
\node[right] at (4,-2.05) {$\mathcal O_{R}(x)$};
\node[right] at (4,-1.15) {$\mathcal O_{R}(x')$};
\node[left] at (-4,-1.15) {$\mathcal O_L(y)$};
\node[left] at (-4,-2.05) {$\tO_R(\by')$};
\end{tikzpicture}
   \caption{}
\end{subfigure}
\begin{subfigure}{0.45\textwidth}
\begin{tikzpicture}[scale=0.72]
\draw[mgray] (-1.60,0.00) -- (6.10,0.00); \node[right, mgray] at (6.10,0.00) {\footnotesize $0$};
\draw[mgray] (-1.60,-2.00) -- (6.10,-2.00); \node[right, mgray] at (6.10,-2.00) {\footnotesize $-\tfrac{\beta}{2}$};
\draw[mgray] (-1.60,-4.00) -- (6.10,-4.00); \node[right, mgray] at (6.10,-4.00) {\footnotesize $-\beta$};
\draw[->] (-1.6,-4.7) -- (-0.4,-4.7); \node[right] at (-0.4,-4.7) {\footnotesize ${\rm Re}\,t$};
% contour of the camera: x on segment 1, excursion down to the left line and back, x' on segment 2, left line, end
\draw[thick] (-1.2,0.08) -- (4.8,0.08) -- (4.8,-1.92) -- (4.5,-1.92) -- (4.5,-0.08) -- (-1.2,-0.08) -- (-1.2,-2.08) -- (5.3,-2.08) -- (5.3,-2.24) -- (-1.2,-2.24) -- (-1.2,-4);
\draw (-1.3,0.08)--(-1.1,0.08) (-1.3,0.16)--(-1.1,0.16);
\draw (-1.3,-4)--(-1.1,-4) (-1.3,-4.08)--(-1.1,-4.08);
\filldraw (0.4,0.08) circle (2pt); \node[above] at (0.4,0.08) {$\mathcal O_R(x)$};
\filldraw (0.6,-0.08) circle (2pt); \node[below] at (0.6,-0.08) {$\mathcal O_R(x')$};
\filldraw (4.65,-1.92) circle (2pt); \node[left] at (4.4,-1.58) {$\tO_R(\by')$};
\filldraw (4.5,-2.24) circle (2pt); \node[below] at (4.5,-2.24) {$\mathcal O_L(y)$};
\node[right] at (4.8,-0.7) {\footnotesize $e^{-\beta H/2}$};
\node[left] at (4.5,-0.7) {\footnotesize $e^{+\beta H/2}$};
\node[left] at (-1.2,-1.1) {\footnotesize $e^{-\beta H/2}$};
\node[left] at (-1.2,-3.1) {\footnotesize $e^{-\beta H/2}$};
\end{tikzpicture}
    \caption{}
\end{subfigure}
    \caption{(a) The two-sided observable which picks up a bulk point in the interior. The operator $\tO_R(\by')$ connected to $X$ by the dashed leg is analytically continued from the right using a simple Fourier transform. (b) The action of the Fourier transform on the top-fold of the otoc pushes it down where it pinches the operator on the left. }
    \label{fig: 1.2}
\end{figure}
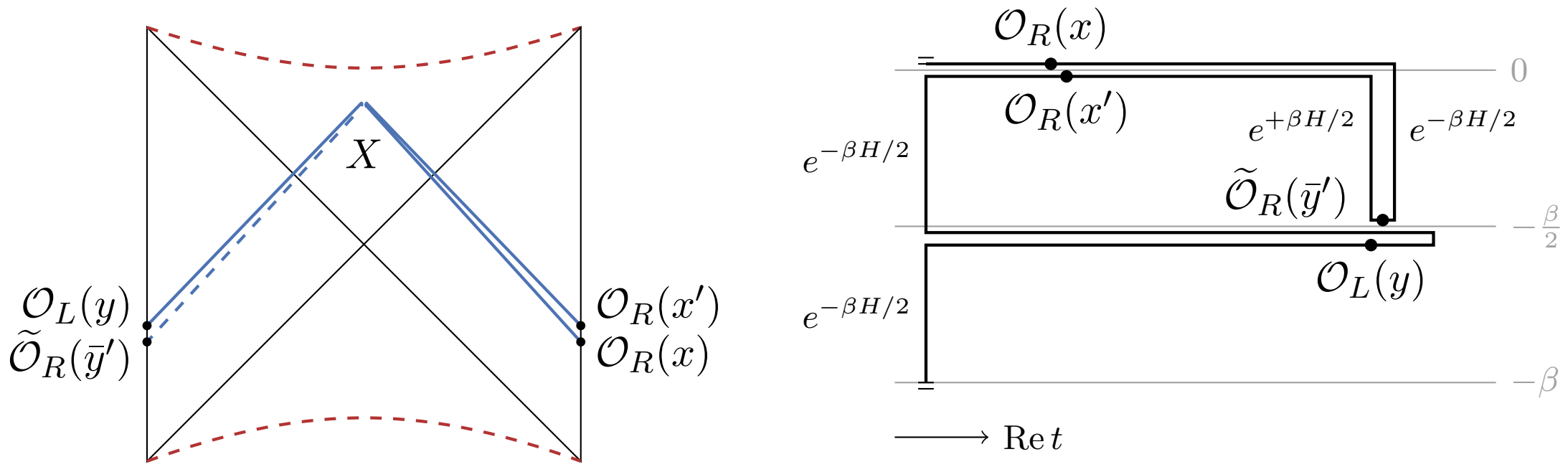

Our approach, following \cite{Caron-Huot:2025hmk, Caron-Huot:2025she, Caron-Huot:2022lff}, is complementary to different approaches to prepare the bulk using the boundary CFT, which go under the name of  bulk reconstruction. The HKLL program constructs perturbative bulk fields from boundary operators via causal reconstruction
\cite{Hamilton:2005ju,Hamilton:2006az,Hamilton:2006fh, Papadodimas:2012aq}. On similar lines, probes of the interior have also been built from two-sided or state-dependent operators \cite{Heemskerk:2012mn,Kourkoulou:2017zaj,deBoer:2018ibj,deBoer:2019kyr,Jafferis:2020ora,Leutheusser:2021qhd,Leutheusser:2021frk} (see also \cite{Kaushal:2025rsw, Kaushal:2025wbn} for Hamiltonian evolution across horizon), while  entanglement-wedge based techniques clarify how subregion reconstruction can extend beyond causal wedges \cite{Ryu:2006bv, Hubeny:2007xt, Almheiri:2014lwa, Engelhardt:2014gca, Dong:2016eik, Jafferis:2015del, Roy:2018ehv,Almheiri:2020cfm, Goto:2026mbj}. Both these approaches require us to keep track of extended regions on the boundary, while the techniques of \cite{Caron-Huot:2025hmk, Caron-Huot:2025she, Caron-Huot:2022lff} are measurements of local bulk physics instead of attempting to directly reconstruct entire bulk subregions using large enough boundary subregions. 

\subsection{Methods and results}
In this paper we design the interior observable by following the construction of \cite{Caron-Huot:2022lff, Caron-Huot:2025hmk, Caron-Huot:2025she}. This probes the bulk using directed wavepackets and nontrivial time-ordering of the operator insertions, using which we build our probe.  The starting point is motivated by the exterior holographic camera setup of \cite{Caron-Huot:2022lff}, where a smeared Fourier transform of a four-point out-of-time-ordered correlator (otoc) focuses the correlator onto a local bulk scattering event.  The smearing here implies that the Fourier transform is weighted by a Gaussian as in \eqref{eq:psi}. Related wavepacket probes of the bulk have been discussed in \cite{Terashima:2023mcr, Tanahashi:2026mia}.

The key idea here is that the smeared Fourier-transform of two closely separated pairs of boundary operator insertions within holographic cameras defines a light transform \cite{Kravchuk:2018htv} whose action is a geodesic folded along a Schwinger-Keldysh fold that propagates into the bulk.  In the high-energy limit, the corresponding wavepacket is
well described by geometric optics, and the transformed correlator provides a sharp signal of scattering about the bulk point
when the geodesics shot from boundary meet at the bulk point.  The boundary wavepackets provide bulk momentum, while the nontrivial operator ordering prevents the near-coincidence limit from collapsing into a product of two two-point functions supported in empty AdS; and instead it picks up a scattering configuration in the bulk at the bulk point.

Motivated from the holographic cameras setup in \cite{Caron-Huot:2022lff}, we propose the following observable for the eternal black hole interior:
\beq \label{eq:camera-def}
\mathcal G(x,p;y,q) = \int {\rm d}\delta x\,{\rm d}\delta y\ \psi_p(\delta x)\,\psi_q(\delta y)\ \langle\Psi|\ \mathcal O_L(y)\ \mathcal O_R(x')\ \tO_R(\by')\ \mathcal O_R(x)\ |\Psi\rangle ,
\eeq
The action of this smeared correlator is drawn in Fig \ref{fig: 1.2}. This is one way to pose a ``two-sided'' holographic camera. Here $|\Psi\rangle$ is the two-sided state, $\mathcal O_R(x)$ and $\mathcal O_R(x')$ are two nearby operators on the right boundary with relative position $\delta x=x'-x$, and $\mathcal O_L(y)$ is an operator on the left boundary. The fourth operator, $\tO_R(\by')=e^{\beta H_R/2}\,\tO_R(-t_{y'},\mathbf y')\,e^{-\beta H_R/2}$ with $\by'\equiv(-t_{y'}-\ic\beta/2,\,\mathbf y')$, is a right operator at complex time: the left operator at the point $y'$, with $\delta y=y-y'$, written as a right operator moved by half of the thermal circle, which is what it equals when it acts directly on the thermofield double state. The Gaussian Fourier kernels $\psi_p$ and $\psi_q$ of \eqref{eq:psi} act on the relative positions within the two pairs and give each pair a shooting momentum, $p$ on the right and $q$ on the left.

The way to read \eqref{eq:camera-def} is the following. In the exterior camera a pair of operators $\mathcal O(x),\mathcal O(x')$ sends a wavepacket of momentum $p$ into the bulk and a pair $\mathcal O'(y),\mathcal O'(y')$, inserted later and out of time order with respect to $\mathcal O(x),\mathcal O(x')$, detects a wavepacket of momentum $q$. The signal peaks when the two geodesics meet at a bulk point $X$, and it is the scattering amplitude at $X$. In the two-sided black hole the pair $\mathcal O_R(x),\mathcal O_R(x')$ does the same: its two legs enter the future interior through the right horizon and supply the modes in the past of an interior point $X$. 

The problem is what constitutes a detector which would collect stuff thrown in. Nothing comes out of the interior to either boundary, and operators on the two boundaries commute, so a pair of left and right operators can be rearranged into a time-ordered product. The leading connected contribution of the two folds cancels identically, and the correlator is blind to $X$. The operator $\tO_R(\by')$ fixes this. It is a right boundary operator, it does not commute with $\mathcal O_R(x)$ and $\mathcal O_R(x')$, and in \eqref{eq:camera-def} it stands between $\mathcal O_R(x)$ and $\mathcal O_R(x')$ on the timefold, where a left operator could never be. Its interaction with $\mathcal O_L(y)$ is therefore out of time order with respect to $\mathcal O_R(x)$ and $\mathcal O_R(x')$, like the pair $\mathcal O'(y),\mathcal O'(y')$ of the exterior camera, and it supplies the modes to the future of $X$. In the bulk its two legs follow the null geodesic that enters the future interior from the left boundary, one of which is an analytic continuation of the right exterior. The signal peaks when this geodesic meets the wavepacket of $\mathcal O_R(x),\mathcal O_R(x')$ at $X$, and it is the scattering amplitude at $X$, since the pair $\mathcal O_L(y),\tO_R(\by')$ can be thought of as a detector.

The construction depends on the state, in the sense of \cite{Papadodimas:2012aq,Papadodimas:2013jku}. The shift in $\tO_R(\by')$ is by half of the thermal circle of the right black hole. In the thermofield double state this is exactly $\beta/2$. In a state with a matter shell between the two exteriors, prepared by Euclidean evolutions of lengths $\beta_L$ and $\beta_R$ around a heavy insertion, it is $\beta_+/2$, with $\beta_+$ the inverse temperature of the black hole on the right of the shell. The operator that plays the detector thus has to be adjusted to the state before the camera can be used, which is what one expects of anything that sees behind a horizon \cite{Papadodimas:2012aq,Papadodimas:2013jku,Papadodimas:2015jra,Harlow:2014yoa}; we return to this in \S\ref{sec:discussion}. 

On the boundary side, the picture works via the Regge limit of the four-point function.  Conformal Regge theory organizes the correlator in the Regge limit in terms of Regge trajectories \cite{Cornalba:2006xm,Costa:2012cb,Caron-Huot:2017vep}.  In two-dimensional CFTs dual to planar BTZ, we motivate \eqref{eq:camera-def} by continuing an exterior Rindler camera along a complex path which moves the two boundary operator insertions at $y$ and $y'$ to the opposite Rindler wedge boundary. The point of this exercise is that the configuration remains a bulk-point configuration all along the path: the four null geodesics have a common intersection point, which moves from the exterior into the future interior. The signature of scattering is that the cross-ratios stay on the sheet of the bulk-point singularity \cite{Maldacena:2015iua, Chandorkar:2021viw}, while the Mandelstam variable $s$ of the exterior amplitude is continued through the complex plane. The endpoint is the exterior correlator with the operators at $y$ and $y'$ both shifted by half of the thermal circle; in the thermofield double state the one at $y$ equals the left operator $\mathcal O_L(y)$, and the endpoint is \eqref{eq:camera-def}.

The resulting observable is sensitive not only to the background interior geometry, but
also to localized objects placed behind the horizon. The legs of $\mathcal O_L(y)$ and $\tO_R(\by')$ reach the interior by different routes. The leg of $\mathcal O_L(y)$ is a propagator from the actual left boundary and sees the state as it is. The leg of $\tO_R(\by')$ is the analytic continuation of a right-boundary propagator and sees only the analytic continuation of the right exterior. An object crossed by one of the four legs shifts it, and the camera records the object as a discontinuity in the dependence of the signal on the shooting data, whenever a leg meets the object before reaching the collision point. We work this out for three cases: a heavy worldline falling in from one side, for timelike shells, which on the boundary are described by partially entangled thermal states \cite{Goel:2018ubv,Balasubramanian:2022gmo, Antonini:2023hdh} and in the bulk by gluing two black-brane geometries across a timelike hypersurface, and for null shocks. For the shell, by Birkhoff's theorem the right exterior, and hence the leg of $\tO_R(\by')$, knows nothing about it, while the leg of $\mathcal O_L(y)$ has to cross it to reach a bulk point on the right of the shell; a WKB probe crossing the shell undergoes refraction, where the coordinate energy changes by
redshift factors, and for a symmetric dust shell in BTZ we obtain the offset of the leg of $\mathcal O_L(y)$ in closed form.

We then analyze null shocks in two-sided black holes \cite{Dray:1984ha,Shenker:2013pqa,Shenker:2013yza,Roberts:2014isa}.
Unlike a timelike shell, a null shock acts by a universal shift of the conjugate null
coordinate.  A shock created by a real-time insertion is seen by every leg that crosses it, including the leg of $\tO_R(\by')$, since its interior portion is the analytic continuation of its exterior portion. Both legs of a fold then acquire the same time-delay phase, which cancels in the forward limit, and what survives is a displacement of the image by the time delay.
As the boundary shooting data are varied, the displacement records the presence and location of
the shock; in the non-forward version of the observable, the shock profile is read off directly from the phase per unit momentum transfer.

The construction is semiclassical and has important limitations.  The WKB approximation
requires sufficiently high energies and sufficiently narrow wavepackets directed almost radially inwards. The camera correlators can therefore detect coarse features of interior objects, but they should not be expected to reconstruct a black hole microstate at this semiclassical level of accuracy. Further, $\tO_R(\by')$ is not an ordinary real-time boundary operator, and in the thermofield double state the observable is an analytic continuation rather than an expectation value of real-time operators; the price for seeing behind the horizon is that the readout is supplied by a complex-time insertion. Backreaction restricts how much of the interior one may access.

The rest of the paper is organized as follows.  In \S \ref{sec:observables} we review exterior
holographic cameras and define the two-sided camera in the thermofield-double
geometry, and describe its WKB interpretation and the associated local flat-space
limit.  In \S \ref{sec:sca-sig} we motivate the observable by continuing an exterior Rindler camera into the interior: we follow the bulk point, the cross-ratios and the Regge signal along the path, explain what the continuation computes; the cross-ratio computation is in Appendix~\ref{app:cross-ratios}. In
\S \ref{sec:objects} we study how localized interior objects are recorded, going from heavy worldlines to null shocks and then to timelike
shells.  In \S \ref{sec:extensions} we discuss extensions to the past interior and the limitations from backreaction, and we conclude in \S\ref{sec:discussion}.

\section{Observables to image the interior}
\label{sec:observables}
Consider the four point correlator over the vacuum with spacelike separated points, i.e., in  the Euclidean signature. Using conformal invariance, we can write it in the standard form in terms of conformal cross ratios.
\beq \label{eq5.1}
 \langle\Omega| \, \mathcal O'(x_3) \, \mathcal O(x_2) \,\mathcal O'(x_4)\, \mathcal O(x_1) \, |\Omega\rangle =
 \frac{1}{x_{12}^{2\Delta_O}x_{34}^{2\Delta_{O'}}} \mathscr{G}(z,\bar{z})
\eeq
For spacelike separated operators time ordering is not important as they can commute past each other. Here the positional dependence of the correlator is captured by cross ratios. 
\beq \label{cratios}
 \frac{x_{12}^2 \ x_{34}^2}{x_{13}^2 \ x_{24}^2}\equiv z\, \bar{z},\qquad
 \frac{x_{14}^2 \ x_{23}^2}{x_{24}^2 \ x_{13}^2}\equiv(1{-}z)\,(1{-}\bar{z}). 
\eeq
With this choice of prefactor, $\mathscr G\to1$ when the pairs $(1,2)$ and $(3,4)$ become coincident, which is the limit relevant for the cameras below. Our interest lies in holographic CFTs. In the scalar sector, the operators with weights $\Delta_O$ and $\Delta_{O'}$ are dual to scalar fields in the bulk AdS$_{d+1}$ spacetime through the mass dimension relation.
\beq \label{eq5.2}
\Delta=\frac{d}{2}+\nu_{\Delta}, \qquad \nu_\Delta = \sqrt{{d^2 \over 4} + m^2 R_{\rm AdS}^2} \eeq
In our work, boundary positions and momenta will be denoted by lower-case labels $(x,p)$, while bulk positions and momenta will be denoted by upper-case labels $(X,P)$.  

\subsection{Cameras in the bulk exterior}
\label{sec:cam_ext}

Let us firstly recap the setting of \cite{Caron-Huot:2022lff} which uses directed wavepackets and nontrivial time ordering to probe bulk locality. We consider the following four-point correlator:
\beq \label{eq:ext-otoc}
\langle\Omega|\ \mathcal O'(y)\ \mathcal O(x')\ \mathcal O'(y')\ \mathcal O(x) \ |\Omega\rangle ,
\eeq
with two nearby insertions of $\mathcal O$ at the points $x,x'$ and two nearby insertions of $\mathcal O'$ at the points $y,y'$, which lie later in time. We write the relative displacements and the centers of the two pairs as
\beq \label{eq:pair-conventions}
\delta x = x'-x,\qquad \delta y = y-y',\qquad x_c=\tfrac12(x+x'),\qquad y_c=\tfrac12(y+y'),
\eeq
so that the two operators adjacent to the ket, $\mathcal O(x)$ and $\mathcal O'(y')$, sit at $x_c-\delta x/2$ and $y_c-\delta y/2$. When we number the four points we use $1,2,3,4$ for $x,x',y,y'$ in this order, which is the numbering of the cross-ratios \eqref{cratios}. With this Lorentzian configuration, the operator insertions describe an out-of-time-ordered correlator (otoc).  The operator insertions in the otoc are taken to lie on the following two-timefold contour:
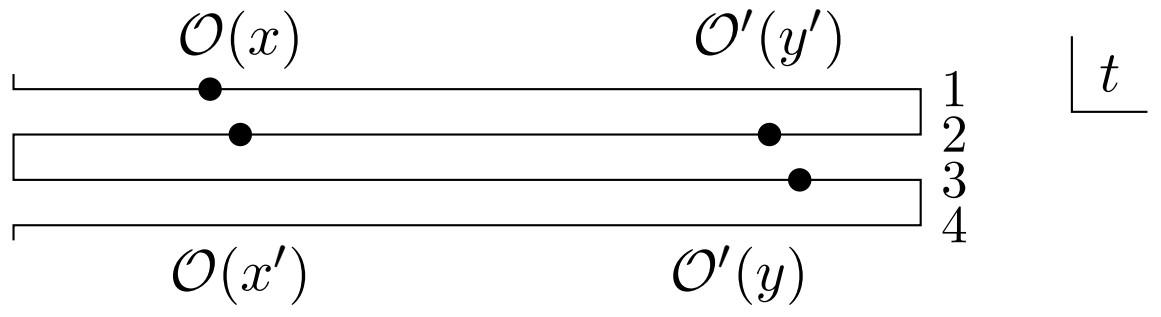
\beq
\centering
 \begin{tikzpicture}[scale=1.0]
 \node [right] at (0,-0.15) {$\quad$}; 
\draw  (-1,0.1)--(-1,0) -- (5,0) -- (5,-0.3) -- (-1,-0.3) -- (-1,-0.6) -- (5,-0.6) -- (5,-0.9) -- (-1,-0.9)--(-1,-1);
\node [above] at (0.5,0) {$\mathcal O(x)$}; 
\node [above] at (4,0) {$\mathcal O'(y')$};
\node [below] at (0.5,-0.9)  {$\mathcal O(x')$};
\node [below] at (3.8,-0.9)  {$\mathcal O'(y)$};
\node [right] at (5,0) {\small $1$}; \node [right] at (5,-0.3) {\small $2$}; \node [right] at (5,-0.6) {\small $3$}; \node [right] at (5,-0.9) {\small $4$};
\filldraw (0.3,0) circle (2pt); \filldraw (4,-0.3) circle (2pt); \filldraw (0.5,-0.3) circle (2pt); \filldraw (4.2,-0.6) circle (2pt);
\draw (6,0.35) -- (6,-0.15) -- (6.5,-0.15);
\node [above] at (6.25,-0.15) {$t$}; 
\end{tikzpicture} \label{fig1.2}
\eeq
The four point correlator has essentially a similar configuration as the chaos correlator in \cite{Maldacena:2015waa}, which takes the form $\langle[\mathcal O'(t), \mathcal O(0)]^2\rangle_\beta$. Here instead of taking the Regge limit as in \cite{Maldacena:2015waa} by going to late times, the holographic cameras of \cite{Caron-Huot:2022lff} works by taking the separations $\delta x$ and $\delta y$ very small. As we will see, this ensures that the cross ratios are small which describe physics in the eikonal limit. 

\subsubsection{Fourier transforms} 
We firstly perform some simple boundary operations that introduce bulk momenta which direct wavepackets towards a bulk point. To this end, we apply a couple of Fourier transforms which introduce shooting momenta $p$ and $q$ near the past boundary points $x$ and future boundary points $y$ respectively:
\beq
\begin{split}
\mathcal G(x,p; y,q) &= \int {\rm d}\delta x \int {\rm d}\delta y\   \psi_{p,\sigma}(\delta x)\, \psi_{q,\sigma}(\delta y) \ 
\langle\Omega|\ \mathcal O'(y)\ \mathcal O(x')\ \mathcal O'(y')\ \mathcal O(x) \ |\Omega\rangle ,
\end{split}
\eeq
at fixed centers $x_c$ and $y_c$, which we denote simply by $x$ and $y$ in the arguments of $\mathcal G$. Here the functions $\psi_p(\delta x)$ include the Fourier kernel along with a Gaussian smearing function. 
 \beq \label{eq:psi}
 \psi_{p}(\delta x) = \frac{{K}_\Delta}{\lc-{p} ^2\rc^{\Delta-\frac{d}{2}}} \exp \lc -\ic p.\delta x -  \frac{1}{2} \sigma^{-1}_{\mu\nu} \, \delta x^\mu \delta x^\nu \rc, \qquad \det \sigma > 0.
 \eeq
 The nontrivial operator ordering ensures that there is a genuine scattering process in the bulk when boundary wavepackets are directed appropriately at the bulk point $X$. This is opposed to the case where the operators with small separations OPE together near the boundary, and therefore the four-point correlator becomes a product of two two-point functions. 
 
The smearing allows us to take the coincidence limit in a controlled limit where the separations between the points systematically go to zero, i.e., $\delta x, \delta y \to 0$. Here we have labelled boundary wavepackets by their conserved boundary momentum $p_\mu$. We also choose ${K}_\Delta$ as follows to simplify the dependence of the two point correlators.
 \beq \label{kdel}
 K_\Delta = \frac{1}{2\pi}
 \frac{4^{\Delta}\Gamma(\Delta)\Gamma(\Delta-\tfrac{d-2}{2})}{(4\pi)^{\frac{d}{2}}}  
 \eeq
From the boundary perspective, the Fourier transforms along with the non-trivial time ordering ensure that the bulk lightcones are crossed in a way such that the cross ratios are folded from the Regge limit to hit the bulk point discontinuity. This is ensured when the boundary shooting momenta are directed appropriately at the bulk point and they locally add up to zero. Under such circumstances, the smeared correlator provides a sharp signal indicating flat-space scattering as a function of boundary shooting momenta.  

\subsubsection{Local physics: eikonal limit and retarded graviton exchange}
Before going into the boundary description, let us first describe what the holographic cameras do in the bulk, within Einstein gravity minimally coupled to a bulk scalar. The above-mentioned operations provide a simple bulk description of the smeared correlator in terms of light transforms along bulk Schwinger Keldysh folds (see for, instance the left panel of Figure \ref{fig:crr}) \cite{Caron-Huot:2022lff}, where the graviton exchange then take place between these two Schwinger Keldysh fold. 

Note that we have normalized the Fourier transform such that the two point contribution is unity. Then the light transform along the Schwinger-Keldysh fold extends into the bulk with its endpoints attached to the boundary, where the action of a single wavepacket takes the form:
\beq
 \int {\rm d}\delta x \, \psi_{p,\sigma}(\delta x) \,  \mathcal O_1(x- \tfrac{\delta x}{2}) \,\mathcal  O_2(x+ \tfrac{\delta x}{2}) \sim 1 + {\ic \over 2} \mathcal L_{x, p}[h_{uu}]
 \label{eq:light-transform}
 \eeq
 Here the term $1$ on rhs is the normalized two-point function, while the term $\mathcal L_{x, p}[h_{uu}]$ is the light transform which is integrated along the entire contour. This term arises from the bulk interaction, i.e. the trilinear coupling of the bulk field $\Phi$ (dual to $\mathcal O$) to the graviton fluctuation $h$. Each $\mathcal L_{x, p}[h_{uu}]$ has fields living on two contour segments, moving forward and back in time (visually represented in the left panel of Figure \ref{fig:crr}).
 
 In terms of the fields $h_i$ living on each contour segment $i$, the light transforms $L[h]$ are defined as follows
  \beq \label{ltransf2}
  \mathcal L_{x, p}[h_{uu}]\equiv \int_{-\infty}^{u_0} {\rm d}u \, (h_{uu,1}-h_{uu,2}), \quad \mathcal L'_{y, p}[h_{vv}]\equiv \int^{\infty}_{v_0} {\rm d} v \, (-h_{vv,2}+h_{vv,3}).
  \eeq
  The two Schwinger Keldysh contours anchored at the boundary endpoints $x$ and $y$ extend right upto the bulk point where exchange dominates. As in the left panel of Fig. \ref{fig:crr}, the high-energy description captures graviton exchange between two null Schwinger-Keldysh folds shot towards the bulk point from the boundary.
  
  \textbf{Folding to retarded exchange:} The connected contribution to the four point correlator is actually expressed in terms of the retarded exchange between the two Schwinger Keldysh folds. This is because the light ray operators $\mathcal L$ in \eqref{ltransf2} have differences of the fields lying on different contours. Insertions of difference fields behave as response, which in turn admits an expression in terms of the retarded graviton propagator $G^h_{\rm ret} (X,Y)$ between two points $X$ and $Y$ on each of the Schwinger-Keldysh folds. 
  
  For the graviton exchange, the connected part of the smeared four-point correlator  simplifies to:
 \beq \label{ltransf1}
  \mathcal G^{\rm con}_\Omega(x,p; y,q) \equiv -2\ic \langle\Omega| \, \mathcal L_{x,p} \, [h_{uu}] \,  \mathcal L'_{y,q}[h_{vv}] \,|\Omega\rangle .
  \eeq 
The non-trivial time-ordering and Fourier transforms implies that correlator describes retarded graviton exchange while still in the eikonal limit \cite{Cornalba:2006xk,Cornalba:2007zb,Camanho:2014apa,Caron-Huot:2022lff}. Note that the two light transforms in \eqref{ltransf2} involve different pairs of contour segments, $(1,2)$ and $(2,3)$. This is what turns the product of two difference fields into a retarded propagator, and it will be the crucial point below.

\subsection{Cameras in the two-sided interior}
\label{cam-2-side}
We are interested in the following question: does a setting exist where we can similarly pick up a bulk point in the black hole interior similar to how the smeared correlator picks up a bulk point in the exterior. The natural guess, motivated from the chaos correlator \cite{Maldacena:2015waa}, is to place one pair of operators on each boundary of the eternal black hole and to Fourier transform in the relative positions within each pair. Here at high energies, the description would become two Schwinger-Keldysh folds shot from the left and the right boundaries meeting at a bulk point in the interior. Since the kinematics is effectively that of two geodesics from either of the boundaries aimed at the bulk point, this offers a much simpler way to locate the bulk point by tuning the shooting momenta while looking for a sharp signal.

We will see in \S\ref{sec:interior-camera-def} that this guess fails for a simple reason, and how it is repaired by the observable \eqref{eq:camera-def}, which the \textit{holographic cameras} of \cite{Caron-Huot:2022lff} motivate: it has the same four insertions, with one of them moved to complex time.

\subsubsection{The two-sided eternal geometry}
\label{subsec:tfd-two-sided-geometry}
We recall the thermofield-double construction and its relation to the eternal black hole geometry \cite{Maldacena:2001kr}.  Let the CFT Hilbert space be $\mathcal H$, with Hamiltonian $H$ and energy eigenstates $H|n\rangle = E_n |n\rangle$. The thermal density matrix at inverse temperature $\beta$ is given by:
\begin{equation}
  \rho_{\beta} =  \frac{1}{Z(\beta)}e^{-\beta H},\qquad Z(\beta)=\operatorname{Tr}_{\mathcal H}e^{-\beta H}.
  \label{eq:tfd-thermal-density-matrix}
\end{equation}
The thermofield-double state is a purification of $\rho_\beta$ in the doubled Hilbert space $\mathcal H_L\otimes \mathcal H_R$, which corresponds to a tensor product of a left CFT and a right CFT. In the energy eigenbasis it takes the form:
\begin{equation}
  |\Psi_{\rm TFD}\rangle=  \frac{1}{\sqrt{Z(\beta)}} \sum_n e^{-\beta E_n/2} |n\rangle_{\rm L}\otimes |n\rangle_{\rm R} .
  \label{eq:tfd-state}
\end{equation}
This is an entangled state of the two CFTs. Tracing out the left Hilbert space gives the thermal density matrix on the right:
\begin{align}
  \rho_R&=  {\rm Tr}_{\mathcal H_L}|\Psi_{\rm TFD}\rangle\langle \Psi_{\rm TFD}| =\frac{1}{Z(\beta)}\sum_n e^{-\beta E_n}|n\rangle_{\rm R}\,\langle n|_{\rm R}
  =\frac{1}{Z(\beta)}e^{-\beta H_{\rm R}}.
  \label{eq:tfd-reduces-to-thermal}
\end{align}
Similarly, tracing out the right Hilbert space gives $\rho_L=e^{-\beta H_L}/Z(\beta)$.
Thus expectation values of right operators in the TFD state reproduce thermal expectation values with the canonical density matrix. The thermofield-doubled description gives a very natural representation of left boundary operators in terms of analytically continued right boundary operators acting upon the state. Operator insertions at left time $t_L$ is represented on the right boundary as follows:
\begin{equation}
 {\cal O}_{L}(t_L,\mathbf{x})|\Psi_{\rm TFD}\rangle
 =
 \widetilde{\cal O}_{R}\left(-t_L+{\mathrm{i}\beta/2},
 \mathbf{x}\right)|\Psi_{\rm TFD}\rangle 
  \qquad
  t_R=-t_L+\ic\beta/2 .
  \label{eq:tfd-left-right-kms}
\end{equation}
Here the tilde denotes the transpose of the operator as it is shifted halfway across the thermal circle.\footnote{For a Hermitian scalar in a theory with an antiunitary time-reversal symmetry, the matrix elements of $\mathcal O$ in the energy basis are real and $\tO=\mathcal O$ as an operator; we keep the tilde to signal the reversed real time. In particular $\tO_R(\by')$ in \eqref{eq:camera-def} is then simply the right operator $\mathcal O_R$ at the complex time $-t_{y'}-\ic\beta/2$.} This relation boils down to the fact that the two CFTs are physically separated by half of the Euclidean thermal circle.  The minus sign in the real time reflects the fact that the exterior Killing time runs in opposite directions on the two boundaries of the eternal black hole. Taking the adjoint, the same relation for an operator acting on the bra reads
\begin{equation}
 \langle\Psi_{\rm TFD}|\,{\cal O}_{L}(t_L,\mathbf{x})
 =
 \langle\Psi_{\rm TFD}|\,\widetilde{\cal O}_{R}\left(-t_L-{\mathrm{i}\beta/2},
 \mathbf{x}\right),
  \label{eq:tfd-left-right-kms-bra}
\end{equation}
so the sign of the imaginary shift depends on which side of the state the operator acts. We will use both forms. A right operator at a complex time $t-\ic\tau$ is understood as $\mathcal O(t-\ic\tau) = e^{\tau H}\mathcal O(t)e^{-\tau H}$, so that \eqref{eq:tfd-left-right-kms} and \eqref{eq:tfd-left-right-kms-bra} simply state that $\mathcal O_L$ and $e^{\mp\beta H_R/2}\,\widetilde{\mathcal O}_R\, e^{\pm\beta H_R/2}$ agree when acting directly on the state.

The bulk dual of $|\Psi_{\rm TFD}\rangle$ is the eternal two-sided AdS black hole, which is the maximal extension of the AdS black hole \cite{Maldacena:2001kr}.  The two CFT Hilbert spaces $\mathcal H_L$ and $\mathcal H_R$ describe the two asymptotic boundaries. The full pure TFD state describes both exteriors together with a connected interior. Geometrically the tracing out of Hilbert space on either side as in \eqref{eq:tfd-reduces-to-thermal} corresponds to restricting the eternal geometry to a single exterior region and tracing out the degrees of freedom behind the horizon.  

\subsubsection{The interior camera}
\label{sec:interior-camera-def}

Let us discuss a few issues here. One guess to construct an observable for the interior is the two-sided correlator of ordinary operators,
\beq \label{eq1.1}
\langle\Psi|\ \mathcal O_L(y)\ \mathcal O_R(x')\ \mathcal O_L(y')\ \mathcal O_R(x)\ |\Psi\rangle ,
\eeq
with the pair $x,x'$ on the right boundary and the pair $y,y'$ on the left, Fourier transformed in the relative positions as in \eqref{eq:psi}. It does not work unfortunately, as the left and right operators act on different factors of $\mathcal H_L\otimes\mathcal H_R$ and commute exactly. Therefore the ordering in \eqref{eq1.1} is only deceptive: the correlator equals $\langle\Psi|\mathcal O_L(y)\mathcal O_L(y')\mathcal O_R(x')\mathcal O_R(x)|\Psi\rangle$. This is an ordinary correlator inside the KMS strip, analogous to a time-ordered rather than an out-of-time-ordered correlator, and real-time Fourier transforms cannot change that. In the bulk language of \S\ref{sec:cam_ext} all four insertions sit on a single Schwinger-Keldysh contour, each pair produces the light transform $h_1-h_2$ of the same two branches, and $\langle(h_1-h_2)(X)\,(h_1-h_2)(Y)\rangle=G^{11}-G^{12}-G^{21}+G^{22}=0$ identically, so the retarded propagator which carries the exterior signal \eqref{ltransf1} never forms\footnote{The same fact is visible in the cross-ratios, which for one pair on each boundary never wind around $z=1$ and so stay on the sheet without the bulk-point singularity, as we check in \S\ref{sec:rindler-cross-ratios}. No boundary insertion lies to the future of an interior point, so no interaction vertex behind the horizon enters the Heisenberg evolution of a real-time boundary operator.}. 

What the exterior camera has and \eqref{eq1.1} lacks is a detector, a pair of operator insertions which can capture stuff coming out of $X$. Fortunately, the thermofield double state supplies one. Acting on the state, a left operator is a shifted right operator, \eqref{eq:tfd-left-right-kms} and \eqref{eq:tfd-left-right-kms-bra}. The camera \eqref{eq:camera-def} uses this representation.

Three of the four insertions are ordinary boundary operators: the pair at $x$ on the right and $\mathcal O_L(y)$ on the left. The fourth is a right operator continued to the left boundary by inserting it at a complex time. We use the notation of Fig \ref{fig: 1.2}: one pair of operators consists of $\mathcal O_R(x)$ and $\mathcal O_R(x')$, the other of the ordinary left operator $\mathcal O_L(y)$ and the right operator $\tO_R(\by')$ at complex time, where
\beq \label{eq:ybar-def}
\by' \equiv \big(-t_{y'}-\ic\tfrac{\beta}{2},\ \mathbf y'\big),\qquad \tO_R(\by') = e^{\beta H_R/2}\,\tO_R(-t_{y'},\mathbf y')\,e^{-\beta H_R/2},
\eeq
is the image under \eqref{eq:tfd-left-right-kms-bra} of the left point $y'$, and the displacements and centers of the pairs are those of \eqref{eq:pair-conventions}, with $\delta y$ measured on the left boundary and transported to the right by the map $t\to-t$. In terms of the smeared operators used below,
\beq \label{eq:camera-smeared}
\mathcal G(x,p;y,q) =  \langle\Psi|\ \mathcal O^\dagger_{y, q,\sigma_y}\ \mathcal O_{x, p,\sigma_x}\ \tO_{\by,q,\sigma_y}\ \mathcal O^\dagger_{x,p,\sigma_x}\ |\Psi\rangle\,,
\eeq
where the subscripts $x,y$ stand for the centers of the pairs. For the interior observable, $p$ denotes the momentum connected to the right boundary, while $q$ denotes the momentum connected to the left boundary. This is consistent with the exterior case, the momenta associated with the future points $q$ is now supplied by the left boundary. The numbering $1,2,3,4$ of \S\ref{sec:cam_ext} refers to $\mathcal O_R(x)$, $\mathcal O_R(x')$, $\mathcal O_L(y)$ and $\tO_R(\by')$ in this order, which is the operator ordering shared with the exterior camera \eqref{eq:ext-otoc}.

Using \eqref{eq:tfd-left-right-kms-bra} for $\mathcal O_L(y)$, the observable is the following thermal trace on the right CFT alone:
\beq \label{eq:camera-trace}
\mathcal G(x,p;y,q) = \frac{1}{Z}\,\operatorname{Tr}\Big[ e^{-\frac{\beta H_R}{2}}\ \tO_R(-t_y)\ e^{-\frac{\beta H_R}{2}}\ \mathcal O_R(x')\ e^{+\frac{\beta H_R}{2}}\ \tO_R(-t_{y'})\ e^{-\frac{\beta H_R}{2}}\ \mathcal O_R(x)\Big],
\eeq
with the smearings and the spatial arguments suppressed. In \eqref{eq:camera-trace} the segment between $\mathcal O_R(x')$ and $\tO_R(-t_{y'})$ runs backward by $\beta/2$: this is the contour drawn in Fig \ref{fig: 1.2}(b), with a Euclidean excursion attached to the right boundary at the time of $\tO_R(\by')$. 

It is useful to read \eqref{eq:camera-trace} as a sequence of operations on a single copy of the CFT, as in Fig \ref{fig: 1.2}(b). Starting from the right, insert $\mathcal O_R(x)$ on the right boundary; evolve by half of the thermal circle, which brings us to the left boundary, and insert $\tO_R(-t_{y'})$ there, which is the right-boundary way of writing a left insertion; evolve back by half of the thermal circle, against the direction of Euclidean evolution, to the right boundary and insert $\mathcal O_R(x')$; evolve again by half of the thermal circle to the left boundary and insert $\tO_R(-t_y)$, which is the ordinary left operator; finally evolve by the remaining half of the circle to close the trace.

\subsubsection{\texorpdfstring{Why $\tO_R(\by')$ sees the interior}{Why the operator at complex time sees the interior}}
\label{sec:why-shifted}

It is not the imaginary shift by itself that makes the difference between \eqref{eq:camera-def} and \eqref{eq1.1}, rather it is the position of $\tO_R(\by')$. A left operator commutes with $\mathcal O_R(x)$ and $\mathcal O_R(x')$ and can always be moved next to the other left operator. The operator $\tO_R(\by')$ is a right operator, it does not commute with $\mathcal O_R(x)$ and $\mathcal O_R(x')$, and in \eqref{eq:camera-def} it sits between them, where a left operator can never be. The pair $\mathcal O_L(y),\tO_R(\by')$ is therefore out of time ordered with respect to the pair $\mathcal O_R(x),\mathcal O_R(x')$, as in the exterior camera, and it plays the role of the detector: in the bulk, its two legs follow the null geodesic from the left boundary into the future interior and supply the modes to the future of $X$, while the two legs of $\mathcal O_R(x)$ and $\mathcal O_R(x')$ enter through the right horizon and supply the modes in its past. The configuration lies outside the KMS strip and, as we show in \S\ref{sec:rindler-cross-ratios}, on the sheet where $\bar z$ has wound around $1$.

Let us understand the physics of the observable in \eqref{eq:camera-def} in more detail. 

\begin{enumerate}
\item Three insertions are ordinary operators and only one is at complex time. The observable is therefore not an expectation value of real-time operators, which is what the argument above requires, but it is a well-defined object in any two-sided state. In the thermofield doubled state, it has a second description (which happens only there). The identity \eqref{eq:tfd-left-right-kms-bra} trades the left operator for a shifted right operator, and \eqref{eq:camera-trace} is then the exterior camera \eqref{eq:ext-otoc} of the right CFT with the times of both $\mathcal O'(y)$ and $\mathcal O'(y')$ continued by $-\ic\beta/2$, i.e. with their whole fold pushed down to the left-boundary line of Fig \ref{fig: 1.2}(b). We use this identity in \S\ref{sec:sca-sig} to compute the signal in the thermofield double state. In any other state the two objects differ, and \eqref{eq:camera-def} is the one we mean by the camera.

\item The operators $\mathcal O_L(y)$ and $\tO_R(\by')$ act on the bulk in different ways. The operator $\mathcal O_L(y)$ is an ordinary left operator; its bulk-to-boundary propagator is the propagator from the actual left boundary in whatever state we are given. The operator $\tO_R(\by')$ is a right operator; its propagator is the analytic continuation of the right bulk-to-boundary propagator through half of the thermal circle. In the thermofield doubled state these agree, while in a different state they need not. In fact, the mismatch will be our diagnostic for objects in the interior in \S\ref{sec:objects}.

\item The construction is state-dependent. The shift in \eqref{eq:ybar-def} is by half of the thermal circle of the right black hole: $\beta/2$ in the thermofield double state, and $\beta_+/2$ in a shell state, with $\beta_+$ the inverse temperature of the black brane on the right of the shell, which is fixed by the preparation data $\beta_L$, $\beta_R$ and the shell. The operator $\tO_R(\by')$ has to be adjusted to the state before the camera is pointed at the interior, in the sense of \cite{Papadodimas:2012aq,Papadodimas:2013jku}.

\item In frequency space the shift \eqref{eq:ybar-def} multiplies each Fourier mode of the real-time operator $\tO_R(-t_{y'},\mathbf y')$ by $e^{\pm\beta\omega/2}$, with the sign fixed by whether the mode raises or lowers the energy; this is the Poisson-kernel transform of \cite{Chakravarty:2025ncy}, and it is the sense in which the leg of $\tO_R(\by')$ in Fig \ref{fig: 1.2}(a) is obtained from the right boundary by a Fourier transform. The weight is what undoes the Boltzmann suppression that a real-time transform across half of the thermal circle would produce \cite{Chakravarty:2025ncy}.
\end{enumerate}

\subsubsection{Folded OPE in the interior}
\label{sec:folded-oscillators}

Before going into the boundary description, let us first describe what the holographic cameras do in the black hole interior, within the framework of Einstein gravity minimally coupled to a bulk scalar. 

The two smeared Fourier transforms in \eqref{eq:camera-def} fold the contour as shown in Fig.~\ref{fig: 1.2}. The pair at $x$ is a same-boundary pair with a small real displacement, as in the exterior camera, and the kernel $\psi_p(\delta x)$ folds the segment supporting $\mathcal O_R(x')$ towards the operator $\mathcal O_R(x)$. For the pair $\mathcal O_L(y),\tO_R(\by')$, using \eqref{eq:tfd-left-right-kms-bra} for $\mathcal O_L(y)$, both members are right operators at the same imaginary time $-\ic\beta/2$, separated by the small real displacement $\delta y$ transported to the right boundary. The pair therefore has the same light-cone nonanalyticity at $\delta y\to 0$ as a same-boundary pair, and the kernel $\psi_q(\delta y)$ selects it in the same way. The only difference from the exterior camera is that the operator $\mathcal O_R(x')$ stands between the two members of the pair, with a backward Euclidean segment on one side of it. This is the structure of the exterior otoc \eqref{fig1.2}, where $\mathcal O(x')$ stands between $\mathcal O'(y')$ and $\mathcal O'(y)$, now with the pair $\mathcal O'(y),\mathcal O'(y')$ moved by half of the thermal circle. At high energy, the two folded pairs
produce two directed bulk wavepackets each evolving a pair of operators, i.e., one associated with the right boundary and the other with the left boundary.

Let us focus on the pair $\mathcal O_L(y),\tO_R(\by')$.  Motivated from the exterior analysis, we can write the pair of insertions as follows:
\begin{equation}
  \mathcal B_{y,q}
  \equiv
  \int \diff^d\delta\,
  \psi_q(\delta)\,
  \mathcal O_L\!\left(y+\frac{\delta}{2}\right)
  \ \cdots\ 
  \tO_R\!\left(\by-\frac{\delta}{2}\right),
  \label{eq:two-sided-folded-bilocal}
\end{equation}
where $\delta^\mu$ is a small real displacement around this configuration, and the dots stand for the operator $\mathcal O_R(x')$ which sits between the two members. The
contour labels of these two operators are specified using
\eqref{eq:camera-def}. 

The Fourier transform probes the limit of small
\emph{real} displacement.  The Gaussian envelope in \eqref{eq:psi}
restricts the integral to a region whose widths are determined by the
wavepacket spread $\sigma_{\mu\nu}$.  The camera correlator in \eqref{eq:camera-def} works in the high frequency regime given by:
\begin{equation}
  |p|R_{\rm AdS}\gg1,
  \qquad
  |p|^2\sigma\gg1,
  \qquad
  \sigma\ll R_{\rm AdS}^2, \qquad \beta|p|\gg1 .
  \label{eq:two-sided-camera-limit}
\end{equation}
The second and third inequalities are best understood eigenvalue by eigenvalue, i.e.,
the eigenvalues of $\sigma_{\mu\nu}$ are large compared to $|p|^{-2}$, but small compared to the AdS length scale.  The first condition ensures that
the bulk propagation is well described by geometrical optics, and the last one that the wavepackets are short compared to the thermal scale.  The remaining
conditions ensure that the wavepacket is localized smaller than AdS length scales where its resolution is controlled by the widths $\sigma_{\mu\nu}$.

To see why the Fourier transform selects small separations, we perform a simple calculation. Consider a contribution to the correlator of the following form:
\begin{equation}
  F(\delta)
  \sim
  \frac{\mathcal A(\hat\delta)}{|\delta|^\alpha}.
  \label{eq:two-sided-small-delta-singularity}
\end{equation}
Note that this contribution is singular at the origin where the relative displacement $\delta$ goes to zero; in Lorentzian signature the singularity sits on the light cone $\delta^2\to0$, with the $\ic\epsilon$ prescription dictated by the operator ordering. Using the wavepacket \eqref{eq:psi}, its Fourier transform is schematically
\begin{align}
  \int \diff^d\delta\,
  \psi_p(\delta)\,F(\delta)
  &\sim
  \int \diff^d\delta\,
  \exp\!\left(
    -\ic p\cdot\delta
    -\frac{1}{2}
    (\sigma^{-1})_{\mu\nu}\delta^\mu\delta^\nu
  \right)
  \frac{\mathcal A(\hat\delta)}{|\delta|^\alpha}
  \nonumber\\
  &=
  |p|^{\alpha-d}
  \int \diff^d u\,
  \exp\!\left(
    -\ic\hat p\cdot u
    -\frac{1}{2|p|^2}
    (\sigma^{-1})_{\mu\nu}u^\mu u^\nu
  \right)
  \frac{\mathcal A(\hat u)}{|u|^\alpha}
  \label{eq:two-sided-fourier-rescaling}
\end{align}
where in the second line we have rescaled $\delta^\mu={u^\mu / |p|}$, and where $\hat p$ is a unit vector. In the regime $|p|^2\sigma\gg1$, the Gaussian factor in the second line is
approximately one for $u^\mu=O(1)$.  Therefore the singular part of the
integral is controlled by
\begin{equation}
  |\delta|\sim\frac{1}{|p|}.
  \label{eq:two-sided-delta-scale}
\end{equation}
Note that $  F(\delta)$ is a singular contribution here. Performing the Fourier transform we can also see that any contribution which is analytic in $\delta$ is exponentially suppressed by the rapidly oscillating Fourier kernel when $p_\mu\sigma^{\mu\nu}p_\nu\gg1$. 

\subsubsection{Regular contributions at the origin}

In contrast to these non-analytic terms at the origin $ \delta =0$, there is a different behaviour if the function $F(\delta)$ is analytic at the origin.
In the high-frequency regime, an analytic contribution  $F(\delta)$ in a finite
complex neighborhood of the real $\delta$ contour may be evaluated by
deforming that contour away from the origin right upto where we \textit{pinch} the nearest complex singularity.  The Fourier transform of such analytic contributions is then exponentially suppressed, with an exponent set by the distance to the nearest complex singularity.  In contrast, a nonanalyticity at  $\delta = 0$ obstructs this deformation and produces a power-law large-momentum tail as in \eqref{eq:two-sided-fourier-rescaling}.  Thus the high-frequency wavepacket selects the nonanalytic folded contribution over terms regular at the origin.  A more detailed discussion of this using specific examples is given in Appendix~\ref{app:fourier-asymptotics}.

\subsection{WKB approximation and flat space limit}
\label{sec:wkb}
Let us consider a translationally invariant black-brane background of the form
\begin{equation}
  \dd s^2
  =
  \frac{1}{z^2}
  \left(
    -f(z)\,\dd T^2
    +
    \frac{\dd z^2}{f(z)}
    +
    \delta_{ab}\,\dd X^a\dd X^b
  \right), \qquad f(z)=1-\frac{z^d}{z_h^d}
  \label{eq:wkb-black-brane-metric}
\end{equation}
where $z \to 0$ takes us near the boundary, and where the horizon is at $z = \zh$. The maximally extended geometry has two exterior right and left regions $z <\zh$, connected through future and past interior regions $z > \zh$. The analytic continuation in \eqref{eq:tfd-left-right-kms} is basically taking the boundary operator from one exterior to the other, where the two exterior boundary times are related by that relation.

We now describe the high-frequency approximation which uses the smeared boundary operators appearing in the camera observable \eqref{eq:camera-def}, and uses it to excite local bulk oscillator modes, i.e. creation and annihilation operators near a bulk point $X$ \cite{Caron-Huot:2025hmk}. This follows since using the GKPW dictionary, a CFT correlator may be written in terms of bulk Witten diagrams. In the high-frequency regime, the external legs of these diagrams are bulk-to-boundary propagators, while internal legs are bulk-to-bulk propagators.  

\subsubsection{The wave equation at high frequency}

Let us consider the external legs first which are related to the scattering process using the bulk-to-boundary propagators. The equation satisfied by the bulk-to-boundary  propagator is same as that of equation of motion of a scalar field, which requires only the minimally-coupled quadratic part of the action. This is given as follows:
\begin{equation}
  \left(\Box - m^2\right)\Phi  = 0,
\qquad
  \Box \Phi=
  \frac{1}{\sqrt{-g}}
  \partial_M
  \left(
    \sqrt{-g}\,
    g^{MN}\partial_N\Phi
  \right).
  \label{eq:covariant-laplacian} 
\end{equation}
For the planar BTZ black hole the solutions are hypergeometric functions, while in higher dimensions they are Heun functions, which we bypass here using the high-frequency limiting case\footnote{The restriction to the phase in \eqref{eq:wkb-bulk-boundary-propagator} in the high energy limit provides a more practical perspective, which is also physically intuitive. The exact solutions are needed only to fix the normalization constants by matching near the boundary, where the geometry is empty AdS.}.  The limit we take is the one of large frequency, where each bulk-to-boundary propagator has the optical form \cite{Caron-Huot:2025hmk, Caron-Huot:2025she} (see also \cite{Dodelson:2023nnr, Dodelson:2023vrw}):
\begin{equation}
  \langle \Phi(X)\,\mathcal O_p^\dagger\rangle_{\Psi}
  =
  \mathcal C_{\Delta,p}^{\pm}
  \sqrt{\mathcal D(p;X)}
  \exp\!\left(\ic S(p;X)\right).
  \label{eq:wkb-bulk-boundary-propagator}
\end{equation}
Here $S(p;X)$ is the Hamilton-Jacobi phase associated with a null geodesic shot from the boundary with momentum $p$ and ending at the bulk point $X$.  The factor $\mathcal D(p;X)$ is the Van Vleck-Morette or the one-loop determinant which basically controls the spread of the wavepacket. Here $\mathcal C_{\Delta,p}^{\pm}$ is a normalization fixed by matching with the near-boundary two-point function.  The sign $\pm$ records the choice of whether the boundary insertion lies to the past or future of the bulk point along the relevant timefold. When the boundary energy is much larger than the boundary spatial momenta, i.e. $\omega\gg|\vec p|$, the geodesics are almost radial, which is the case we mostly use below.

\subsubsection{Local physics near the bulk point}
\textbf{Null geodesics and local bulk momenta:} The phase $S(p;X)$ is highly useful, it is the Legendre transform of the Hamilton-Jacobi function $S(x;X)$ (i.e. the on-shell action evaluated for the geodesic), where $x^\mu$ is the boundary endpoint of the geodesic and $X^M$ is the bulk point \cite{Caron-Huot:2025hmk, Caron-Huot:2025she}.
\begin{equation}
  S(p;X)
  =
  p_\mu x^\mu
  -
  S(x;X),
  \label{eq:wkb-legendre-transform}
\end{equation}
Here the boundary momentum conjugate to $x^\mu$ is $p_\mu$.  This is a highly useful function, for instance the boundary position $x$ and the bulk momenta $P$ at the bulk point $X$ can be extracted using derivative of the generating function $S(p;X)$. These take the following form:
\begin{equation}
  x^\mu(p;X)
  =
  \frac{\partial S(p;X)}{\partial p_\mu},
  \qquad
  P_M(p;X)
  =
  \frac{\partial S(p;X)}{\partial X^M}.
  \label{eq:wkb-canonical-relations}
\end{equation}
The expression for $P_M(p;X)$ defines the bulk covariant momentum $P_M$ at the point $X$.  Note that this is the same momenta which appears while taking $\lambda$ as an affine parameter along the null geodesic, using which the bulk momenta is given by $P_M  = g_{MN} \frac{\dd X^N}{\dd\lambda}$. The reduction for null geodesics for a massless scalar is then given by:
\begin{equation}
  g^{MN}(X)P_M P_N=0.
  \label{eq:wkb-null-condition}
\end{equation}

Let us now employ cross ratios $z$ and $\bar z$ as defined in \eqref{cratios}. These cross ratios are a function of the boundary coordinates, which in turn are a function of bulk point $X$ using \eqref{eq:wkb-canonical-relations}. From the boundary perspective, we can show that the local bulk momenta at $X$ takes the following compact form in terms of the cross ratio difference $\delta z = z -\bar{z}$:\cite{Chakravarty:2025ncy}:
\beq \label{bulk mom formula}
P_{jM} = \lc \frac{p_{1\mu} \xi^\mu}{\frac{\partial (\delta z)}{\partial ( \delta x_1^\nu )} \xi^\nu }\rc  \frac{\partial (\delta z)}{\partial ( \delta x_j^\mu)} \ \frac{\partial x_j^\mu}{\partial X^M}.
\eeq
where $\xi$ is an arbitrary vector on the boundary, and $j$ denotes the external leg while $M$ is the bulk Lorentz index.  This follows from a simple chain rule of derivatives.

\subsubsection{Plane wave phases}

Taking the WKB approximation at high energy allows us to conveniently solve the eom, and the Hamilton-Jacobi form described above allows us to extract plane-wave like phases in the vicinity of the bulk point $X$. This is because when we expand the Hamilton-Jacobi phase around a bulk point $X$, it obeys a simple relation.
\begin{equation}
  S(p;X+\delta X)
  =
  S(p;X)
  +
  P_M(p;X)\,\delta X^M
  +
  O(\delta X^2).
  \label{eq:wkb-phase-expansion}
\end{equation}
This simple property leads to the case where the bulk-to-boundary propagator near $X$ takes a simple expanded form:
\begin{equation}
  \langle \Phi(X+\delta X)\,\mathcal O_p^\dagger\rangle_{\Psi}
  =
  \mathcal C_{\Delta,p}^{\pm}
  \sqrt{\mathcal D(p;X)}
  \exp\!\left(
    \ic S(p;X)
    +
    \ic P_M(p;X)\delta X^M
  \right)
  \left(1+O(\delta X^2)\right).
  \label{eq:wkb-local-plane-wave}
\end{equation}
We can now use this form to obtain the flat space limit and extend it to the black hole interior.

\subsubsection{Local flat space physics: kinematics} 
The boundary wavepacket propagates to the bulk point by following the geodesic, where its description is mentioned earlier. In a sufficiently small neighborhood of $X$, the bulk wavepacket looks like a local plane wave with momentum $P_M(p;X)$.

To understand the flat space limit more precisely, we work with a basis of local bulk oscillators near $X$. We choose the convention of \cite{Caron-Huot:2023vxl, Caron-Huot:2023ikn}: let $a_{X,P}$, $a^\dagger_{X,P}$ denote local modes in the past of $X$, and let $b_{X,P}$, $b^\dagger_{X,P}$ denote local modes in the future of $X$. Now locally, a quantum field can be expressed as a bunch of these oscillators which gives the following identity:
\begin{equation}
  \langle \Phi(X+\delta X)\,a^\dagger_{X,P}\rangle
  =
  \exp\!\left(\ic P_M\delta X^M\right),
  \label{eq:wkb-local-oscillator-normalization}
\end{equation}
Here we have used the standard one-particle normalization. 

We are now in a position to relate the boundary Fourier transformed operators to bulk emission and absorption operators. Comparing \eqref{eq:wkb-local-plane-wave} with \eqref{eq:wkb-local-oscillator-normalization} gives the following dictionary relating boundary emission operators to bulk emission operators:
\begin{equation}
 \mathcal  O_p^\dagger|\Psi\rangle
  \simeq
  \sqrt{\mathcal D(p;X)}
  \exp\!\left(\ic S(p;X)\right)
  \times
  \begin{cases}
    \mathcal C^+_{\Delta,p}\,
    a^\dagger_{X,P}|0\rangle,
    &
   \mathcal  O_p^\dagger \text{ lies in the past of } X,
    \\[0.5em]
    \mathcal C^-_{\Delta,p}\,
    b^\dagger_{X,P}|0\rangle,
    &
    \mathcal O_p^\dagger \text{ lies in the future of } X.
  \end{cases}
  \label{eq:wkb-boundary-bulk-dictionary}
\end{equation}
Here $|0\rangle$ denotes the local flat-space vacuum in the tangent-space approximation near
$X$. This is the plane wave limit of the general wavepacket dictionary found in \cite{Caron-Huot:2025hmk, Caron-Huot:2025she}.  A similar dictionary holds for annihilation operators, with the smeared boundary operator mapped to the corresponding annihilation oscillator. 

\textbf{Exterior cameras:} Applying the WKB approximation to the cameras reduces the
problem to geodesics shot from the boundary to the bulk point. Within the more general scattering we denote the local bulk momenta associated with the two legs of the $x$-fold by $P_1$ and $P_2$, corresponding to the insertions at $x$ and $x'$ respectively.

The cameras work in the forward scattering limit $P_1 \to P_2$. The two smeared boundary operators in the folded pair at $x$ source two bulk wavepackets near the bulk point, one associated with emission and the
other with absorption.  This gives the schematic form:
\begin{equation}
  \mathcal B_{x,p}
  \ \longrightarrow\
  a^\dagger_{X,P}\,a_{X,P},
  \label{eq:bilocal-to-two-oscillators}
\end{equation}
where $P$ is the local bulk momenta associated with the two WKB legs of the folded pair. Note that \eqref{eq:bilocal-to-two-oscillators} is a schematic representation of the
process: it suppresses the WKB phases, normalization factors, and the
contour dependence of the emission and absorption operators. 

The same argument applies to the second folded pair, defined by the
relative displacement
\begin{equation}
  \delta y=y-y'
  \label{eq:left-folded-relative-displacement}
\end{equation}
of \eqref{eq:pair-conventions}. It produces a second pair of local bulk oscillator insertions near the
bulk point, where we denote the local bulk momenta associated with the two legs of the $y$-fold by $Q_3$ and $Q_4$, corresponding respectively to the insertions at $y$ and $\by'$, following the numbering of \S\ref{sec:interior-camera-def}. Thus, we have the following momenta labels:
\begin{equation} 
x\leftrightarrow P_1, \qquad x'\leftrightarrow P_2, \qquad y\leftrightarrow Q_3, \qquad \by'\leftrightarrow Q_4. 
\end{equation}
Since the camera works in the forward limit, we have $Q_4 = Q_3$.

\textbf{Interior cameras:} Using the flat-space limit discussed above, the two folded transforms in the interior camera produce four local bulk oscillator insertions near the interaction point $X$ behind the horizon. The two legs of the $x$-fold are right bulk-to-boundary propagators continued through the future horizon along infalling null geodesics, as in \cite{Chakravarty:2025ncy}. For the pair $\mathcal O_L(y),\tO_R(\by')$, the leg of $\mathcal O_L(y)$ is the propagator from the actual left boundary into the future interior, while the leg of $\tO_R(\by')$ is the analytic continuation of the right propagator through half of the thermal circle. In the thermofield doubled state these two legs are the same function, since the maximally extended geometry is the analytic continuation of either exterior, and near $X$ both are local plane waves with the same momentum $Q$. The four-point camera observable therefore factorizes into a flat-space-like local scattering process near $X$, along the lines of \cite{Caron-Huot:2025hmk,Caron-Huot:2025she}. Each smeared boundary insertion supplies one external leg of this local scattering process. 

The wavepacket supplied by $\mathcal O_L(y)$ and $\tO_R(\by')$ physically moves into the interior with a future-directed momentum $Q^{\rm phys}$, since nothing comes out of the interior to a boundary detector. In the correlator, however, the pair $\mathcal O_L(y),\tO_R(\by')$ plays the role that $\mathcal O'(y),\mathcal O'(y')$ played in the exterior camera, namely that of a detector, so its leg enters the local amplitude with momentum $Q=-Q^{\rm phys}$. We denote by $\tilde a_{X,Q}, \tilde a^\dagger_{X,Q}$ the local modes supplied by $\mathcal O_L(y)$ and $\tO_R(\by')$, so that the correlator \eqref{eq:camera-def} factorizes onto the local amplitude
\beq \label{eq:local-amplitude-camera}
\langle\, \tilde a_{X,Q_3}\ a_{X,P_2}\ \tilde a^\dagger_{X,Q_4}\ a^\dagger_{X,P_1}\,\rangle ,
\eeq
with the corresponding local momenta satisfying
\begin{equation} \label{eq:local-momentum-conservation}
P_1+Q_4 = P_2+Q_3 . 
\end{equation} 
As discussed before for the exterior case, the interior camera considered here further specializes to the forward-eikonal configuration: 
\begin{equation} 
P_1\to P_2, \qquad Q_4\to Q_3. 
\end{equation} 
Thus, the Fourier-transformed boundary operators are tuned so that their associated null geodesics meet at $X$, while their local momenta satisfy momentum conservation there. In the forward limit the conservation condition is automatic, and the sharpness of the signal comes from the requirement that the two geodesics actually meet, i.e. from the vanishing impact parameter. In \S\ref{sec:contact} we relax the forward identifications while retaining local momentum conservation, thereby obtaining a more general four-wavepacket observable that probes nonforward local scattering behind the horizon.

\subsubsection{Retarded graviton exchange in eikonal limit}
\label{sec:retarded-exchange}
With this discussion of basic kinematics, let us now move onto the physics of the graviton exchange. Starting with the Regge limit, the foldings take us to a situation similar to \cite{Caron-Huot:2022lff} where the physics is given by the exchange of a retarded graviton, mediated by the retarded graviton propagator. 

This can be drastically simplified in the flat-space like limit discussed above near the point $X$ where we have pointed both the two boundary geodesics at $X$. Recall that in flat space, the $t$-channel amplitude for a spin$-j$ exchange goes roughly as follows:
\beq
\mathcal M(s,t) \sim {s^{j} \over t}
\label{eq:grav_exchange}
\eeq
Prior to folding, in the language of \cite{Chandorkar:2021viw}, this is causally-Regge configuration which is dominated by exchange in the Regge limit. As described above upon Fourier transform, the bilocal operators OPE together, and we encounter folding to a bulk point configuration where the bulk exchange is dominated by the retarded propagator \cite{Caron-Huot:2022lff}. In the language of \cite{Chandorkar:2021viw}, this is causally-scattering configuration.

In the exterior cameras setup, the geodesic approximation for the correlator has maximum support where the Schwinger Keldysh contours almost intersect and thus give the flat space limit. In the eikonal approximation, this amounts to the following contribution \cite{Caron-Huot:2022lff}, written in terms of the eikonal phase of the local graviton exchange:
\beq \label{eq:eikonal-exterior}
\mathcal G_{\rm con}(x,p; y,q) \approx -\ic\,\delta(s,b), \qquad \delta(s,b) = 
 \begin{cases}
 \dfrac{4\pi G_{\rm N}\, C_d\, s}{b^{d-3}}; \qquad \, \, \, \, d\neq 3\\[2mm]
 -2 G_{\rm N}\, s \log b; \qquad d = 3
 \end{cases}, \quad \text{where} \quad C_d= \frac{\Gamma(\tfrac{d-3}{2})}{4\pi^{\frac{d-1}{2}}} 
\eeq
Here $C_d/b^{d-3}$ is the Fourier transform of $1/(-t)=1/q_\perp^2$ to the $(d-1)$-dimensional transverse impact parameter space, the impact parameter $b$ and the center of mass energy $s$ are given by:
\beq \label{sb}
b =    \lc g_{\mu \nu}\Delta X^{\mu} \Delta X^{\nu} \rc^{\frac{1}{2}} ; \qquad s = -2 g_{\mu \nu}P^{\mu} Q^{\nu}, 
\eeq
with $\Delta X$ the transverse separation between the two geodesics at closest approach, and with both quantities evaluated using the local momenta at the bulk point. In AdS$_3$ the transverse space is one-dimensional and curved on the scale of the image, and the flat-space kernel is replaced by the exact transverse propagator $\tfrac12 e^{-|\phi|}$ which we obtain in \S\ref{sec:sca-sig}.

For the interior cameras the structure is identical: in the thermofield doubled state the contour \eqref{eq:camera-trace} has the same two-timefold form as \eqref{fig1.2}, so the leading connected contribution is again the retarded exchange \eqref{eq:eikonal-exterior}, evaluated with the local momenta at the interior bulk point. The one difference is kinematical. The leg supplied by $\mathcal O_L(y)$ and $\tO_R(\by')$ enters with $Q=-Q^{\rm phys}$, and since the eikonal phase $\delta(s,b)$ is linear in $s=-2g_{\mu\nu}P^\mu Q^\nu$, this flips the sign of the signal,
\beq \label{eq2.15}
\mathcal G_{\rm con}\approx-\ic\,\delta(s,b)\Big|_{Q=-Q^{\rm phys}}=+\ic\,\delta(s_{\rm phys},b),\qquad s_{\rm phys}=-2g_{\mu\nu}P^\mu Q^{{\rm phys}\,\nu}>0,
\eeq
where $s_{\rm phys}$ is the center of mass energy of the physical collision of the two infalling wavepackets. It is positive, as it should be: both wavepackets fall into the interior as ordinary future-directed wavepackets. The reversed momentum only records the role of $\mathcal O_L(y)$ and $\tO_R(\by')$ in the correlator, where they act as the detector of the exterior camera, and its one effect is the sign of the signal, which is opposite to that of the exterior camera. We confirm this sign flip explicitly in \S\ref{sec:rindler-camera-signal}, and interpret it in \S\ref{sec:what-continuation-computes}. No Boltzmann factors appear, as explained in \S\ref{sec:why-shifted}.

\subsection{Observables where contact interactions dominate}
\label{sec:contact}
We now consider an extension of the interior camera which probes a generic
local scattering amplitude rather than its forward eikonal limit.  This is
obtained by independently smearing the four external operator insertions (and is similar to the non-forward discussion of \cite{Chakravarty:2025ncy} but now the correlator is supported on both boundaries).
For suitable local kinematics, a bulk contact interaction can dominate over
graviton exchange.  For simplicity, we consider a real massless scalar field
$\Phi$ propagating on the background geometry $g_{MN}$, with action
\begin{equation}
  S_{\Phi}
  =
  \int {\rm d}^{d+1}X\,\sqrt{-g}\,
  \left(
    -\frac{1}{2}g^{MN}\partial_M\Phi\,\partial_N\Phi
    -
    \frac{\lambda}{4!}\Phi^4
  \right).
  \label{eq:massless-phi4-action-local}
\end{equation}
Here we have a contact interaction term in the bulk Lagrangian. A simple dimensional analysis tells us that the contact interaction will dominate over the graviton exchange (produced by linearization of the Einstein-Hilbert action) in the following perturbative limit:
\beq
{\mathcal M}_{\rm grav}(s,t) \ll {\mathcal M}_{\rm contact} (s,t)\quad  \implies \quad G_{\rm N}\, s^{\frac{d-1}{2}}\, {s \over |t|} \ll \lambda\, s^{d-3 \over 2} \ll 1,
\label{eq:condition_contact}
\eeq
where $s$ and $t$ are Mandelstam variables evaluated locally about the bulk point $X$, and where we have made all quantities dimensionless using $[G_{\rm N}]=1-d$ and $[\lambda]=3-d$. As we have discussed, these variables are controlled in turn by the boundary shooting momenta. Let us systematically study how to extract this case here.

\subsubsection{Interior modes and scattering experiment}
\begin{figure}[t!]
    \centering
\begin{tikzpicture}[x=0.75pt,y=0.75pt,yscale=-1.5,xscale=1.5]
%uncomment if require: \path (0,432); %set diagram left start at 0, and has height of 432

%Straight Lines [id:da0004243024950587593] 
\draw    (419,107) -- (571,262) ;
%Straight Lines [id:da7508642819472269] 
\draw    (570,108) -- (419,261) ;
%Curve Lines [id:da44156152232657775] 
\draw [thick]   (573,244) .. controls (576,222) and (624,209) .. (626,247) ;
\draw  [dashed, thick]  (573,143) .. controls (576,167) and (624,175) .. (626,143) ;
%Curve Lines [id:da8192625246566455] 
\draw[thick]    (425,244.1) .. controls (428,221.2) and (377,219.1) .. (377,249.1) ;
\draw  [dashed, thick]  (425,143) .. controls (428,167) and (377,175) .. (377,143) ;
%Straight Lines [id:da9952277553569533] 
\draw [thick, mblue!50  ,draw opacity=1 ]   (509,143.13) -- (492,157) ;
%Straight Lines [id:da3483122642895101] 
\draw [thick, mblue,   decoration={markings,mark=at position 0.5 with \arrow{<}},postaction=decorate ,draw opacity=1 ]   (492,157) -- (551,207) -- (580,232) ;
%Straight Lines [id:da7703577705042084] 
\draw [thick, mblue, dashed,  decoration={markings,mark=at position 0.5 with \arrow{<}},postaction=decorate ,draw opacity=1 ]   (492,157) -- (464,186.98) -- (421,233.1) ;
%singularities: following two lines
%Curve Lines [id:da020377881959623534] 
\draw [thick,  dashed, color={rgb, 255:red, 210; green, 48; blue, 48 }  ,draw opacity=1 ]   (427,105) .. controls (471,133) and (521,134) .. (561,104) ;
%Curve Lines [id:da9162005172270737] 
\draw [thick, dashed, color={rgb, 255:red, 222; green, 27; blue, 27 }  ,draw opacity=1 ]   (428,262) .. controls (468,232) and (528,233) .. (558,263) ;
%Straight Lines [id:da41494779422121253] 
\draw [thick, mblue!50  ,draw opacity=1 ]   (470,144) -- (492,157) ;
%Straight Lines [id:da23299761126656782] 
\draw[dashed]    (518,141) -- (498,141) (484,141.13) -- (464,141.13) ;
%Curve Lines [id:da41724592799531446] 
\draw [thick, mblue!50, decoration={markings,mark=at position 0.75 with \arrow{>}},postaction=decorate ,draw opacity=1 ]   (470,144) .. controls (471,138.13) and (474,142.13) .. (486,148.13) ;
%Curve Lines [id:da7235658474407228] 
\draw [thick, mblue!50  ,decoration={markings,mark=at position 0.7 with \arrow{>}},postaction=decorate ,draw opacity=1 ]   (509,143.13) .. controls (508,137.13) and (496,148.13) .. (491,151.13) ;
%Straight Lines [id:da38377813024339014] 
\draw [thick, mblue!50  , decoration={markings,mark=at position 0.5 with \arrow{>}},postaction=decorate ,draw opacity=1 ]   (503,156.73) -- (606,222.88) ;
%Straight Lines [id:da2967276672653598] 
\draw [thick, mblue!50  , decoration={markings,mark=at position 0.5 with \arrow{<}},postaction=decorate ,draw opacity=1 ]   (403,226.88) -- (482,157.88) ;

% Text Node
\draw (490,163.1) node [anchor=north west][inner sep=0.75pt]  [font=\small] [align=left] {$X$};
% Text Node
\draw (387,241) node [anchor=north west][inner sep=0.75pt]  [font=\footnotesize] [align=left] {$\displaystyle H_{L}$};
% Text Node
\draw (602,239) node [anchor=north west][inner sep=0.75pt]  [font=\footnotesize] [align=left] {$\displaystyle H_{R}$};

\end{tikzpicture}
 \caption{The non-forward correlator \eqref{eq:blandcorrelator}. The four smearings prepare local modes with momenta $P_1,P_2$ from $\mathcal O_R(x),\mathcal O_R(x')$ and $Q_3,Q_4$ from $\mathcal O_L(y)$ and dashed $\tO_R(\by')$, the latter drawn at its mirror point on the left, which meet and conserve momentum at the interior point $X$.}
    \label{mainfig000}
\end{figure}
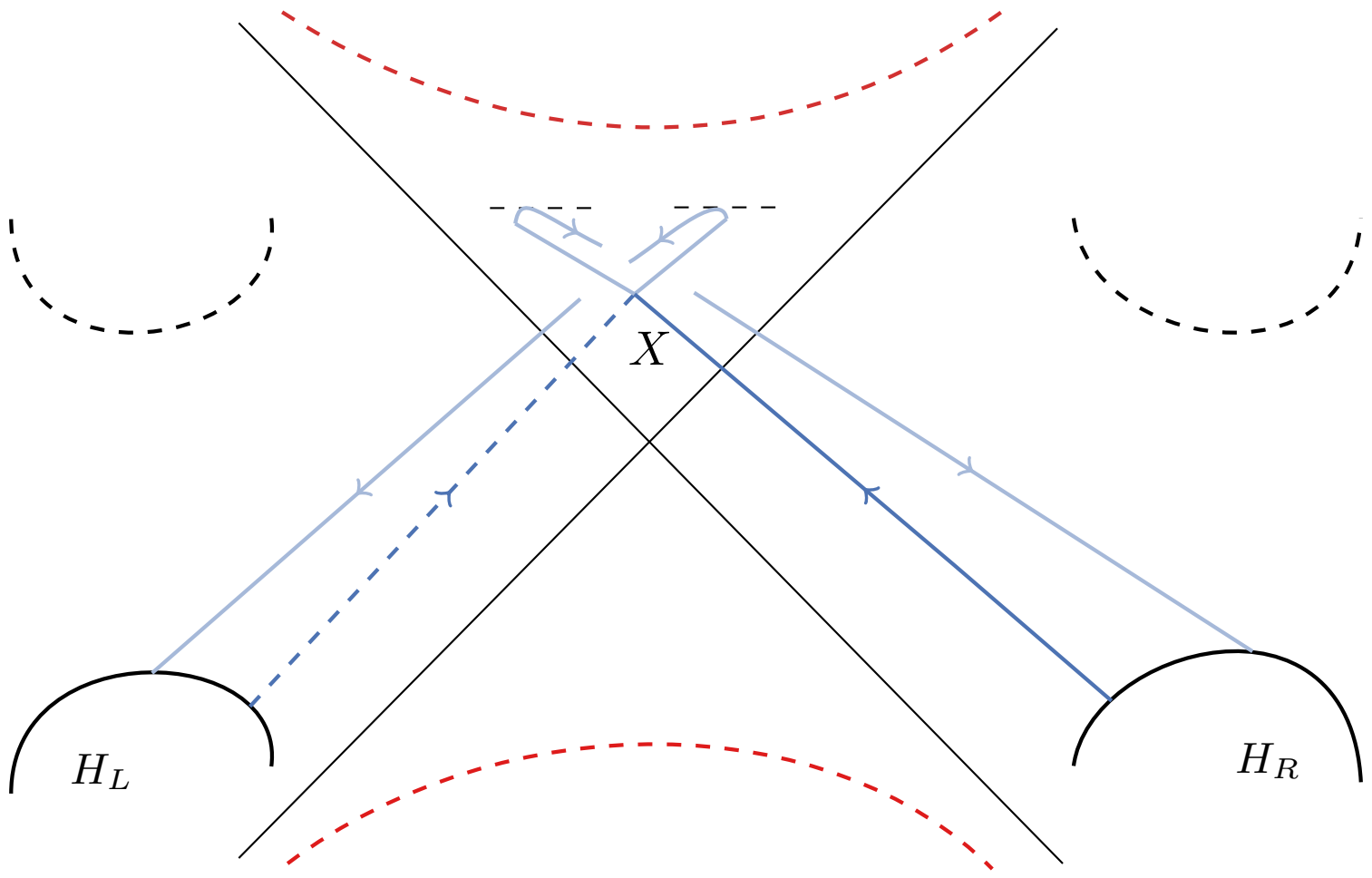

Let us systematically build up to the interior scattering experiment in Fig. \ref{mainfig000}, where our goal is to design a scattering experiment which gives us a flat space like amplitude near $X$. We use smeared Fourier transforms to which introduce directed wavepackets \cite{Caron-Huot:2025hmk}:
\begin{equation} \label{psips}
    \tilde \psi_{p,\sigma}(\delta x) =
    \exp(-\ic p_\mu \delta x^\mu - \frac{1}{2} \sigma^{-1}_{\mu\nu} \, \delta x^\mu \delta x^\nu), \qquad \det \sigma > 0,
\end{equation}
The above wavepacket $\tilde \psi$ is same as the wavepacket $\psi$ introduced earlier, we use different labels only to indicate that we have simply stripped off the numerical prefactor. The four smeared operators are the ordinary ones,
\beq \label{finform}
\mathcal O_{x,p,\sigma} = \int \diff^d\delta x\ \tilde\psi^*_{p,\sigma}(\delta x)\ \mathcal O(x+\delta x), \qquad \mathcal O^\dagger_{x,p,\sigma} = \int \diff^d\delta x\ \tilde\psi_{p,\sigma}(\delta x)\ \mathcal O(x+\delta x),
\eeq
on either boundary, together with the smeared version of the right operator at complex time,
\beq \label{eq:smeared-shifted}
\tO_{\by,q,\sigma} = \int \diff^d\delta y\ \tilde\psi^*_{q,\sigma}(\delta y)\ \tO_R(\by+\delta y), \qquad \by = \big(-t_y-\ic\tfrac{\beta}{2},\mathbf y\big),
\eeq
where, as in \eqref{eq:camera-def}, the displacement $\delta y$ is transported from the left boundary to the right by $t\to-t$. As in \S\ref{sec:interior-camera-def}, the operator \eqref{eq:smeared-shifted} is the right-boundary representation \eqref{eq:tfd-left-right-kms-bra} of a smeared left operator, inserted where it is not adjacent to the state\footnote{The reader may wonder whether $\tO_R(\by')$ can be traded for a real-time smearing of right operators, as in the Poisson-kernel transform of \cite{Chakravarty:2025ncy}. It cannot, for the reason explained in \S\ref{sec:interior-camera-def}: such a transform is a linear combination of real-time insertions and reproduces \eqref{eq:tfd-left-right-kms} only when it acts directly on the state. The transform of \cite{Chakravarty:2025ncy} is a convenient way of writing a left operator as a right operator when it is adjacent to the state, which is the situation of the radar experiment there, but not of $\tO_R(\by')$ here.}. We can also have other variants of this experiment involving different correlators. The main restriction for the most general scattering amplitude involving a boundary correlator is that on the bulk timefold near the point $X$, we require at least one past operator and one future operator for a scattering experiment to take place. This also allows us to easily generalize to $n$-point correlators. For instance we consider the correlator: 
\beq 
\langle \Psi | \ \mathcal O_L(y_3) \ \mathcal O_R(x'_2)\ \tO_R(\by'_4) \ \mathcal O_R(x_1) \ | \Psi \rangle 
\label{eq:blandcorrelator} 
\eeq 
with the four points numbered as in \S\ref{sec:interior-camera-def}. This is basically the correlator \eqref{eq:camera-def} with the replacements
$$x\to x_1, \qquad x'\to x'_2, \qquad y\to y_3, \qquad y'\to y'_4,$$ 
and with the four insertions smeared independently. Upon smearing \eqref{eq:blandcorrelator} with wavepackets $\tilde \psi/\tilde \psi^*$ given in \eqref{psips} appropriately, this correlator describes processes as in Fig. \ref{mainfig000} with an emission and absorption operator on each side. 

Performing the four Fourier transforms on each of the operator locations in \eqref{eq:blandcorrelator}, we introduce the following emission and absorption modes: 
\beq 
\label{2sidedmap} 
\begin{split} 
\mathcal O_R(x_1) &\mapsto \mathcal O^\dagger_{x_1,p_1,\sigma_1}, \qquad \mathcal O_L(y_3) \mapsto \mathcal O^\dagger_{y_3,q_3,\sigma_3}, \\ \tO_R(\by'_4) &\mapsto \tO_{\by_4,q_4,\sigma_4}, \qquad \mathcal O_R(x'_2) \mapsto \mathcal O_{x_2,p_2,\sigma_2}. 
\end{split} 
\eeq 
As explained in \S\ref{sec:why-shifted}, $\tO_R(\by'_4)$ carries no Boltzmann suppression.

\subsubsection{Factorization into flat space like scattering near bulk point}

Let us now develop the story better for contact interaction. Here for convenience, we can work with a real massless bulk scalar field $\Phi$ with the action given in \eqref{eq:massless-phi4-action-local}. 
At leading order in $\lambda$, the connected four-point function receives the contact Witten
diagram contribution
\begin{equation} 
\mathcal G^{(4)}_{\rm contact}(x_1,x'_2,y_3,\by'_4) = -\ic\lambda \int {\rm d}^{d+1}X\,\sqrt{-g(X)}\, K(X;x_1)K(X;x'_2)K(X;y_3)K(X;\by'_4), 
\label{eq:phi4-contact-witten-diagram} 
\end{equation}
where $K(X;x_i)$ is the bulk-to-boundary propagator for the massless scalar. Performing a Fourier transform and working in the high frequency WKB limit, we obtain a similar relation for the propagators \cite{Caron-Huot:2025hmk, Caron-Huot:2025she}:
\begin{equation}
  K(X;p)
  \sim
  \mathcal C
  \sqrt{\mathcal D(p,X)}
  \exp\!\left(\ic S(p,X)\right),
  \label{eq:wkb-bulk-to-boundary-leg}
\end{equation}
For the leg of $\tO_R(\by'_4)$, $K(X;\by'_4)$ is the analytic continuation of the right propagator through $-\ic\beta/2$, which in the thermofield doubled state is the left propagator from the point $y'_4$. Using the above description, we can find the bulk point using four geodesics shot in from the boundary. Hence locating the bulk point governed by four Hamilton-Jacobi functions $S_i(p_i,X)$ and $S_j(q_j,X)$, which provide the phases in \eqref{eq:wkb-bulk-to-boundary-leg}. This again leads to a dictionary between smeared boundary operators acting on the state and local excited modes acting on the flat space vacuum as in \eqref{eq:wkb-boundary-bulk-dictionary}.

In the bulk, using these locally excited modes acting on the flat space vacuum excited using smearing operations,  the correlator \eqref{eq:blandcorrelator} factorizes to the following local flat-space like amplitude about the bulk point:
\beq
 \langle  \, \tilde a_{X,Q_3} \, a_{X,P_2} \, \tilde a^\dagger_{X,Q_4}\,   a^\dagger_{X,P_1} \,   \rangle
\eeq
This allows us to write down a factorization formula for the smeared four-point correlator in terms of the amplitude: 
\beq \label{eq:factorization-contact}
\begin{split} 
\langle\Psi| \ & \mathcal O^\dagger_{y_3,q_3,\sigma_3}\ \mathcal O_{x_2,p_2,\sigma_2}\ \tO_{\by_4,q_4,\sigma_4}\ \mathcal O^\dagger_{x_1,p_1,\sigma_1} |\Psi\rangle \\ &\approx \int \qty[\prod_{i=1}^{2}\diff P_i\,\tilde\psi_i] \int \qty[\prod_{j=3}^{4}\diff Q_j\,\tilde\psi_j]\, \sqrt{-g}\,(2\pi)^{d+1}\, \delta^{d+1}\qty(\textstyle{P_1-P_2+Q_4-Q_3})\, \ic\cM(\{P_i,Q_j\}), 
\end{split} 
\eeq 
where the measure involves the momentum-space wavepackets $\tilde\psi_i(P_i)$ of \cite{Caron-Huot:2025hmk}, and where for our case $\ic\cM=-\ic\lambda$ following the action in \eqref{eq:massless-phi4-action-local}. The momenta $Q_j$ of the legs of $\mathcal O_L(y_3)$ and $\tO_R(\by'_4)$ enter with the opposite sign, $Q_j = -Q_j^{\rm phys}$, as explained in \S\ref{sec:wkb}.

\subsubsection{Forward limit} 
The original two-sided camera is recovered in the forward limit
\begin{equation}
  P_1\to P_2,
  \qquad
  Q_4\to Q_3.
  \label{eq:contact-forward-limit}
\end{equation}
The momentum transfer is then small, and the contact-dominance condition
\eqref{eq:condition_contact} need not hold.  In a holographic theory, the
leading connected contribution in this regime is instead the eikonal
graviton exchange discussed above.  Thus the two-sided camera is the
forward-scattering limit of the more general four-wavepacket local
scattering experiment.

Using this, the factorization formula for the correlator takes the following form:
\beq \label{eq:factorization-forward}
\begin{split}
\langle\Psi| \ &  \mathcal O^\dagger_{y, q,\sigma_y}\ \mathcal O_{x, p,\sigma_x}\ \tO_{\by,q,\sigma_y}\ \mathcal O^\dagger_{x,p,\sigma_x}\ |\Psi\rangle\\
 & \approx \int \qty[\prod \diff P_i\,\tilde \psi_i] \int \qty[\prod \diff Q_j\,\tilde \psi_j]  \sqrt{-g} \, (2\pi)^{d+1}\ \delta^{d+1}\qty(\textstyle{P_1 - P_2 + Q_4 - Q_3})\, \ \ic \cM(\{P_i, Q_j\}), \\
\end{split}
\eeq
where $\ic \cM$ is essentially the graviton exchange \eqref{eq:grav_exchange} in the limit $P_1 \to P_2$ and $Q_4 \to Q_3$. This is the factorization formula in the coincidence limit. Since $\mathcal M\sim s^2/t$ is singular in the forward limit, the formula is to be understood after the transform to impact parameter space, which produces the eikonal phase \eqref{eq:eikonal-exterior}. The factorization basically states that the primary contribution arises from the exchange involving the local Minkowski space retarded graviton propagator. 

 \section{Pushing the bulk point into the interior}
 \label{sec:sca-sig}
This section motivates the observable \eqref{eq:camera-def} from the exterior camera. We start from an exterior scattering process in AdS-Rindler, which is planar BTZ, and continue the insertion times of $\mathcal O_R(y)$ and $\mathcal O_R(y')$ of the exterior camera along a complex path which moves them to the opposite boundary. The point we want to establish is that the configuration remains a bulk-point configuration along the whole path: the four null geodesics keep a common intersection, which moves from the exterior into the future interior; the cross-ratios keep the relative winding which puts the exterior camera on the sheet of the bulk-point singularity; a contact Witten diagram keeps its bulk-point singularity; and the leading Regge signal stays finite and focused. The endpoint of the path is the exterior correlator with both $\mathcal O_R(y)$ and $\mathcal O_R(y')$ shifted by half of the thermal circle, i.e. replaced by $\tO_R(\by)$, $\tO_R(\by')$. It is not by itself a two-sided observable. In the thermofield double state, and only there, $\tO_R(\by)$ equals the left operator $\mathcal O_L(y)$ by \eqref{eq:tfd-left-right-kms-bra}, and the endpoint is the camera \eqref{eq:camera-def}. Since this identity is exact, the calculation below is a calculation of the camera itself in the thermofield double state, within the leading Regge approximation; in any other state it only motivates the camera as the right object to consider. In \S\ref{sec:objects} we use the camera in other states, where its interior signal rests on the bulk description of \S\ref{sec:wkb}.
\subsection{Basic conformal Regge theory}
\label{sec:crt}

Let us now understand the behaviour of the correlator on the second sheet in the Regge limit. To go to the Regge limit we need to cross lightcones which takes one of the cross ratios crosses over to the second sheet in which they approach $z, \bar{z} \to 0$. This is described by conformal Regge theory where the four point correlator is understood as sum over exchanges of Regge trajectories \cite{Costa:2012cb, Costa:2023wfz}. 

Subtracting off contributions from the two point correlators, the function $\mathscr{G}(z,\bar{z})$ in \eqref{eq5.1} takes the following form.
\beq \label{eq5.3}
\mathscr{G}(z,\bar{z})-1\approx 2\pi \ic  \int_{0}^{\infty} \frac{\diff \nu}{2\pi} \, \rho(\nu) \,  \alpha(\nu) \lc z \bar{z} \rc^{\frac{1-J(\nu)}{2}} \, \mathcal{P}_{\frac{2-d}{2}+\ic\nu} \lc \tfrac{z+\bar{z}}{2\sqrt{z\bar{z}}} \rc
 \eeq
Following the notation of \cite{Caron-Huot:2022lff}, the various symbols on the rhs of \eqref{eq5.3} are explained below.
 \begin{enumerate}
     \item The measure $\rho (\nu)$ in the integral is given by
     \beq \label{rhonu}
      \rho(\nu) =\frac{2^{d-2}}{\Omega_{d-1}}
      q_{\frac{2-d}{2}+ \ic \nu}q_{\frac{2-d}{2}- \ic \nu} \quad\mbox{where}\quad q_J\equiv \frac{(d-2)_J}{(\tfrac{d-2}{2})_J}\,
     \eeq
     where $(x)_n$ denotes the Pochhammer symbol $(x)_n=\Gamma(x+n)/\Gamma(x)$.
     \item The symbol $\mathcal{P}_J(\eta)$ is the Gegenbauer polynomial for integer $J$ and is given by the following hypergeometric function.
     \beq \label{pjnu}
    \mathcal{P}_J(\eta) = {}_2F_1\big({-}J,d{-}2+J,\tfrac{d-1}{2},\tfrac{1-\eta}{2}\big)\,.
     \eeq
     \item The quantity $\alpha(\nu)$ can be derived by studying the Regge behaviour of the corresponding conformal partial wave expansion. 
 \end{enumerate}
In two boundary dimensions, which is the case we work out explicitly below, the harmonic function reduces to $\mathcal P_{\ic\nu}(\cosh\phi)=\cos(\nu\phi)$, and the half-line integral with the measure $\rho(\nu)$ combines into an integral of $e^{\ic\nu\phi}$ over the whole real line with no additional factor. This is the form in which the thermal CFT$_2$ camera is written in eq.~(4.4) of \cite{Caron-Huot:2022lff}, with the same $\widetilde\alpha(\nu)$ as in general $d$, and it is the form we use in \S\ref{sec:rindler-camera-signal}.

\subsubsection{The graviton Regge trajectory}

Consider for instance, the graviton exchange in $AdS_{d+1}$ between two bulk scalar fields with weights $\Delta_O$ and $\Delta_{O'}$. Here $\alpha(\nu)$ which is a function of the Reggeon spin $J(\nu)$ takes the form \cite{Costa:2012cb, Caron-Huot:2022lff}:
     \beq \label{alpha2}
      \alpha_{J=2}(\nu) = -\frac{8\, G_{\rm N} \, R_{\rm AdS}^{1-d}}{\nu^2+\tfrac{d^2}{4}} \, 
    \frac{\gamma_{2\Delta_O+1}(\nu) \,\gamma_{2\Delta_{O'}+1}(\nu)}{\Gamma(\Delta_O) \,\Gamma(\Delta_O-\frac{d-2}{2}) \,\Gamma(\Delta_{O'}) \,\Gamma(\Delta_{O'}-\frac{d-2}{2})}\ 
     \eeq
     The poles in $\gamma$ are the double twist exchanges, while the pole at $\ic \nu = \frac{d}{2}$ indicates the stress tensor exchange. In the strong coupling limit where we have a supergravity description in the bulk, $\alpha_{J=2}(\nu)\sim e^{-\pi|\nu|}$ at large $\nu$. 
     
Applying the Fourier transforms upon \eqref{eq5.3} turns it into its momenta space counterpart, which is given below.
\beq
\mathcal G(x,p; y,q)-1\approx -\ic  \int_{0}^{\infty} \frac{\diff \nu}{2\pi} \, \rho(\nu) \,  \widetilde{\alpha}(\nu) \lc \abs{p} \abs{q} \rc^{J(\nu)-1} \, \mathcal{P}_{\frac{2-d}{2}+\ic \nu} \lc \hat p . \, \mathcal{I}_{x-y} . \hat q  \rc e^{-{\nu^2 \over 2}\sigma_0^2}
\label{eq:mom_space_crt}
\eeq
This takes the form of an impact parameter amplitude given in \cite{Cornalba:2006xm, Meltzer:2017rtf, Kulaxizi:2017ixa}. We now explain the above formula term by term.
\begin{enumerate}
    \item In the argument of the Gegenbauer polynomial, the hatted quantities such as $\hat p$ denote unit vectors. The quantity $\mathcal{I}_{x-y}$ denotes a reflection in the $x-y$ direction.
    \beq
    \mathcal{I}_{x-y} \equiv \mathcal{I}_{xy}^{\mu \nu} = \eta^{\mu \nu} - {2 (xy)^\mu (xy)^\nu \over (xy)^2}
    \eeq
    \item The quantity $\sigma_0$ arises as the contribution from the Gaussian smearing. For a diagonal matrix $\sigma_{\mu \nu}$, this takes the form:
    \beq \label{sigmapq}
\sigma_0^2 = {1 \over \sigma_{p}^2 p^2}  + {1 \over \sigma_{q}^2 q^2}
   \eeq
where $\sigma_p,\sigma_q$ denote the widths of the two wavepackets. This serves to tune the camera to give a sharp signal in our regime of interest $\sigma_t \abs{  p_t}> 1, \,\sigma_x \abs{ \vec p} >1$.
 
    \item The subtraction of unity on lhs denotes the contribution from the two point correlator after Fourier transformation.
    \item The term $\widetilde{\alpha}(\nu)$ is given by
    \beq \label{alpha3}
\widetilde{\alpha}(\nu) = - \alpha(\nu) \frac{2\pi}{4^{J(\nu) -1}} \, 
    \frac{\Gamma(\Delta_O) \,\Gamma(\Delta_O-\frac{d-2}{2}) \,\Gamma(\Delta_{O'}) \,\Gamma(\Delta_{O'}-\frac{d-2}{2})}{\gamma_{2\Delta_O+J(\nu)-1}(\nu) \,\gamma_{2\Delta_{O'}+J(\nu)-1}(\nu)}\ 
    \eeq
\end{enumerate}
Substituting \eqref{alpha2} in \eqref{alpha3}, we find a huge simplification wherein most factors cancel. The power of the Fourier transform is to cancel out all double trace exchanges sitting inside the gamma functions. For the graviton exchange, we find a very simple expression. 
\beq \label{alpha4}
\widetilde{\alpha}(\nu) = {4 \pi G_{\rm N} R^{1-d} \over \nu^2 +{d^2 \over 4}}
\eeq
 The Fourier transformed amplitude essentially depends on the curvature of the leading Regge trajectory. As in \cite{Caron-Huot:2022lff}, we can parametrize this behaviour in terms of an effective variance dictating the sharpness of the four point correlator $\mathcal G(x,p; y,q)$.
\beq
\sigma^2_{\rm eff} \gtrsim \sigma_0^2 + \frac{\log\big(|p||q|\,R_{\rm AdS}^2\big)}{\left(\Delta_{\rm gap}^{\rm higher-spin}\right)^2}
\eeq
where the last term is the transverse broadening coming from the curvature of the leading Regge trajectory, which for $\mathcal N=4$ super Yang-Mills at strong coupling is the term $\log(\ldots)/\sqrt\lambda$ in eq.~(3.29) of \cite{Caron-Huot:2022lff}. At strong coupling, the Regge trajectory is almost flat \cite{Costa:2012cb}, and hence we obtain a sharp signal since $\Delta_{\rm gap}^{\rm higher-spin}$ is very large. For small coupling, $\Delta_{\rm gap}^{\rm higher-spin}$ is small and hence we start observing effects of higher spin operators. This implies that the camera is only able to resolve bulk objects larger than the string scale, where the broadening effect kicks in.

\subsection{Continuing the exterior camera along a complex path}
\label{sec:rindler-continuation}
\label{sec:planar-btz}

\begin{figure}[t]
    \centering
    \begin{subfigure}{1.0\textwidth}
        \centering
        \begin{tikzpicture}[scale=1.0, xscale=2.0]
            \draw
                (-1.0,2.25) -- (-1.0,-2.75)
                (1,-2.75) arc (0:360:1 and 0.3)
                -- (1,2.25) arc (0:360:1 and 0.3);
            \draw[dashed]
                (1,-1.25) arc (0:360:1 and 0.3)
                (1,0.5) arc (0:360:1 and 0.3);
            \draw[mgray,thick]
                (1,1.2) -- (0,-0.3)
                (0,-0.9) -- (1,1.2)
                (-1,1.2) -- (0,-0.3)
                (0,-0.9) -- (-1,1.2)
                (1,-2.3) -- (0,-0.3)
                (0,-0.9) -- (1,-2.3)
                (-1,-2.3) -- (0,-0.3)
                -- (0,-0.9) -- (-1,-2.3);
            \draw[mred,thick,->]
                (.9,-0.1)
                .. controls (0.7,-0.6) and (-0.3,-1.2) ..
                (-0.9,-1.25);
            \draw[mblue,thick]
                (1,-1.25) -- (0.5,-0.5)
                to[out=200,in=-150] (0.5,-0.55)
                -- (0.97,-1.36)
                (1,0.5) -- (0.5,-0.44)
                to[out=200,in=-150] (0.5,-0.49)
                -- (0.99,0.39);
            \node[right] at (1,-1.25) {$Q$};
            \node[right] at (1,0.39) {$P(0)$};
            \node[left] at (-1,-1.25)
                {$P(\beta/2)$};
            \node[right] at (2,0)
                {$\mapsto \qquad$};
        \end{tikzpicture}
        \begin{tikzpicture}[scale=1.0, xscale=2.0]
            \draw
                (-1.0,2.25) -- (-1.0,-2.75)
                (1,-2.75) arc (0:360:1 and 0.3)
                -- (1,2.25) arc (0:360:1 and 0.3);
            \draw[dashed]
                (1,-1.25) arc (0:360:1 and 0.3)
                (1,0.5) arc (0:360:1 and 0.3);
            \draw[mgray,thick]
                (1,1.2) -- (0,-0.3)
                (0,-0.9) -- (1,1.2)
                (-1,1.2) -- (0,-0.3)
                (0,-0.9) -- (-1,1.2)
                (1,-2.3) -- (0,-0.3)
                (0,-0.9) -- (1,-2.3)
                (-1,-2.3) -- (0,-0.3)
                -- (0,-0.9) -- (-1,-2.3);
            \draw[mblue,thick]
                (1,-1.25) -- (0.02,0.45)
                to[out=200,in=-150] (0.02,0.4)
                -- (0.97,-1.36)
                (-1,-1.25) -- (-0.02,0.45)
                to[out=-350,in=50] (-0.02,0.4)
                -- (-0.97,-1.36);
            \node[right] at (1,-1.25) {$Q$};
            \node[left] at (-1,-1.25)
                {$P(\beta/2)$};
        \end{tikzpicture}
        \caption{}
    \end{subfigure}

    \begin{subfigure}{1.0\textwidth}
        \centering
        \begin{tikzpicture}[scale=0.8]
            \draw (-1,0.1) -- (-1,0) -- (5,0) -- (5,-0.2) -- (-1,-0.2) -- (-1,-0.4) -- (5,-0.4) -- (5,-0.6) -- (-1,-0.6) -- (-1,-0.7);
            \filldraw (0.3,0) circle (2pt); \node[above] at (0.3,0) {$\mathcal O_R(x)$};
            \filldraw (4.4,0) circle (2pt); \node[above] at (4.4,0) {$\mathcal O_R(y')$};
            \filldraw (0.5,-0.2) circle (2pt); \node[below] at (0.5,-0.6) {$\mathcal O_R(x')$};
            \filldraw (4.6,-0.6) circle (2pt); \node[below] at (4.6,-0.6) {$\mathcal O_R(y)$};
            \node at (6.3,-1.2) {$\mapsto$};
            \draw[white] (0,-2.7) circle (0.1pt);
        \end{tikzpicture}\quad
        \begin{tikzpicture}[scale=0.8]
            \draw (-1,0.1) -- (-1,0) -- (1.5,0) .. controls (3.5,0) and (3.5,-1.9) .. (5.5,-1.9) -- (5.5,-2.1)
                  .. controls (3.5,-2.1) and (3.5,-0.2) .. (1.5,-0.2) -- (-1,-0.2) -- (-1,-0.4) -- (1.5,-0.4)
                  .. controls (3.5,-0.4) and (3.5,-2.3) .. (5.5,-2.3) -- (5.9,-2.3) -- (5.9,-2.5) -- (5.5,-2.5)
                  .. controls (3.5,-2.5) and (3.5,-0.6) .. (1.5,-0.6) -- (-1,-0.6) -- (-1,-0.7);
            \draw[dashed] (4.4,-2.2) -- (6.7,-2.2);
            \filldraw (0.3,0) circle (2pt); \node[above] at (0.3,0) {$\mathcal O_R(x)$};
            \filldraw (0.5,-0.2) circle (2pt); \node[below] at (0.5,-0.6) {$\mathcal O_R(x')$};
            \filldraw (5.5,-2.0) circle (2pt); \node[right] at (5.6,-1.8) {$\tO_R(\by')$};
            \filldraw (5.9,-2.4) circle (2pt); \node[right] at (6.0,-2.6) {$\tO_R(\by)$};
            \draw[<->] (8.0,0) -- (8.0,-2.2); \node[right] at (8.0,-1.1) {$\tfrac{\beta}{2}$};
        \end{tikzpicture}
        \caption{}
    \end{subfigure}
    \caption{(a) The exterior camera in Rindler AdS$_3$ (left) and its continuation along the complex path \eqref{eq:rindler-time-continuation}, which pushes the bulk point into the future interior and $\mathcal O_R(y)$, $\mathcal O_R(y')$ to the left wedge (right). (b) The exterior contour (left) and the contour at the endpoint (right). The parts of the two timefolds carrying $\mathcal O_R(x)$ and $\mathcal O_R(x')$ are untouched; their ends, carrying $\mathcal O_R(y')$ and $\mathcal O_R(y)$, are pushed down together by $\beta/2$, where these two operators become $\tO_R(\by')$ and $\tO_R(\by)$ and meet on the dashed line. In the thermofield double state $\tO_R(\by)$ equals $\mathcal O_L(y)$, and this is the contour of Fig \ref{fig: 1.2}(b).}
    \label{fig:crr}
\end{figure}
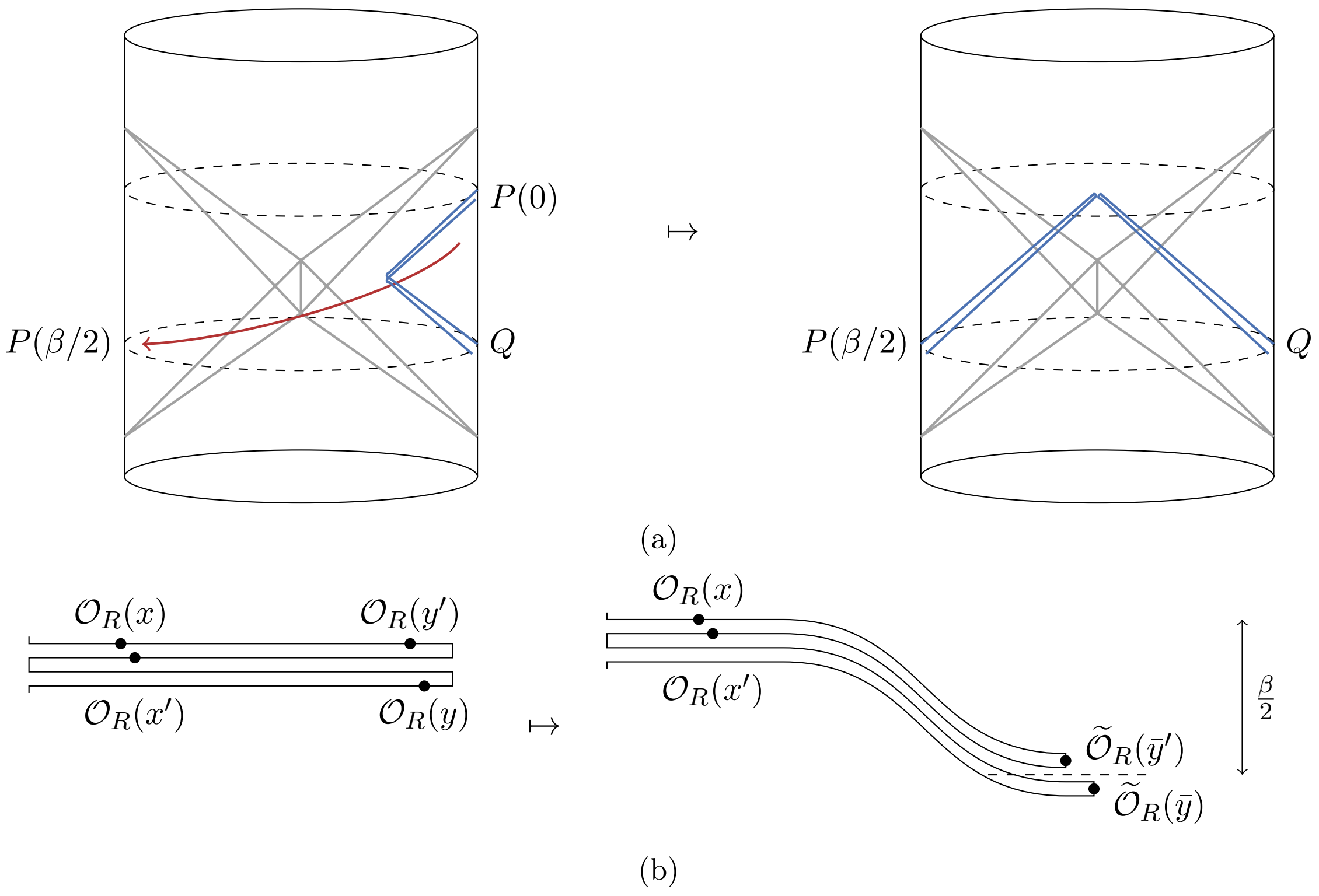

To study this we consider physics in the Rindler wedge. We consider AdS$_3$-Rindler spacetime and apply the conformal Regge formulae described above and its Fourier transform from \S~\ref{sec:crt} (specifically \eqref{eq:mom_space_crt}). The calculation below establishes the behaviour of the continued exterior camera using
semiclassical gravity as described earlier in \S~\ref{cam-2-side}. 

Let us set the AdS radius to one for convenience. The metric of AdS$_3$-Rindler spacetime is given by:
\begin{equation}
    \diff s^2
    =
    \frac{1}{z^2}
    \left(
        -(1-k^2z^2)\,\diff t^2
        +\frac{\diff z^2}{1-k^2z^2}
        +\diff x^2
    \right),
    \qquad
    k=\frac{2\pi}{\beta}.
    \label{eq:rindler-metric}
\end{equation}
Here the boundary is at $z=0$, and the right exterior is the region $0<z<1/k$, while the horizon is at $z=1/k$.

Let us now consider the exterior camera of \S~\ref{sec:cam_ext}.
We can place the pair $\mathcal O_R(x),\mathcal O_R(x')$ at zero boundary time and zero spatial position, and the pair $\mathcal O_R(y),\mathcal O_R(y')$ at time $T>0$ and zero spatial position. Here we take the shooting momentum $p$ of $\mathcal O_R(x),\mathcal O_R(x')$ to be radial, such that the shooting angle $\phi_1=0$. The shooting angle of $\mathcal O_R(y),\mathcal O_R(y')$ is parametrized by $\phi_2$, and the sharpest signal will be at the radial value $\phi_2=0$.
Note that these momenta choices leave vectors integrated over in the Fourier transforms unchanged.

Since we are in a scattering configuration, the exterior correlator lies on the second sheet of cross ratios with the operator ordering of eqn \eqref{fig1.2} (this can be cleanly demonstrated by starting with an Euclidean configuration and then tracking the cross ratios as we move to a scattering configuration similar to that in \cite{Chandorkar:2021viw, Maldacena:2015iua}).

\subsubsection{A path from right to left boundary} We then continue the insertion times $t_3=t_y$ and $t_4=t_{y'}$ of $\mathcal O_R(y)$ and $\mathcal O_R(y')$ along a path as follows:
\begin{equation}
    t_{3,4}\longrightarrow t_{3,4}-\ic\theta,
    \qquad
    0\leq\theta\leq\frac{\beta}{2},
    \label{eq:rindler-time-continuation}
\end{equation}
Note that while following this path, we keep $\mathcal O_R(x)$ and $\mathcal O_R(x')$ fixed. This is depicted in Figure \ref{fig:crr}. 

In boundary lightcone coordinates, we can nicely parametrize Rindler coordinates using the exponential map $    \rho=e^{-k(t-x)}, \ \bar\rho=e^{k(t+x)}$.
Consequently, along the path \eqref{eq:rindler-time-continuation}, the coordinates of $\mathcal O_R(y)$ and $\mathcal O_R(y')$ obey:
\begin{equation}
    \rho_{3,4}(\theta)
    =
    e^{\ic k\theta}\rho_{3,4}(0),
    \qquad
    \bar\rho_{3,4}(\theta)
    =
    e^{-\ic k\theta}\bar\rho_{3,4}(0).
\end{equation}
In particular, at $\theta=\beta/2$, both exponentials become $-1$, where $\mathcal O_R(y)$ and $\mathcal O_R(y')$ both reach the opposite boundary Rindler wedge. The endpoint of the complex path of the correlator is understood as the continuation
along this path. We stress what the endpoint is. It is a correlator of the right CFT alone, in which both operators $\mathcal O_R(y)$ and $\mathcal O_R(y')$ of the exterior camera have become $\tO_R(\by)$ and $\tO_R(\by')$ at imaginary time $-\ic\beta/2$. On the contour, the parts of the timefolds carrying $\mathcal O_R(x)$ and $\mathcal O_R(x')$ are untouched while their ends, carrying $\mathcal O_R(y')$ and $\mathcal O_R(y)$, are pushed down to ${\rm Im}\,t=-\beta/2$, as drawn in Fig \ref{fig:crr}(b): the first timefold descends to $\tO_R(\by')$ and comes back, the second descends to $\tO_R(\by)$ and continues around the thermal circle. It is not by itself a two-sided observable. The link with the camera is the identity \eqref{eq:tfd-left-right-kms-bra} of the thermofield double state: $\tO_R(\by)$ is adjacent to the bra, so there it can be traded for the ordinary left operator $\mathcal O_L(y)$ at left time $-T$, while $\tO_R(\by')$, which is not adjacent to the state, stays a right operator at complex time. In the thermofield double state the endpoint therefore equals the camera \eqref{eq:camera-def}, by \eqref{eq:camera-trace}, and this is the only way in which the results below bear on it.

The operators $\mathcal O_R(y)$ and $\mathcal O_R(y')$ have to be moved in the same direction. One might instead move them in opposite directions, $\mathcal O_R(y)$ down by $\beta/2$ and $\mathcal O_R(y')$ up by $\beta/2$, so that they again meet at the left boundary, now through the identification $\tau\sim\tau+\beta$ of the thermal circle. The two endpoints sit at the same points of the thermal circle, but they are reached on different sheets. In the opposite-direction continuation the center of the pair $\mathcal O_R(y),\mathcal O_R(y')$ does not move, and only the separation within the pair changes, by $-\ic\beta$. In cross-ratio space the numerator factors $\sinh\frac{ku_{34}}{2}$ and $\sinh\frac{kv_{34}}{2}$ then turn by $+\pi$ and $-\pi$ respectively, since $u_{34}$ and $v_{34}$ have opposite signs for a spacelike separation. So $z$ turns by $+\pi$, $\bar z$ by $-\pi$, and the offset of $2\pi\ic$ between $\log\bar z$ and $\log z$, which puts the exterior camera on the bulk-point sheet (\S\ref{sec:rindler-cross-ratios}), is lost at the endpoint. Following the contact diagram of \S\ref{sec:rindler-cross-ratios} along this path, the numerator \eqref{eq:contact-numerator} goes from $-4\pi^2$ to zero, and the bulk-point singularity disappears. Moving both operators down instead turns $z$ and $\bar z$ by $+\pi$ each, keeps the offset and keeps the numerator at $-4\pi^2$, as shown in Fig \ref{fig:crpath}.

\subsubsection{Tracking the bulk point along the path}
\label{sec:btz-bulk-point}

For radial null geodesics, evaluating the geodesic equation using the metric \eqref{eq:rindler-metric} gives us:
\begin{equation}
    \frac{\diff t}{\diff z}
    =
    \pm\frac{1}{1-k^2z^2}.
\end{equation}
Using this, the geodesic from boundary $t=0$ to the bulk point, and the geodesic traced backward from $t = T$ to the bulk point satisfy the following equations respectively:
\begin{equation}
    t=\frac{1}{k}\operatorname{arctanh}(kz),
    \qquad
    t=T-\frac{1}{k}\operatorname{arctanh}(kz),
\end{equation}
 Their bulk intersection gives us the location of the bulk point, which lies at ${t^\star}=\frac{T}{2}$ and ${z^\star}=\frac{1}{k}\tanh\frac{kT}{2}$ \footnote{This unfortunate notation for intersection point with $z^\star$ should not be confused with the Tortoise coordinate $z_\star$.}. Since $T>0$ this point lies outside the horizon.

Let us now follow the bulk point along the path from right boundary to the left boundary using \eqref{eq:rindler-time-continuation}. The bulk point
trajectory along this path is given by:
\begin{equation}
    {t^\star}(\theta)=\frac{T-\ic\theta}{2},
    \qquad
    {z^\star}(\theta)
    =
    \frac{1}{k}
    \tanh\frac{k(T-\ic\theta)}{2}.
    \label{eq:continued-radial-intersection}
\end{equation}
In particular, we see that the radial coordinate ${z^\star}$ takes the following form:
\begin{equation}
    {
    {z^\star}\!\left(\frac{\beta}{2}\right)
    =
    \frac{1}{k}\coth\frac{kT}{2}
    >
    \frac{1}{k}.
    }
    \label{eq:interior-radial-intersection}
\end{equation}
This is an interior value of ${z^\star}$. This is more physically clear upon introducing and studying Kruskal coordinates in the right exterior. These are defined as follows
\begin{equation}
     U
    =
    -e^{-kt}\sqrt{\frac{1-kz}{1+kz}},
    \qquad
     V
    =
    e^{kt}\sqrt{\frac{1-kz}{1+kz}}.
    \label{eq:rindler-kruskal}
\end{equation}
Then as before, we continue these coordinates along the same path. Substituting the intersection coordinates using \eqref{eq:continued-radial-intersection} gives us the following values:
\begin{equation}
     U_\star(\theta)
    =
    -e^{-k(T-\ic\theta)},
    \qquad
     V_\star(\theta)=1.
    \label{eq:intersection-kruskal-path}
\end{equation}
Using this we obtain the following values for the bulk point following the complexified trajectory:
\begin{equation}
     U_\star\!\left(\frac{\beta}{2}\right)
    =
    e^{-kT}>0,
    \qquad
     V_\star\!\left(\frac{\beta}{2}\right)=1>0.
    \label{eq:interior-kruskal-endpoint}
\end{equation}
These signs place bulk point evaluated at $\beta / 2$, i.e. the endpoint of the complex path, in the future interior.

The Kruskal coordinates have another advantage, it makes it clear that the complexified bulk point remains regular along the path since the coordinates satisfy:
\begin{equation}
    \left|
     U_\star(\theta) V_\star(\theta)
    \right|
    =
    e^{-kT}<1,
    \qquad
    kz=\frac{1+ U V}
             {1- U V}.
\end{equation}
Note here that neither $1+ U_\star V_\star$ or
$1- U_\star V_\star$ vanishes along the path. Thus the
bulk point for exterior scattering continues to a real bulk point in the future interior,
while $\mathcal O_R(y)$ and $\mathcal O_R(y')$ are transported along the path from the right boundary to the left boundary. In the bulk, the leg from $\mathcal O_R(y),\mathcal O_R(y')$ is the null ray at constant $U=e^{-kT}$ which enters the future interior from the left boundary, while the leg from $\mathcal O_R(x),\mathcal O_R(x')$ is the null ray at constant $V=1$ which enters from the right. The two rays cross at $X$, which confirms the picture of Fig \ref{fig: 1.2}(a).

The metric \eqref{eq:rindler-metric} is that of the planar BTZ black hole, so the point found above is the continued bulk point of planar BTZ, reached while $\mathcal O_R(y)$ and $\mathcal O_R(y')$ arrive at its left boundary. The same holds at finite separation within the pairs, as we now show.

Let us use the spatial positions in eqn  \eqref{eq:spatial-camera-separations}, with $\mathcal O_R(x),\mathcal O_R(x')$ at $t=0$ and $\mathcal O_R(y),\mathcal O_R(y')$ at time $T>2a$.
Retain the $\ic \epsilon$ prescription of
eqn  \eqref{eq:explicit-regulators} when continuing the
correlator. To locate the limiting geometric bulk point,
first solve the null conditions for the real insertion
coordinates.

A bulk point in the planar BTZ exterior is connected by a null geodesic
to a boundary insertion $(t_j,x_j)$ when the following condition is satisfied:
\begin{equation}
    \cosh[k(x-x_j)]
    -
    \sqrt{1-k^2z^2}\,
    \cosh[k(t-t_j)]
    =
    0.
    \label{eq:btz-null-condition}
\end{equation}
This is the vanishing condition for the denominator of
the bulk-to-boundary propagator. The square root is
positive in the right exterior, and its branch is then
fixed by continuation.

Let us now evaluate this at spatial position zero and bulk time $T/2$, where the four null conditions coincide. Their common solution satisfies the following equation:
\begin{equation}
    k^2{z^\star}^2(0)
    =
    1-
    \frac{\cosh^2(ka)}{\cosh^2(kT/2)}.
    \label{eq:btz-separated-exterior-point}
\end{equation}
Since $T>2a$, this point is real and outside the horizon.

Under eqn  \eqref{eq:rindler-time-continuation}, the bulk
time continues as in eqn  \eqref{eq:continued-radial-intersection}, while the spatial position remains zero. All four null
conditions upon the continuation are solved by
\begin{equation}
    k^2{z^\star}^2(\theta)
    =
    1-
    \frac{\cosh^2(ka)}
         {\cosh^2[k(T-\ic\theta)/2]}.
    \label{eq:btz-separated-point-path}
\end{equation}
The radial branch is fixed by
eqn  \eqref{eq:btz-separated-exterior-point}.
For intermediate values of $\theta$, this solution is
generally a point in the complexified BTZ geometry.
Its initial and final endpoints are real Lorentzian
bulk points, as seen using the Kruskal coordinates below.

The denominator in
eqn  \eqref{eq:btz-separated-point-path} does not vanish,
because the argument of the hyperbolic cosine has
strictly positive real part. The radial coordinate
does not vanish either. For $0<\theta<\beta/2$,
\begin{equation}
    \operatorname{Im}
    \cosh^2\frac{k(T-\ic\theta)}{2}
    =
    -\frac12\sinh(kT)\sin(k\theta)<0.
\end{equation}
It therefore cannot equal the positive real number
$\cosh^2(ka)$. At the initial endpoint of the complex path, we have $T>2a$ which removes the equality case. At the final endpoint of the complex path, using $k\beta=2\pi$,
the denominator becomes $-\sinh^2(kT/2)$ and is
nonzero for $T>0$.

The branch starting at the positive exterior value
therefore continues regularly to
\begin{equation}
    k^2{z^\star}^2\!\left(\frac{\beta}{2}\right)
    =
    1+
    \frac{\cosh^2(ka)}{\sinh^2(kT/2)}
    >
    1.
    \label{eq:btz-separated-interior-point}
\end{equation}
Following the same branch in
eqn  \eqref{eq:rindler-kruskal} gives
\begin{equation}
\begin{aligned}
     U_\star
    =
    e^{-kT/2}
    \sqrt{\frac{k{z^\star}-1}{k{z^\star}+1}}>0, \qquad
     V_\star
    =
    e^{kT/2}
    \sqrt{\frac{k{z^\star}-1}{k{z^\star}+1}}>0.
\end{aligned}
    \label{eq:btz-separated-interior-kruskal}
\end{equation}
Thus the endpoint of the complex path is a real point in the future interior, also at finite separation between the pair of operator insertions. This establishes a common solution of the four continued
bulk-to-boundary propagator singularity conditions along
the complex path. At the endpoint of the complex path, the solution becomes a real future-interior bulk point. Taking $a\to0$
recovers eqn  \eqref{eq:interior-radial-intersection}.

This calculation does not turn the external
branches upon the analytic continuation into four ordinary future-directed signals between the interior and the boundaries. Their
prescriptions remain those inherited from the original exterior
experiment, which is why the legs of $\mathcal O_R(y)$ and $\mathcal O_R(y')$ enter the amplitude at the endpoint with reversed momentum, \S\ref{sec:what-continuation-computes}.

\subsubsection{Continuation of the camera signal}
\label{sec:rindler-camera-signal}

We next show that the leading exterior camera signal
remains well defined under the path discussed above. Consider the CFT$_2$ configuration at a finite temperature, where the past and the future pair of operators in the correlator \eqref{fig1.2} are taken to be centred about $(0,0)$ and $(T,0)$ respectively.

Using this and taking two Fourier-transforms of the conformal Regge representation as in \eqref{eq:mom_space_crt} gives us the connected part of the Green function as follows:
\begin{equation}
\begin{aligned}
    G_{\mathrm{con}}\bigl(0,p;T,q\bigr)
    \simeq{}&
    -\ic\int_{-\infty}^{\infty}\frac{\diff\nu}{2\pi}\,
    \widetilde{\alpha}(\nu)
    \left(
        \frac{4|p||q|}{k^2}
        \sinh^2\frac{kT}{2}
    \right)^{J(\nu)-1}
    e^{\ic\nu\phi_2-\nu^2\sigma_0^2/2}.
    \label{eq:thermal-regge-camera}
\end{aligned}
\end{equation}
Here $\sigma_0$ is the angular resolution used in \eqref{sigmapq}. A time-dependent angular shift term vanishes because the boundary pairs of past and future operator insertions are centred about equal spatial positions. The quantity raised to the power $J(\nu)-1$ is the local center of mass energy at the bulk point: for the radial rays above, $s = 4\omega_p\omega_q\, {z^\star}^2/f(z^\star)$ which with $kz^\star = \tanh(kT/2)$ gives $s = \tfrac{4|p||q|}{k^2}\sinh^2\tfrac{kT}{2}$, in agreement with \eqref{sb}. It grows like $e^{kT}$ at large $T$, as expected from the chaos correlator.

For Einstein graviton exchange, we have the two-dimensional formulae:
\begin{equation}
    J(\nu)=2,
    \qquad
    \widetilde{\alpha}(\nu)
    =
    \frac{4\pi G_N}{\nu^2+1}.
\end{equation}
Using this, the CFT correlator takes the following form:
\begin{equation}
\begin{aligned}
    G_{\mathrm{con}}\bigl(0,p;T,q\bigr)
    \simeq{}&
    -\frac{16\pi\ic G_N|p||q|}{k^2}
    \sinh^2\frac{kT}{2} \    \int_{-\infty}^{\infty}\frac{\diff\nu}{2\pi}\,
    \frac{e^{\ic\nu\phi_2-\nu^2\sigma_0^2/2}}
         {\nu^2+1}.
    \label{eq:exterior-einstein-camera}
\end{aligned}
\end{equation}
We can smoothly evaluate this formula along the path
$T\to T-\ic\theta$. The above integral is independent of $T$ since its dependence is a multiplicative factor.
The endpoint of the complex path therefore gives us the following contribution:
\begin{equation}
{
\begin{aligned}
\left.
    G_{\mathrm{con}}\bigl(0,p;T-\ic\theta,q\bigr)
    \right|_{\theta=\beta/2}\simeq
    \frac{16\pi\ic G_N|p||q|}{k^2}
    \cosh^2\frac{kT}{2}
    \int_{-\infty}^{\infty}\frac{\diff\nu}{2\pi}\,
    \frac{e^{\ic\nu\phi_2-\nu^2\sigma_0^2/2}}
         {\nu^2+1}.
\end{aligned}
}
\label{eq:continued-einstein-camera}
\end{equation}
For $\sigma_0>0$, the integral is convergent. Consequently the leading contribution has a
nice and clean continuation along our chosen path. At the radial angle $\phi_2=0$,
\begin{equation}
    \int_{-\infty}^{\infty}\frac{\diff\nu}{2\pi}\,
    \frac{e^{-\nu^2\sigma_0^2/2}}{\nu^2+1}>0.
\end{equation}
Thus  the contribution along this path gives
a nonzero signal at the interior bulk point
in eqn  \eqref{eq:interior-radial-intersection}. Note the sign flip between \eqref{eq:exterior-einstein-camera} and \eqref{eq:continued-einstein-camera}, anticipated in \eqref{eq2.15}. Along the path the factor $\tfrac{4|p||q|}{k^2}\sinh^2\tfrac{k(T-\ic\theta)}{2}$ continues to $-s_{\rm phys}$, where $s_{\rm phys}=\tfrac{4|p||q|}{k^2}\cosh^2\tfrac{kT}{2}$ is precisely the center of mass energy of the two infalling rays at the interior point \eqref{eq:interior-radial-intersection}.

\subsubsection{Sharp resolution limit and angular dependence}
In the sharp-resolution limit of \eqref{eq:continued-einstein-camera} where we impose $\sigma_0\to0$, the Gaussian factor in the integral tends to one. Since the integrand is bounded as follows:
\begin{equation}
    \left|
    \frac{e^{\ic\nu\phi_2-\nu^2\sigma_0^2/2}}
         {\nu^2+1}
    \right|
    \leq \frac{1}{\nu^2+1},
\end{equation}
and since the right-hand side is a simple integrable function, we can simply take the limit inside the integral.

Upon taking the limit inside the integral, it can be evaluated by residues. For $\phi_2>0$, we can close the contour in the upper half-plane,
where $e^{\ic\nu\phi_2}$ decays. The only enclosed pole
is at $\nu=\ic$, giving us the expression:
\begin{equation}
    \int_{-\infty}^{\infty}\frac{\diff\nu}{2\pi}\,
    \frac{e^{\ic\nu\phi_2}}{\nu^2+1}
    =
    \ic\,
    \underset{\nu=\ic}{\operatorname{Res}}
    \frac{e^{\ic\nu\phi_2}}{\nu^2+1}
    =
    \frac12 e^{-\phi_2}.
\end{equation}
For $\phi_2<0$, we can close the contour in the lower half-plane. The contour is clockwise and encloses the pole at $\nu=-\ic$, so as before, similarly we obtain $\frac12 e^{\phi_2}$. Combining both the cases, we obtain
\begin{equation}
{
    \int_{-\infty}^{\infty}\frac{\diff\nu}{2\pi}\,
    \frac{e^{\ic\nu\phi_2}}{\nu^2+1}
    =
    \frac12 e^{-|\phi_2|}.
    }
\end{equation}
Consequently we obtain the Fourier transformed amplitude at the endpoint of the complex path:
\begin{equation}
\begin{aligned}
    &\left.
    G_{\mathrm{con}}\bigl(0,p;T-\ic\theta,q\bigr)
    \right|_{\theta=\beta/2,\;\sigma_0\to0} \simeq
    \frac{8\pi\ic G_N|p||q|}{k^2}
    \cosh^2\frac{kT}{2}\,
    e^{-|\phi_2|}.
\end{aligned}
\label{eq:continued-angular-image}
\end{equation}
This signal for scattering at a bulk point has a nonzero jump in its first angular derivative at $\phi_2=0$. The cusp at $\phi_2=0$ is the signature of the bulk point: it is the one-dimensional transverse propagator $\tfrac12 e^{-|\phi_2|}$ evaluated at zero impact parameter, which replaces the flat-space kernel $C_d/b^{d-3}$ of \eqref{eq:eikonal-exterior} in AdS$_3$. At finite $\sigma_0$ the cusp is smoothed over an angular width $\sigma_0$, so \eqref{eq:continued-angular-image} is the image in the limit of perfect resolution.

\subsection{Cross-ratios and the bulk-point discontinuity}
\label{sec:rindler-cross-ratios}

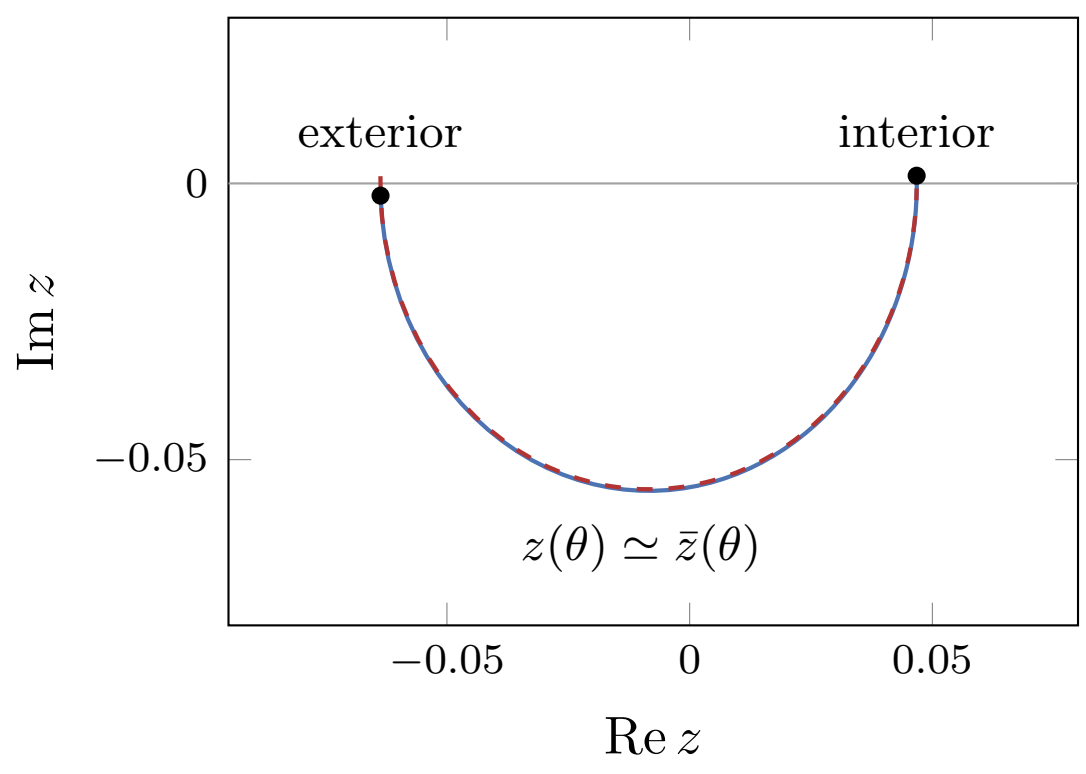
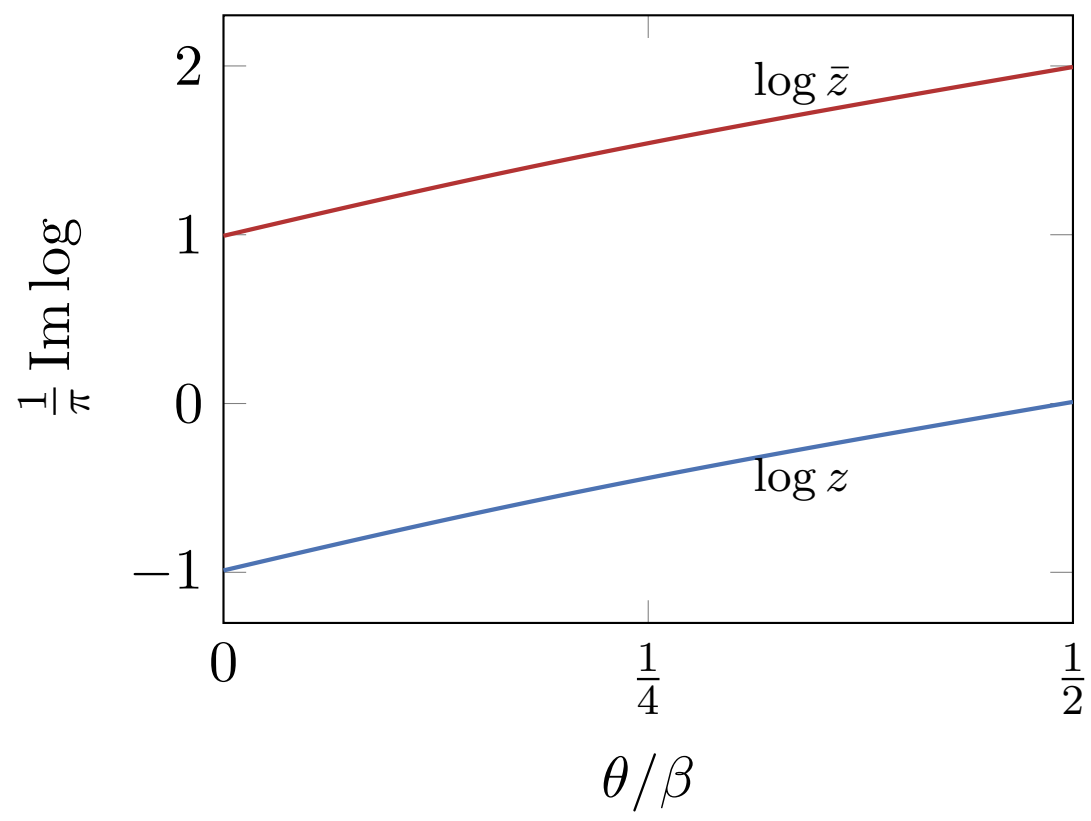
\begin{figure}[t]
\centering
\begin{subfigure}{0.48\textwidth}
\centering
\begin{tikzpicture}
\begin{axis}[width=7.2cm,height=5.6cm, xlabel={${\rm Re}\,z$}, ylabel={${\rm Im}\,z$}, xmin=-0.095,xmax=0.08, ymin=-0.08,ymax=0.03, xtick={-0.05,0,0.05}, ytick={-0.05,0}, scaled ticks=false, tick label style={font=\scriptsize, /pgf/number format/fixed, /pgf/number format/precision=2}, ylabel style={font=\small}, xlabel style={font=\small}]
\addplot[mgray, thin] coordinates {(-0.095,0) (0.08,0)};
\addplot[mblue, thick] coordinates {(-0.06368,-0.00221) (-0.06339,-0.00612) (-0.06283,-0.00999) (-0.06201,-0.01380) (-0.06092,-0.01754) (-0.05958,-0.02117) (-0.05800,-0.02468) (-0.05620,-0.02806) (-0.05418,-0.03127) (-0.05197,-0.03432) (-0.04958,-0.03719) (-0.04703,-0.03986) (-0.04433,-0.04233) (-0.04151,-0.04459) (-0.03859,-0.04663) (-0.03558,-0.04847) (-0.03251,-0.05009) (-0.02938,-0.05149) (-0.02622,-0.05268) (-0.02304,-0.05366) (-0.01985,-0.05444) (-0.01667,-0.05502) (-0.01352,-0.05541) (-0.01040,-0.05561) (-0.00732,-0.05564) (-0.00429,-0.05550) (-0.00132,-0.05520) (0.00157,-0.05474) (0.00440,-0.05415) (0.00714,-0.05341) (0.00981,-0.05256) (0.01238,-0.05158) (0.01486,-0.05050) (0.01725,-0.04931) (0.01955,-0.04803) (0.02175,-0.04666) (0.02385,-0.04522) (0.02586,-0.04370) (0.02777,-0.04211) (0.02958,-0.04046) (0.03129,-0.03875) (0.03291,-0.03699) (0.03443,-0.03519) (0.03586,-0.03334) (0.03720,-0.03145) (0.03844,-0.02954) (0.03959,-0.02759) (0.04065,-0.02561) (0.04163,-0.02361) (0.04251,-0.02159) (0.04331,-0.01955) (0.04403,-0.01749) (0.04466,-0.01543) (0.04521,-0.01334) (0.04567,-0.01125) (0.04605,-0.00915) (0.04636,-0.00705) (0.04658,-0.00494) (0.04672,-0.00283) (0.04678,-0.00071) (0.04676,0.00140)};
\addplot[mred, thick, dashed] coordinates {(-0.06370,0.00131) (-0.06365,-0.00261) (-0.06332,-0.00651) (-0.06271,-0.01037) (-0.06185,-0.01417) (-0.06072,-0.01788) (-0.05934,-0.02149) (-0.05773,-0.02497) (-0.05590,-0.02830) (-0.05386,-0.03148) (-0.05163,-0.03449) (-0.04922,-0.03731) (-0.04666,-0.03994) (-0.04397,-0.04237) (-0.04115,-0.04459) (-0.03824,-0.04660) (-0.03524,-0.04839) (-0.03218,-0.04998) (-0.02907,-0.05135) (-0.02593,-0.05251) (-0.02277,-0.05346) (-0.01961,-0.05422) (-0.01647,-0.05478) (-0.01334,-0.05515) (-0.01025,-0.05534) (-0.00720,-0.05536) (-0.00421,-0.05521) (-0.00127,-0.05491) (0.00160,-0.05445) (0.00439,-0.05386) (0.00711,-0.05314) (0.00974,-0.05229) (0.01229,-0.05133) (0.01475,-0.05026) (0.01712,-0.04909) (0.01940,-0.04783) (0.02158,-0.04648) (0.02367,-0.04505) (0.02566,-0.04355) (0.02756,-0.04198) (0.02936,-0.04036) (0.03107,-0.03867) (0.03268,-0.03694) (0.03421,-0.03516) (0.03564,-0.03333) (0.03697,-0.03147) (0.03822,-0.02957) (0.03938,-0.02764) (0.04045,-0.02569) (0.04144,-0.02371) (0.04233,-0.02170) (0.04315,-0.01968) (0.04388,-0.01764) (0.04452,-0.01558) (0.04509,-0.01351) (0.04557,-0.01143) (0.04597,-0.00934) (0.04629,-0.00724) (0.04653,-0.00514) (0.04669,-0.00303) (0.04677,-0.00092)};
\addplot[only marks, mark=*, mark size=1.5pt, black] coordinates {(-0.06368,-0.00221) (0.04676,0.00140)};
\node[above] at (axis cs:-0.0637,0.003) {\footnotesize exterior};
\node[above] at (axis cs:0.0468,0.003) {\footnotesize interior};
\node[below] at (axis cs:-0.01,-0.058) {\footnotesize $z(\theta)\simeq\bar z(\theta)$};
\end{axis}
\end{tikzpicture}
\caption{}
\end{subfigure}
\begin{subfigure}{0.48\textwidth}
\centering
\begin{tikzpicture}
\begin{axis}[width=7.2cm,height=5.6cm, xlabel={$\theta/\beta$}, ylabel={$\tfrac{1}{\pi}\,{\rm Im}\log$}, xmin=0,xmax=0.5, ymin=-1.3,ymax=2.3, ytick={-1,0,1,2}, xtick={0,0.25,0.5}, xticklabels={$0$,$\tfrac14$,$\tfrac12$}]
\addplot[mblue, thick] coordinates {(0.00000,-0.98896) (0.00833,-0.96938) (0.01667,-0.94982) (0.02500,-0.93028) (0.03333,-0.91077) (0.04167,-0.89132) (0.05000,-0.87193) (0.05833,-0.85261) (0.06667,-0.83337) (0.07500,-0.81421) (0.08333,-0.79516) (0.09167,-0.77621) (0.10000,-0.75737) (0.10833,-0.73865) (0.11667,-0.72005) (0.12500,-0.70158) (0.13333,-0.68325) (0.14167,-0.66505) (0.15000,-0.64699) (0.15833,-0.62907) (0.16667,-0.61129) (0.17500,-0.59366) (0.18333,-0.57617) (0.19167,-0.55882) (0.20000,-0.54162) (0.20833,-0.52456) (0.21667,-0.50763) (0.22500,-0.49084) (0.23333,-0.47419) (0.24167,-0.45767) (0.25000,-0.44129) (0.25833,-0.42502) (0.26667,-0.40889) (0.27500,-0.39287) (0.28333,-0.37696) (0.29167,-0.36117) (0.30000,-0.34549) (0.30833,-0.32991) (0.31667,-0.31444) (0.32500,-0.29906) (0.33333,-0.28377) (0.34167,-0.26856) (0.35000,-0.25345) (0.35833,-0.23841) (0.36667,-0.22344) (0.37500,-0.20854) (0.38333,-0.19371) (0.39167,-0.17894) (0.40000,-0.16423) (0.40833,-0.14957) (0.41667,-0.13496) (0.42500,-0.12039) (0.43333,-0.10586) (0.44167,-0.09136) (0.45000,-0.07689) (0.45833,-0.06245) (0.46667,-0.04803) (0.47500,-0.03363) (0.48333,-0.01924) (0.49167,-0.00485) (0.50000,0.00953)};
\addplot[mred, thick] coordinates {(0.00000,0.99345) (0.00833,1.01304) (0.01667,1.03261) (0.02500,1.05216) (0.03333,1.07168) (0.04167,1.09115) (0.05000,1.11057) (0.05833,1.12992) (0.06667,1.14919) (0.07500,1.16838) (0.08333,1.18747) (0.09167,1.20647) (0.10000,1.22535) (0.10833,1.24411) (0.11667,1.26276) (0.12500,1.28127) (0.13333,1.29966) (0.14167,1.31791) (0.15000,1.33602) (0.15833,1.35399) (0.16667,1.37182) (0.17500,1.38951) (0.18333,1.40705) (0.19167,1.42445) (0.20000,1.44171) (0.20833,1.45882) (0.21667,1.47579) (0.22500,1.49263) (0.23333,1.50933) (0.24167,1.52589) (0.25000,1.54232) (0.25833,1.55863) (0.26667,1.57481) (0.27500,1.59087) (0.28333,1.60681) (0.29167,1.62264) (0.30000,1.63835) (0.30833,1.65396) (0.31667,1.66947) (0.32500,1.68489) (0.33333,1.70020) (0.34167,1.71543) (0.35000,1.73058) (0.35833,1.74564) (0.36667,1.76063) (0.37500,1.77555) (0.38333,1.79040) (0.39167,1.80519) (0.40000,1.81992) (0.40833,1.83459) (0.41667,1.84922) (0.42500,1.86380) (0.43333,1.87834) (0.44167,1.89285) (0.45000,1.90733) (0.45833,1.92177) (0.46667,1.93620) (0.47500,1.95061) (0.48333,1.96500) (0.49167,1.97939) (0.50000,1.99377)};
\node[right] at (axis cs:0.3,-0.45) {\footnotesize $\log z$};
\node[right] at (axis cs:0.3,1.9) {\footnotesize $\log\bar z$};
\end{axis}
\end{tikzpicture}
\caption{}
\end{subfigure}
\caption{Cross-ratios along the path \eqref{eq:rindler-time-continuation}, for $kT=3$, $ka=\tfrac12$, $k\epsilon=0.006$. (a) $z$ and $\bar z$ run together from the exterior to the interior bulk-point value, never through $0$ or $1$, starting on opposite sides of the negative axis. (b) Their logarithms keep the offset $2\pi\ic$, so the endpoint stays on the sheet of the exterior camera.}
\label{fig:crpath}
\end{figure}

The bulk point has a sharp boundary counterpart, and it is worth following it too. When the light cones of the four insertions meet at a common bulk point, the correlator is singular at $z=\bar z$. On the principal sheet, where the Euclidean correlator lives, nothing happens there; the singularity appears on the sheet reached when $\bar z$ has gone once around $1$ relative to $z$ \cite{Gary:2009ae,Maldacena:2015iua}. In the bulk this is the statement that the four light cones meet at a point and that the boundary operators are ordered so that an interaction at that point contributes.

To follow the cross-ratios we use the configuration of Appendix~\ref{app:cross-ratios}: the two members of each pair sit at spatial positions $\pm a$, the pair $x,x'$ at $t=0$ and the pair $y,y'$ at $t=T>2a$, with the $\ic\epsilon$ prescription of the ordering \eqref{fig1.2}. Along the path \eqref{eq:rindler-time-continuation}, and up to terms of order $\epsilon$, the two cross-ratios are equal,
\begin{equation}
z(\theta)\simeq\bar z(\theta)\simeq-\frac{\sinh^2(ka)}{\sinh^2\frac{k(T-\ic\theta)}{2}-\sinh^2(ka)},
\label{eq:z-path-main}
\end{equation}
which runs from the negative exterior value at $\theta=0$ to $\sinh^2(ka)/\big(\cosh^2\frac{kT}{2}+\sinh^2(ka)\big)\in(0,1)$ at $\theta=\beta/2$, where the denominator has become $-\cosh^2\frac{kT}{2}-\sinh^2(ka)$. The terms of order $\epsilon$ are what keep track of the sheet. The exterior camera sits on the bulk-point sheet. Starting from a Euclidean configuration and moving the pair $y,y'$ to its Lorentzian position, the two cross-ratios end up on opposite sides of the cut along the negative real axis, and their logarithms, followed continuously from the Euclidean configuration, differ by
\beq \label{eq:exterior-offset-main}
\log\bar z-\log z=2\pi\ic+O(\epsilon),\qquad \log(1-\bar z)-\log(1-z)=2\pi\ic+O(\epsilon),
\eeq
where $\epsilon$ denotes terms arising from the $\ic \epsilon$ contributions to the operator ordering. The second relation is the winding around $1$ which defines the sheet, and the first is its trace on the negative axis.

The question now is whether these offsets survive upon continuing, and it turns out that they do. Along the path \eqref{eq:rindler-time-continuation} the two cross-ratios move together, with $z-\bar z=O(\epsilon)$, from the negative to the positive real axis through the lower half plane, Fig \ref{fig:crpath}(a). Neither of them passes through $0$, $1$ or $\infty$, so the sheet cannot change, and the logarithms keep their offset, Fig \ref{fig:crpath}(b). The endpoint pinches the same singularity as the exterior camera, now with the bulk point in the future interior.

A contact Witten diagram makes this quantitative. For external dimensions $\Delta_i=1$ its dependence on the cross-ratios is $N(z,\bar z)/(z-\bar z)$, up to an analytic prefactor, with
\begin{equation}
    N(z,\bar z)    =
    2\operatorname{Li}_2(z)
    -2\operatorname{Li}_2(\bar z)
    +
    \bigl[\log(1-z)-\log(1-\bar z)\bigr]
    \bigl[\log z+\log\bar z\bigr].
    \label{eq:contact-numerator}
\end{equation}
On the principal sheet $N$ vanishes at $z=\bar z$ and the diagram is regular there. On the sheet of the exterior camera, the offsets \eqref{eq:exterior-offset-main} together with the matching branch of $\operatorname{Li}_2(\bar z)$ give $N=-4\pi^2+O(\epsilon)$, and since $\partial_\theta N=O(\epsilon)$ along the path, this value is carried all the way to the interior endpoint,
\begin{equation}
    N\big(z(\theta),\bar z(\theta)\big)=-4\pi^2+O(\epsilon),\qquad 0\leq\theta\leq\tfrac{\beta}{2},
    \qquad\text{so that}\qquad
    G_{\rm con}\propto-\frac{4\pi^2\lambda}{z-\bar z}\quad\text{at }\theta=\tfrac{\beta}{2}.
    \label{eq:N-path-main}
\end{equation}
The real-time two-sided correlator \eqref{eq1.1} behaves differently. It is reached from the Euclidean region with one pair on each boundary from the start, i.e. with the pair $y,y'$ placed at imaginary time $-\ic\beta/2$ before its real time is dialled up. Then $\sinh^2\frac{k(sT-\ic\beta/2)}{2}=-\cosh^2\frac{ksT}{2}$ never vanishes, the cross-ratios stay real and inside $(0,1)$ all along, the offsets \eqref{eq:exterior-offset-main} never form, and $N=0$. This is the cross-ratio form of the statement in \S\ref{sec:interior-camera-def} that a correlator of ordinary left and right operators does not see the interior. The opposite-direction continuation discussed in \S\ref{sec:rindler-continuation} also ends with $N=0$.

The cross-ratio argument settles the analytic structure, not the size of the coefficient of the singularity, which is what the Regge calculation of \S\ref{sec:rindler-camera-signal} gives. Likewise, the contact diagram is good for checking the sheet but it does not compute the graviton contribution. The explicit computation, with the regulators and the branches of all logarithms and dilogarithms, is in Appendix~\ref{app:cross-ratios}.

The identification with the two-sided observable of \S\ref{sec:interior-camera-def} goes through the thermofield double state: the operator content at the endpoint of the path of \eqref{eq:rindler-time-continuation} is that of the single-sided trace \eqref{eq:camera-trace}, as explained below \eqref{eq:rindler-time-continuation}, and \eqref{eq:camera-trace} equals the two-sided camera \eqref{eq:camera-def} in that state by \eqref{eq:tfd-left-right-kms-bra}. Since the Fourier transforms act on the relative coordinates $\delta x$, $\delta y$ at fixed pair centers while the continuation acts on the center of the pair $y,y'$, the two operations commute as long as the smeared integrand is analytic in $T$ along the path at fixed real $\delta y$, which is guaranteed by the absence of singularities along the path, and the Gaussian weights keep the $\delta y$ integral uniformly convergent. The continued exterior camera is therefore the two-sided camera in the thermofield double state. What we have not established is the uniform validity of the Regge approximation along the whole path beyond the leading graviton and the contact contributions; we return to this in \S\ref{sec:what-continuation-computes}.

\subsection{Interpretation of the analytic continued path}
\label{sec:what-continuation-computes}

The results of this section have a simple interpretation. The variable that controls the Regge formula \eqref{eq:thermal-regge-camera} is the center of mass energy of the exterior camera, which along the path becomes
\begin{equation}
    s(\theta) = \frac{4|p||q|}{k^2}\sinh^2\frac{k(T-\ic\theta)}{2},
    \qquad
    \sinh\!\Big(\frac{kT}{2}-\ic\frac{k\theta}{2}\Big) = \sinh\frac{kT}{2}\cos\frac{k\theta}{2} - \ic\cosh\frac{kT}{2}\sin\frac{k\theta}{2}.
    \label{eq:s-path}
\end{equation}
As $\theta$ runs from $0$ to $\beta/2$, the argument of sinh falls monotonically from $0$ to $-\pi/2$, so the argument of $s$ falls from $0$ to $-\pi$: the path rotates $s$ by $e^{-\ic\pi}$ through the lower half-plane, without touching the real axis in between, and at the endpoint $s=e^{-\ic\pi}s_{\rm phys}$, with $s_{\rm phys}>0$ the center of mass energy of the physical collision. At fixed momentum transfer, $t<0$, the local flat-space amplitude is analytic off the real $s$ axis, so the path stays inside its domain of analyticity. This is why the continued signal remains finite and focused in \eqref{eq:continued-einstein-camera}. The phase $e^{-\ic\pi}$ is the reversed momentum of \eqref{eq2.15}, and nothing more. For pure graviton exchange it gives $s^{J-1}\to-s_{\rm phys}$, which is the sign flip between \eqref{eq:exterior-einstein-camera} and \eqref{eq:continued-einstein-camera}. Away from the Einstein gravity limit the trajectory is curved, $J(\nu)<2$, and the same path gives
\begin{equation}
    s^{J(\nu)-1}\ \longrightarrow\ e^{-\ic\pi(J(\nu)-1)}\,s_{\rm phys}^{J(\nu)-1},
\end{equation}
which is the usual signature factor of Regge theory. This is a concrete prediction for the stringy corrections to the interior image: the broadening of the signal discussed below \eqref{alpha4} comes together with a $\nu$-dependent phase.

Physically both wavepackets, one from each boundary collide at $X$, with center of mass energy $s_{\rm phys}$. The observable computes the amplitude for that collision, with the momentum of one leg reversed: the leg supplied by $\mathcal O_L(y)$ and $\tO_R(\by')$, which in the exterior experiment is a detector, keeps that role in the correlator while the physical wavepacket moves inward. The detection which no boundary operator can provide is supplied by the backward Euclidean segment in \eqref{eq:camera-trace}, which effectively places $\tO_R(\by')$ to the future of $X$. This is the precise sense in which the observable sees behind the horizon, and the precise price paid for it. It is also why the construction cannot be reduced to real-time operators: as explained in \S\ref{sec:interior-camera-def}, such operators are never downstream of $X$.

At the endpoint of the complex path of the planar BTZ continuation, $\mathcal O_R(x)$ and $\mathcal O_R(x')$ remain on the right boundary, and $\tO_R(\by)$ and $\tO_R(\by')$ are right operators at imaginary time $-\ic\beta/2$, with $\tO_R(\by)$ adjacent to the bra and $\tO_R(\by')$ between $\mathcal O_R(x)$ and $\mathcal O_R(x')$. In the thermofield double state $\tO_R(\by)$ is, by \eqref{eq:tfd-left-right-kms-bra}, the ordinary left operator $\mathcal O_L(y)$ at left time $-T$, while $\tO_R(\by')$ stays a right operator at complex time: this is the boundary arrangement of the two-sided camera \eqref{eq:camera-def}, by the identity \eqref{eq:camera-trace}. Therefore, for the thermofield double state of planar BTZ, and in that state only, the camera observable has the connected contribution \eqref{eq:continued-einstein-camera}, with the angular image \eqref{eq:continued-angular-image}, and with interaction point \eqref{eq:btz-separated-interior-point}. In the radial coincidence
limit, the interaction point is the future-interior point identified in
Section~\ref{sec:btz-bulk-point}. In the language of \S\ref{sec:wkb}, the signal is $+\ic\,\delta(s_{\rm phys},b)$ of \eqref{eq2.15}, with $s_{\rm phys}=\tfrac{4|p||q|}{k^2}\cosh^2\tfrac{kT}{2}$ and with the transverse kernel $\tfrac12 e^{-|\phi_2|}$.

We also state clearly what has been established. We computed the analytic continuation of the single-sided exterior camera  along the path. In the thermofield double state this continuation coincides with the two-sided observable \eqref{eq:camera-def}, so the interior signal \eqref{eq:continued-einstein-camera} is a statement about the two-sided camera in that state, within the approximations listed below. For any other state, the two-sided observable is defined by \eqref{eq:camera-def} with its ordinary left operator, it is not the continuation of a right-CFT correlator, and its signal is argued for through the bulk description of \S\ref{sec:wkb} rather than through the Regge continuation. This is the form in which we use it in \S\ref{sec:objects}.

These statements are perturbative. They rest on the leading graviton exchange and on the contact diagram, and they do not establish an exact finite-$N$ singularity or uniform control of the full correlator along the complex path. The spatially compact BTZ black hole needs more care, since its boundary correlator is a torus correlator rather than the thermal-cylinder correlator used here, and we restrict to planar BTZ.

More generally, in a two-sided black-hole
state with a semiclassical bulk description, suppose
that the wavepacket integrals admit a contributing
saddle at which the associated bulk wavepackets meet
in the shared interior. If their local momenta satisfy
momentum conservation and the approximations of
Section~2 apply, the observable \eqref{eq:camera-def} has a sharply focused connected
contribution governed by the corresponding local scattering amplitude, with the momenta of the legs of $\mathcal O_L(y)$ and $\tO_R(\by')$ reversed as in \eqref{eq2.15}. For the thermofield double state of planar BTZ this is what we have shown. What remains assumed is the uniform validity of the Regge approximation along the complete complex path beyond the leading graviton and the contact contributions, and the absence of competing saddles, which we discuss in \S\ref{sec:extensions}. We do not claim equality of the complete correlators along the path: disconnected contributions and other scattering configurations need not agree, and only the endpoints of the path correspond to real bulk points.

Once its smearing and operator prescriptions are
specified, the two-sided camera \eqref{eq:camera-def} is a boundary
observable in its own right, defined in any two-sided state; in the thermofield double state it equals the analytic continuation \eqref{eq:camera-trace}, which is what we have computed. What the planar BTZ calculation provides is its interpretation as a measurement of the scattering of two infalling wavepackets behind the horizon, together with an explicit example against which deformations of the state can be tested. This is what we do next.

\section{Localized objects in the interior}
\label{sec:objects}
We now turn to states which contain \textit{something} behind the horizon. We work with the planar BTZ black hole of \S\ref{sec:planar-btz}, with $k=2\pi/\beta=r_h$ in units of the AdS radius, and with the Kruskal coordinates \eqref{eq:rindler-kruskal}. The two folds of the camera are then simply two radial null rays: the fold of $\mathcal O_R(x),\mathcal O_R(x')$ is the ray $V=V_x$ coming in from the right boundary, and the fold of $\mathcal O_L(y),\tO_R(\by')$ is the ray $U=U_y$ coming in from the left. If the pair $\mathcal O_R(x),\mathcal O_R(x')$ is at right time $t_R$ and the pair $\mathcal O_L(y),\tO_R(\by')$ at left time $t_L$, the labels are
\beq \label{eq:kruskal-labels}
V_x = e^{k t_R}, \qquad U_y = e^{k t_L}, \qquad X=(U_y,V_x),
\eeq
which reduces to \eqref{eq:interior-kruskal-endpoint} for $t_R=0$ and $t_L=-T$. The meeting point is in the future interior as long as $U_yV_x<1$. In these variables the thermofield double signal \eqref{eq:continued-angular-image} reads
\beq \label{eq:camera-signal-kruskal}
\mathcal G^{\rm con}_0(U_y,V_x) \simeq 2\pi\ic\,G_{\rm N}\, s_{\rm phys}\, e^{-|\phi_2|}, \qquad s_{\rm phys} = \frac{|p||q|}{k^2}\,\frac{(1+U_yV_x)^2}{U_yV_x},
\eeq
where we used $(1+e^{-kT})^2/e^{-kT}=4\cosh^2(kT/2)$. Only the boost invariant combination $U_yV_x$ appears, as it should.

Everything in this section follows from two simple observations. The signal \eqref{eq:factorization-forward} is a product of four WKB legs and a local amplitude, and in the forward limit the two legs of a fold come with opposite phases, $e^{\ic S_{\rm emit}}e^{-\ic S_{\rm abs}}$. So:
\begin{enumerate}

\item If an object is crossed by both legs of a fold, their phases cancel and the bulk Schwinger Keldysh fold introduced by the smeared operators meets the other Schwinger Keldysh fold at a different point than in the TFD state. The signal is just the undeformed one, \eqref{eq:camera-signal-kruskal}, evaluated at the displaced labels.

\item If an object is crossed by only one leg of a fold, the two legs no longer run along the same null ray, and the fold does not close for the shooting data of the undeformed geometry. We can still arrange the collision at $X$: the leg which crosses the object is offset by some $\delta t$, and shifting the Lorentzian time of its operator by $\delta t$, keeping the Euclidean shift $\beta_+/2$ of $\tO_R(\by')$ fixed, brings the two legs back onto the same ray. The signal then picks up the phase and the attenuation of the one leg which crossed the object. If we do not make the shift, the signal is suppressed once the mismatch $\delta t$ exceeds the width of the wavepacket.
\end{enumerate}
In both cases the signal jumps as the shooting data cross the values at which a leg first meets the object on its way to $X$, and the jump is smoothed only over the width of the wavepacket. Locating these jumps is how the camera tracks an object. Which of the two cases occurs depends on the routes of the four legs explained in \S\ref{sec:interior-camera-def}. The two legs of $\mathcal O_R(x)$ and $\mathcal O_R(x')$ are ordinary right propagators. They see the actual geometry along the ray from the right boundary to $X$, and they see it twice, once on the bra and once on the ket, so anything crossed by their fold simply displaces it. The leg of $\tO_R(\by')$ is the continuation of a right propagator through half of the thermal circle. By Birkhoff's theorem it is fixed by the right exterior alone, so it sees the eternal black brane $(+)$ no matter what lies behind the horizon. The leg of $\mathcal O_L(y)$ is the propagator from the actual left boundary and sees the state as it is. So an object crossed by the fold of $\mathcal O_L(y),\tO_R(\by')$ simply displaces this fold only when the leg of $\tO_R(\by')$ sees it too. This happens for a shock created by a real-time right insertion, whose interior portion is the analytic continuation of its exterior portion. Otherwise, as for a shell prepared in the Euclidean past or for an object dropped in from the left, only the leg of $\mathcal O_L(y)$ sees it, and the two legs of the fold no longer run along the same ray.

\subsection{Heavy worldlines}
\label{sec:timelike-objects}
\label{sec:dropped-objects}

The simplest object to put behind the horizon is a heavy particle on a radial timelike geodesic, for instance one inserted on the left side after the state has been prepared, $|\Psi\rangle=D_L|\Psi_{\rm TFD}\rangle$. Since the camera legs are radial null rays, which leg meets the particle, and where, is just a question about the geodesic in Kruskal coordinates. Let us work this out first; the same formulas will describe the shell of \S\ref{sec:timelike-shells}.

At fixed transverse position, the planar BTZ metric in the Kruskal coordinates \eqref{eq:rindler-kruskal} is $\dd s^2=-4\,\dd U\,\dd V/(1+UV)^2$, which is just AdS$_2$. It is convenient to think of it as the hyperboloid $-T_1^2-T_2^2+Y^2=-1$, with
\begin{equation}
  T_1=\frac{1-UV}{1+UV}=\frac{r}{r_h},\qquad T_2=\frac{U+V}{1+UV},\qquad Y=\frac{V-U}{1+UV},
  \label{eq:ads2-embedding}
\end{equation}
where $r=1/z$ and $r_h=k$. The boost $U\to e^{-c}U$, $V\to e^{c}V$, which translates the Killing time, acts linearly on $(T_2,Y)$, and timelike geodesics are simply obtained as the sections of the hyperboloid by planes through the origin. The geodesic turning around at radius $r_0$ on the time-symmetric slice $T_2=0$ of the left exterior is the plane $Y=-T_1\sin\phi_0$, i.e.
\begin{equation}
  U-V=\sin\phi_0\,(1-UV),\qquad \cos\phi_0=\frac{r_h}{r_0},
  \label{eq:geodesic-kruskal}
\end{equation}
A turning point at left time $t_0$ is obtained by replacing $U\to e^{-kt_0}U$ and $V\to e^{kt_0}V$. For a particle with energy $E$ per unit mass, $r_0=\sqrt{r_h^2+E^2}$ and $\sin\phi_0=E/r_0$. Along \eqref{eq:geodesic-kruskal} we have
\begin{equation}
  V=\frac{U-\sin\phi_0}{1-U\sin\phi_0},\qquad \frac{\dd V}{\dd U}=\frac{1+V\sin\phi_0}{1-U\sin\phi_0}>0 ,
  \label{eq:radial-geodesic}
\end{equation}
so both coordinates grow along the curve, as they should for a future directed curve. The geodesic comes out of the past singularity at $U=V=-1$, crosses the left future horizon $V=0$ at $U=\sin\phi_0$, and ends up in the future singularity $UV=1$ at $U=V=1$. In the limit $E\to\infty$ it becomes the null ray $U=1$. A particle created at the boundary with no smearing in time is therefore a shock, which we treat in \S\ref{sec:interior-null-shocks}.

Now place the camera. The fold of $\mathcal O_R(x),\mathcal O_R(x')$ is the ray $V=V_x$ and the fold of $\mathcal O_L(y),\tO_R(\by')$ is the ray $U=U_y$, and they meet at $X=(U_y,V_x)$ as in \eqref{eq:kruskal-labels}. The geodesic starts out in the past of $X$, and since the future singularity is not in the past of an interior point, it has to leave the past of $X$ through one of these two rays. Using \eqref{eq:radial-geodesic}, it meets the ray $V=V_x$ at $U=(V_x+\sin\phi_0)/(1+V_x\sin\phi_0)$ and the ray $U=U_y$ at $V=(U_y-\sin\phi_0)/(1-U_y\sin\phi_0)$. The first crossing happens before $X$ when $U_y-V_x>\sin\phi_0(1-U_yV_x)$, the second when the opposite inequality holds, so exactly one of the two happens. The two cases are separated by the shooting data for which the geodesic passes right through $X$,
\begin{equation}
  e^{kt_L}-e^{kt_R}=\sin\phi_0\,\big(1-e^{k(t_L+t_R)}\big),\qquad t_L+t_R<0 .
  \label{eq:worldline-image}
\end{equation}
This curve in the plane of the two shooting times is the image of the interior part of the worldline, and this is what the camera measures. When $U_y-V_x>\sin\phi_0(1-U_yV_x)$ the worldline leaves through the ray $V=V_x$. Both legs of $\mathcal O_R(x)$ and $\mathcal O_R(x')$ cross it, both are ordinary right propagators, and their fold is simply displaced as a whole. In the other case it leaves through the ray $U=U_y$. Now only the leg of $\mathcal O_L(y)$ crosses it while the leg of $\tO_R(\by')$, which is fixed by the right exterior does not.

In practice one tracks the worldline as follows. Fix the left shooting time $t_L$ and scan $t_R$. The signal is smooth until $X$ hits the worldline, which by \eqref{eq:radial-geodesic} happens at
\begin{equation}
  e^{kt_R}=\frac{e^{kt_L}-\sin\phi_0}{1-e^{kt_L}\sin\phi_0}
  \label{eq:worldline-scan}
\end{equation}
for a turning point at left time $0$, as long as $e^{kt_L}>\sin\phi_0$. Doing this at two values of $t_L$ fixes $\sin\phi_0$, and hence the energy of the particle, together with the left time $t_0$ of its turning point through the boost. The position of the jump is purely kinematic. 

As we will see the worldline calculation models many aspects of the physics of extended objects to a good degree. For specific extended objects its size tells us how heavy the object is, and depends on the field of the object. The next two subsections compute it, for a shock as the shift of its null coordinate, and for a shell through the gluing of the two sides.

\subsection{Null shocks}
\label{sec:interior-null-shocks}

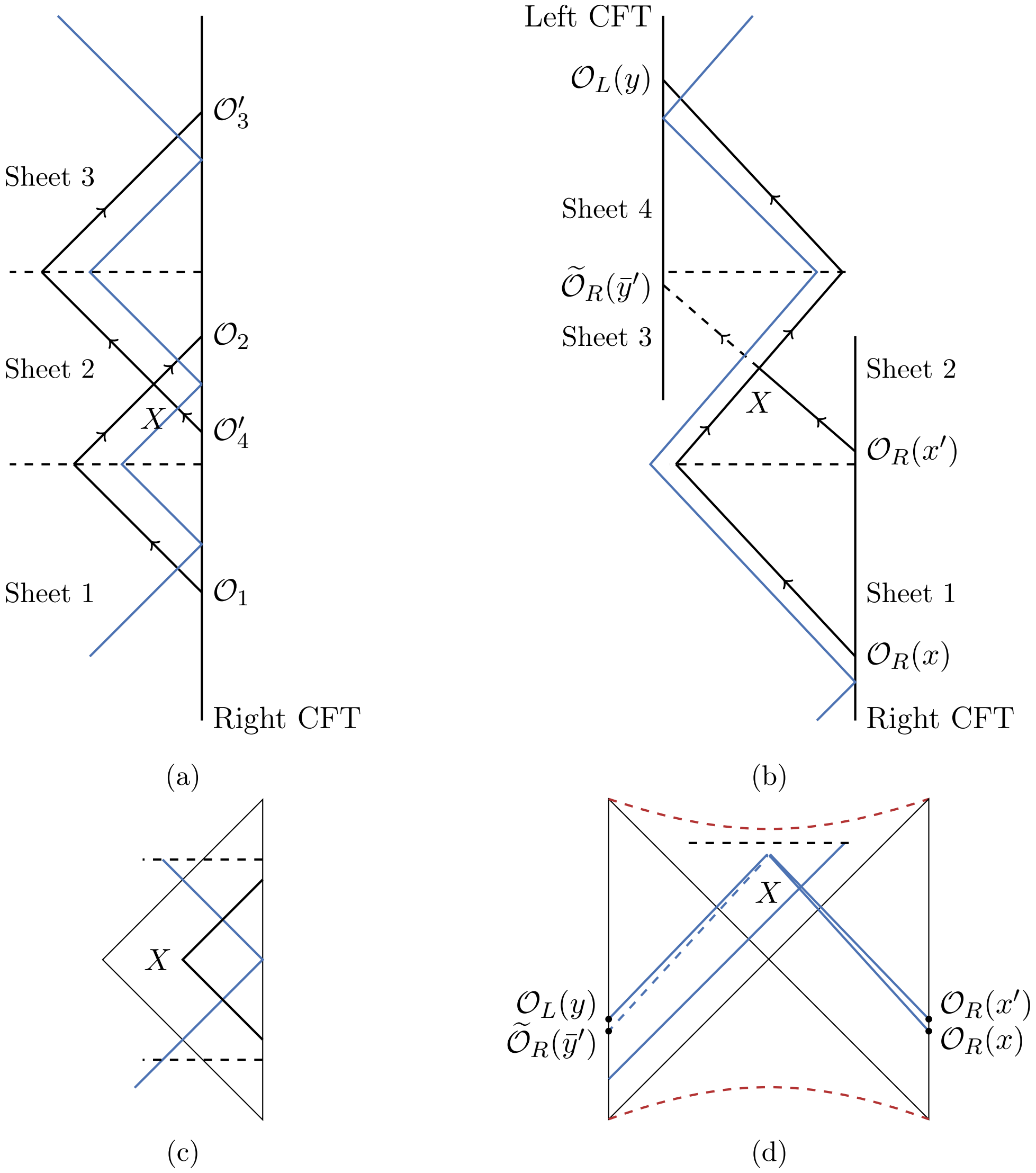
\begin{figure}
\centering
\begin{subfigure}{0.45\textwidth}
\centering
\begin{tikzpicture}[scale=0.8]
\draw[thick] (0,-1) -- (0,10) ;
\draw[thick, dashed] (0,3) -- (-3,3) (0,6) -- (-3,6);
\draw[thick,decoration={markings,mark=at position 0.4 with \arrow{>}},postaction=decorate ,draw opacity=1 ] (0,1) --(-2,3);
\draw[thick,decoration={markings,mark=at position 0.4 with \arrow{>}},postaction=decorate ,draw opacity=1 ] (-2,3) -- (-0.75,4.25);
\draw[thick,decoration={markings,mark=at position 0.4 with \arrow{>}},postaction=decorate ,draw opacity=1 ] (0,3.5) -- (-0.75,4.25);
\draw[thick,decoration={markings,mark=at position 0.4 with \arrow{>}},postaction=decorate ,draw opacity=1 ]  (-0.75,4.25)--(0,5);
\draw[thick,decoration={markings,mark=at position 0.4 with \arrow{>}},postaction=decorate ,draw opacity=1 ] (-0.75,4.25) -- (-2.5,6);
\draw[thick,decoration={markings,mark=at position 0.4 with \arrow{>}},postaction=decorate ,draw opacity=1 ] (-2.5,6) -- (0,8.5);
\draw[thick,mblue ] (0,4.25) -- (-1.25,3) --(0, 1.75)--(-1.75,0) (0,4.25) -- (-1.75,6)--(0,7.75)--(-2.25,10);
\node [below] at (-0.75,4.05) {$X$};
\node [right] at (0,1) {$\mathcal O_1$}; 
\node [right] at (0,3.5) {$\mathcal O'_4$};
\node [right] at (0,5)  {$\mathcal O_2$};
\node [right] at (0,8.5)  {$\mathcal O'_3$};
\node [left] at (-1.5,1) {\small Sheet $1$}; 
\node [left] at (-1.5,4.5) {\small Sheet $2$}; 
\node [left] at (-1.5,7.5) {\small Sheet $3$}; 
\node [right] at (0,-1) {Right CFT};
\end{tikzpicture}
\caption{}
\end{subfigure}
\begin{subfigure}{0.45\textwidth}
\centering
\begin{tikzpicture}[scale=0.8]
\draw[thick] (0,-1) -- (0,5) (-3,4)--(-3,10);
\draw[thick, dashed] (0,3) -- (-2.9,3) (-0.15,6) -- (-3,6);
\draw[thick,decoration={markings,mark=at position 0.4 with \arrow{>}},postaction=decorate ,draw opacity=1 ] (0,0) --(-2.8,3);
\draw[thick,decoration={markings,mark=at position 0.4 with \arrow{>}},postaction=decorate ,draw opacity=1 ] (-2.8,3) -- (-1.5,4.5);
\draw[thick,decoration={markings,mark=at position 0.4 with \arrow{>}},postaction=decorate ,draw opacity=1 ] (0,3.2) -- (-1.5,4.5);
\draw[thick, dashed, decoration={markings,mark=at position 0.4 with \arrow{>}},postaction=decorate ,draw opacity=1 ]  (-1.5,4.5)--(-3,5.8);
\draw[thick,decoration={markings,mark=at position 0.4 with \arrow{>}},postaction=decorate ,draw opacity=1 ] (-1.5,4.5) -- (-0.2,6);
\draw[thick,decoration={markings,mark=at position 0.4 with \arrow{>}},postaction=decorate ,draw opacity=1 ] (-0.2,6) -- (-3,9);
\draw[thick,mblue ] (-0.6,-1)-- (-0,-0.4)--(-3.2,3) --(-0.6,6) -- (-3, 8.4) -- (-1.6,10);
\node [below] at (-1.5,4.3) {$X$};
\node [right] at (0,0) {$\mathcal O_R(x)$};
\node [right] at (0,3.2) {$\mathcal O_R(x')$};
\node [left] at (-3,5.8)  {$\tO_R(\by')$};
\node [left] at (-3,9)  {$\mathcal O_L(y)$};
\node [right] at (0,1) {\small Sheet $1$}; 
\node [right] at (0,4.5) {\small Sheet $2$}; 
\node [left] at (-3,5) {\small Sheet $3$}; 
\node [left] at (-3,7) {\small Sheet $4$}; 
\node [right] at (0,-1) {Right CFT};
\node [left] at (-3,10) {Left CFT};
\end{tikzpicture}
\caption{}
\end{subfigure}
\begin{subfigure}{0.45\textwidth}
\centering
\begin{tikzpicture}[scale=0.5]
\draw (4,-4) -- (0,0) --(4,4) --cycle;
\draw [mblue, thick] (0.8,-3.2) --(4,0) -- (1.5,2.5);
\draw[dashed, thick] (4,2.5) -- (1,2.5) (4,-2.5) -- (1,-2.5);
\draw[thick] (4,2) --(2,0) --(4,-2);
\node [left] at (2,0) {$X$};
\end{tikzpicture}
\caption{}
\end{subfigure}
\begin{subfigure}{0.45\textwidth}
\centering
\begin{tikzpicture}[scale=0.5]
\node (I)    at ( 4,0)   {};
\node (II)   at (-4,0)   {};
\path  (II) +(90:4)  coordinate (IItop) +(-90:4) coordinate (IIbot) +(0:4) coordinate (IIright);
\draw (IItop) -- (IIright) -- (IIbot)--cycle;
\draw[dashed, mred, thick]  (-4,4) ..controls (-1.0,3) and (1.0,3).. (4,4);
\path (I) +(90:4)  coordinate (Itop) +(-90:4) coordinate (Ibot) +(180:4) coordinate (Ileft);
\draw  (Ileft) -- (Itop) -- (Ibot) -- (Ileft) -- cycle;
\draw[dashed, mred, thick] (IIbot) to[out=+20,in=-200] (Ibot) ;
\draw[mblue, thick] (-4,-3.0) -- (1.9,2.9);
\draw[thick, dashed] (-2,2.9)--(2,2.9);
\draw[mblue, thick] (4,-1.5) -- (0.05,2.6) to[out=200,in=-150] (0.05, 2.55) -- (4,-1.8);
\draw[mblue, thick] (-4,-1.5) -- (-0.05,2.6) to[out=-350,in=50] (-0.05, 2.55);
\draw[mblue, thick, dashed] (-0.2, 2.35) -- (-4,-1.8); 
\node[below] at (0,2.2) {$X$};
\filldraw (4,-1.5) circle (2pt); \filldraw (4,-1.8) circle (2pt); \filldraw (-4,-1.5) circle (2pt); \filldraw (-4,-1.8) circle (2pt); 
\node[right] at (4,-2.05) {$\mathcal O_{R}(x)$};
\node[right] at (4,-1.15) {$\mathcal O_{R}(x')$};
\node[left] at (-4,-1.15) {$\mathcal O_L(y)$};
\node[left] at (-4,-2.05) {$\tO_R(\by')$};
\end{tikzpicture}
\caption{}
\end{subfigure}
\caption{Null shocks (blue). (a,c) Exterior camera, on the unfolded contour and in the Penrose diagram: all four legs cross the shock. (b,d) Interior camera, with the dashed leg supplied by $\tO_R(\by')$; a shock created early on the left runs parallel to the legs of $\mathcal O_L(y)$ and $\tO_R(\by')$ and is crossed by both legs of $\mathcal O_R(x)$ and $\mathcal O_R(x')$, which are displaced by the same amount. Note that in (c,d) the shock gets reflected off the dashed horizontal line which indicates a future timefold.}
\label{fig:object1}
\end{figure}

 We now discuss null shocks, along the lines of \cite{Shenker:2013pqa, Shenker:2013yza}. Its effect on the bulk to boundary propagators is interesting: crossing the shock shifts its conjugate null coordinate by a finite amount, and each WKB propagator which crosses the shock records this shift as a time delay. In the language of \S\ref{sec:dropped-objects}, a shock is analogous to the worldline of a particle created at the boundary at time $t_0$ in the limit $E\to\infty$ but with a transverse profile as well. Its image \eqref{eq:worldline-image} collapses to the line $t_L=t_0$, so the jump sits at a fixed shooting time, and its size is the time delay computed below.

\subsubsection{Interior null coordinates and shock matching}
\label{subsec:interior-null-coordinates}

For simplicity, let us consider the case of planar BTZ. Let us introduce the tortoise coordinate $z_\ast(z)
  =
  \int^z \frac{\dd z'}{f(z')},$  
and define boundary-Rindler-like null coordinates $  u=T-z_\ast,
\ \ 
  v=T+z_\ast .$
Then the black brane metric becomes
\begin{equation}
  \dd s^2
  =
  \frac{1}{z^2}
  \left(
    -f(z)\dd u\dd v
    +
    \dd  X^2
  \right),
  \label{eq:shock-null-metric}
\end{equation}
where $z=z(u,v)$ is determined implicitly through $z_\ast=(v-u)/2$. This form remains useful inside the horizon, even though $T$ and $z$ exchange their causal
roles. Let us now place a null shock at  $u=u_s$. We choose a convention that crossing the shock shifts the conjugate null coordinate as
\begin{equation}
  v_+=v_-+h(X),
  \label{eq:shock-v-shift}
\end{equation}
where $h(X)$ is the transverse shock profile.  Since $T=\frac{u+v}{2}$, a geodesic crossing the shock at transverse position $X_s$ experiences the local Killing-time shift
\begin{equation}
  \Delta T_s
  =
  \frac12 h(X_s).
  \label{eq:shock-time-delay}
\end{equation}
Thus the time delay is the null-coordinate shift expressed in
the original $T$-coordinate.  For a shock which runs along the horizon itself, such as the one created by a highly boosted early perturbation in \cite{Shenker:2013pqa}, the matching is more conveniently written as a constant shift of the Kruskal coordinate. The Killing-time delay for a ray crossing the shock at a given point is then obtained from the same formula with $h$ the local value of that shift expressed in the $v$ coordinate.

Note that while we can employ a different choice of coordinate, the
matching condition \eqref{eq:shock-v-shift} and the time delay
\eqref{eq:shock-time-delay} are invariant statements. For instance, the geometry may also be obtained by introducing a continuous coordinate $V$,
\begin{equation}
  V=v+h(X)\Theta(u-u_s).
  \label{eq:shock-continuous-coordinate}
\end{equation}
Using this we have the differential:
\begin{equation}
  \dd V
  =
  \dd v
  +
  h(X)\delta(u-u_s)\dd u
  +
  \Theta(u-u_s)\partial_Xh(X)\dd X.
  \label{eq:shock-dV}
\end{equation}
Substituting this into \eqref{eq:shock-null-metric} produces a
$\delta(u-u_s)\dd u^2$ term. 

\subsubsection{Propagator phase from a single shock}
\label{subsec:shock-propagator-phase}

Using the WKB form of a bulk-to-boundary propagator as described before, we can record the null shock.  With the convention
\eqref{eq:shock-v-shift}, a scalar field satisfies
\begin{equation}
  \Phi_+(u_s,v,X)
  =
  \Phi_-(u_s,v-h(X),X).
  \label{eq:shock-field-matching}
\end{equation}
Therefore a propagator that crosses the shock obeys
\begin{equation}
  K(u_s^+,v,X;x_b)
  =
  K_{<}(u_s^-,v-h(X),X;x_b).
  \label{eq:shock-propagator-matching}
\end{equation}
Let us now go to the momentum space where the propagator $K_{<}$ is decomposed into $p_v$-modes,
\begin{equation}
  K_{<}
  =
  \int\frac{\dd p_v}{2\pi}\,
  \widetilde{K}_{<}(p_v,X;x_b)
  \ee^{-\ic p_v v},
  \label{eq:shock-pv-mode}
\end{equation}
Using this form, the shift in the propagator is a phase that records the shockwave metric $h$. Hence each crossing gives an additional phase contribution to the propagator, which is given by
\begin{equation}
  \widetilde{K}_{<}
 \ \  \mapsto \ \ 
  \ee^{\ic p_vh(X_s)} \ 
  \widetilde{K}_{<}.
  \label{eq:shock-eikonal-phase}
\end{equation}
Equivalently, the WKB action receives the additive shift $ \Delta S_s=p_vh(X_s).$ Combining this additive shift to the Hamilton-Jacobi action and using \eqref{eq:shock-time-delay}, the same effect can be interpreted as the phase accumulated across the path with the shockwave because the shock delays the geodesic by $\Delta T_s=h(X_s)/2$.  In other words, the presence of the null shock is recorded in each leg which crosses it as a time-delay phase.

For several null shocks $u=u_s^{(m)}$, with shifts $h_m(X)$, a null geodesic crossing them at positions $X_s^{(m)}$ accumulates
\begin{equation}
  v_{\rm out}
  =
  v_{\rm in}
  +
  \sum_m h_m(X_s^{(m)}),
  \qquad
  \Delta T_{\rm tot}
  =
  \frac12
  \sum_m h_m(X_s^{(m)}).
  \label{eq:shock-multiple-time-delay}
\end{equation}
Therefore the change in the action is given by:
\begin{equation}
  \Delta S_{\rm tot}
  =
  \sum_m p_{v}h_m(X_s^{(m)}),
  \label{eq:shock-multiple-phase}
\end{equation}
where $p_v$ is the same for all crossings since it is conserved across each shock, as shown below. This additivity is the reason that why the camera knows about shocks using the geodesic description of the propagator at high energy. Note that the picture of the shocks intersecting with the interior camera is similar to the shell picture of Fig \ref{fig:ampint4pt}(a) below. 

\subsubsection{Transverse kick}
\label{subsec:shock-transverse-kick}

The transverse part of the matching follows by differentiating the shifted phase.  Near the shock,
write the incoming phase locally as
\begin{equation}
  S_<
  =
  -p_v v+p_X^{(-)}X+\cdots .
  \label{eq:shock-incoming-phase}
\end{equation}
As discussed, since after the crossing, $  S_+
  =
  -p_v v+p_X^{(-)}X+p_vh(X)+\cdots,$ therefore we have the following relations for the different momentum components:
\begin{equation}
  p_v^{(+)}=p_v^{(-)},
  \qquad
  p_X^{(+)}
  =
  p_X^{(-)}+p_v\partial_Xh(X_s).
  \label{eq:shock-kick}
\end{equation}
Thus the shock preserves the null momentum conjugate to $v$, but changes the transverse momentum
by the gradient of the time-delay profile.  A spatially uniform shock produces a pure time delay with no transverse refraction; an inhomogeneous shock both time delays and bends the ray.

For example, if we consider the following profile of the shockwave:
\begin{equation}
  h(X)=h_0+h_1(X-X_0)+O((X-X_0)^2),
  \label{eq:shock-linear-profile}
\end{equation}
then the change in action and consequently the transverse momenta take the following form:
\begin{equation}
  \Delta S_s
  =
  p_vh_0+p_vh_1(X_s-X_0)+\cdots,
  \qquad
  p_X^{(+)}
  =
  p_X^{(-)}+p_vh_1+\cdots.
  \label{eq:shock-linear-result}
\end{equation}
A localized shock profile instead gives a position-dependent kick.  The time-delay profile
$h(X)$ is maximal where the phase shift is largest, while the transverse kick is controlled by
$\partial_Xh$.

\subsubsection{Shock imprint on the two-sided camera}
\label{subsec:shock-camera-imprint}

Let us now see which legs a shock affects. The camera has two physical Schwinger-Keldysh folds, one for the $P$-ray and one for the $Q$-ray, but four WKB legs once the contour is doubled. A null shock only affects the legs whose geodesics actually cross it, and which legs these are depends on how the shock was made.

Consider first a shock created by an early insertion $W_L(-t_w)$ on the left boundary, as in Fig.~\ref{fig:object1}. The perturbation falls in along the null ray $U=U_s$ with $U_s=e^{-kt_w}$, which for large $t_w$ hugs the right future horizon $U=0$ from the inside. Its stress tensor shifts the conjugate Kruskal coordinate of every ray crossing it \cite{Shenker:2013pqa},
\begin{equation}
  V\ \to\ V+\alpha(X),\qquad \alpha\propto G_{\rm N}\,E_w\,e^{2\pi t_w/\beta},
  \label{eq:kruskal-shock-shift}
\end{equation}
where $E_w$ is the energy of the perturbation and the transverse profile $\alpha(X)$ is the Kruskal version of the profile $h(X)$ in \eqref{eq:shock-v-shift}. The rays crossing the sheet $U=U_s$ are those at constant $V$, i.e. the two legs of $\mathcal O_R(x)$ and $\mathcal O_R(x')$, which come in from the right boundary along $V=V_x$. The legs of $\mathcal O_L(y)$ and $\tO_R(\by')$ run along constant $U$, parallel to the shock, and never meet it. Both legs of $\mathcal O_R(x)$ and $\mathcal O_R(x')$ are ordinary right propagators, and the shock is there both on the bra and on the ket, so it only displaces their fold. The fold is shifted as a whole, $V_x\to V_x+\alpha$, its two legs still coincide after the crossing, and the collision takes place at the displaced point $(U_y,V_x+\alpha)$. The signal is therefore
\begin{equation}
  \mathcal G^{\rm con}_{\rm shock}(t_L,t_R)\simeq\mathcal G^{\rm con}_0\big(U_y,\,V_x+\alpha\big)
  =\mathcal G^{\rm con}_0\Big|_{t_R\to t_R+\delta t_s},\qquad \delta t_s=\frac{\beta}{2\pi}\log\!\big(1+\alpha\,e^{-kt_R}\big),
  \label{eq:shock-camera-first-order}
\end{equation}
with $\mathcal G^{\rm con}_0$ the function \eqref{eq:camera-signal-kruskal}. The shock shows up as a translation of the signal in the time of the pair whose fold crosses it, by the Shenker-Stanford time delay $\delta t_s$ \cite{Shenker:2013pqa}. Since \eqref{eq:camera-signal-kruskal} depends on the labels only through the center of mass energy $s_{\rm phys}\propto(1+U_yV_x)^2/U_yV_x$, at fixed shooting data the translation is seen as a change of intensity, which grows with $t_w$ just as the shock does. An inhomogeneous shock also kicks the crossing legs, by \eqref{eq:shock-kick}, $P_X\to P_X+P_V\,\partial_X\alpha(X_s)$. This rotates the local momentum of the legs of $\mathcal O_R(x)$ and $\mathcal O_R(x')$ at $X$ and moves the cusp of the angular image \eqref{eq:continued-angular-image} away from $\phi_2=0$ by the corresponding angle, $\delta\phi_2\simeq P_V\partial_X\alpha/|P|$.

A shock created by an early insertion $W_R(-t_w)$ on the right boundary is the mirror image of this. It falls in along $V=V_s=e^{-kt_w}$, hugs the left future horizon, shifts $U\to U+\alpha$ for the rays crossing it, and is crossed by the legs of $\mathcal O_L(y)$ and $\tO_R(\by')$ while running parallel to those of $\mathcal O_R(x)$ and $\mathcal O_R(x')$. Here the leg of $\mathcal O_L(y)$ sees the shock through the state and the leg of $\tO_R(\by')$ sees it through the analytic continuation of the right exterior, and the two agree, because the interior portion of the sheet $V=V_s$ is the analytic continuation of its exterior portion, which runs near the past horizon of the right exterior. So the legs of $\mathcal O_L(y)$ and $\tO_R(\by')$ are both shifted by the same $\alpha$, their fold is again displaced as a whole, and the signal is $\mathcal G^{\rm con}_0(U_y+\alpha,V_x)$, a translation in $t_L$ by $\frac{\beta}{2\pi}\log(1+\alpha e^{-kt_L})$. In both cases one fold crosses the shock, with both of its legs.

The non-forward observable of \S\ref{sec:contact} does not have this cancellation between the two legs of a fold. Before taking the coincidence limit, each leg $a$ which crosses the shock carries the factor
\begin{equation}
  \exp\!\left(
    \ic p_{v,a}h(X_{s,a})
  \right),
  \label{eq:shock-leg-factor}
\end{equation}
and a pair of legs with momenta $P_1\neq P_2$ crossing the shock contributes
\begin{equation}
  \exp\!\left(
    \ic\,(p_{v,1}-p_{v,2})\,h(X_{s})
  \right),
  \label{eq:shock-nonforward-phase}
\end{equation}
so the shock profile $h$ can be read off directly as the phase per unit momentum transfer, as the shooting data scan the crossing point $X_s$. So as we vary the boundary shooting data, the appearance of the translation \eqref{eq:shock-camera-first-order}, or of the phase \eqref{eq:shock-nonforward-phase}, tells us that there is a null shock in the interior and where it is. Unlike a shell, a shock created in real time is also visible from the exterior it passes through. The camera images its interior portion, and what it extracts is the analytic continuation of exterior data, which fits with the fact that both legs of the crossing fold see it.

For multiple shocks, or multiple image geodesics, the leading WKB evaluation of the propagator is a sum over the shocks encountered along each path,
\begin{equation}
  K_{\rm shock}
  \sim
  \sum_{\gamma}
  \cA_\gamma
  \exp\!\left(
    \ic S_\gamma^{(0)}
    +
    \ic
    \sum_{s\in\gamma}
    p_{v}h_s(X_s)
  \right).
  \label{eq:shock-multiple-saddles}
\end{equation}
The single-crossing regime relevant for the leading camera saddle keeps one dominant term in this sum. Near caustics the Van Vleck prefactor has to be treated more carefully as in \cite{Caron-Huot:2025hmk, Caron-Huot:2025she, Chakravarty:2025ncy}, but the additive time-delay phase along each WKB path remains the main physical principle.

\subsection{Timelike shells}
\label{sec:timelike-shells}

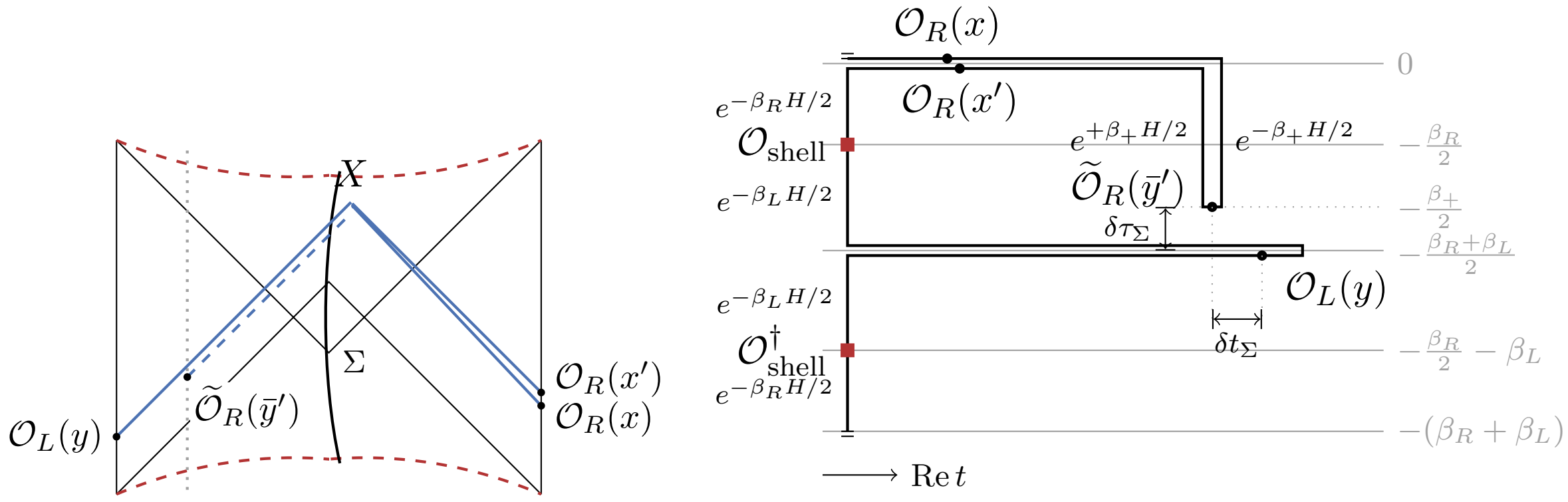
\begin{figure}[t!]
\centering
\begin{subfigure}{0.44\textwidth}
\centering
\begin{tikzpicture}[scale=0.44]
\def\dd{0.8}
\begin{scope}
\clip (-5.3,-4.1) rectangle (0,4.1);
\draw (-4.8,-4) -- (-4.8,4);
\draw (-\dd,0) -- (-4.8,4) (-\dd,0) -- (-4.8,-4) (-\dd,0) -- (3.2,4) (-\dd,0) -- (3.2,-4);
\draw[dashed, mred, thick] (-4.8,4) ..controls (-2.2,2.9) and (0.6,2.9).. (3.2,4);
\draw[dashed, mred, thick] (-4.8,-4) ..controls (-2.2,-2.9) and (0.6,-2.9).. (3.2,-4);
\end{scope}
\begin{scope}
\clip (0,-4.1) rectangle (5.3,4.1);
\draw (4.8,-4) -- (4.8,4);
\draw (\dd,0) -- (4.8,4) (\dd,0) -- (4.8,-4) (\dd,0) -- (-3.2,4) (\dd,0) -- (-3.2,-4);
\draw[dashed, mred, thick] (-3.2,4) ..controls (-0.6,2.9) and (2.2,2.9).. (4.8,4);
\draw[dashed, mred, thick] (-3.2,-4) ..controls (-0.6,-2.9) and (2.2,-2.9).. (4.8,-4);
\end{scope}
\draw[black, thick] (0.25,-3.3) .. controls (-0.18,-1.4) and (-0.18,1.4) .. (0.25,3.3);
\node[right] at (0.0,-1.0) {\footnotesize $\Sigma$};
\draw[mblue, thick] (4.8,-1.7) -- (0.55,2.55) to[out=200,in=-150] (0.55,2.5) -- (4.8,-2.0);
\draw[mgray, thick, dotted] (-3.2,-3.9) -- (-3.2,3.9);
\draw[mblue, thick] (-4.8,-2.7) -- (0.5,2.6);
\draw[mblue, thick, dashed] (-3.2,-1.35) -- (0.4,2.25);
\node[above] at (0.5,2.65) {$X$};
\filldraw (4.8,-1.7) circle (2pt); \filldraw (4.8,-2.0) circle (2pt); \filldraw (-4.8,-2.7) circle (2pt); \filldraw (-3.2,-1.35) circle (2pt);
\node[right] at (4.8,-2.3) {\small $\mathcal O_{R}(x)$};
\node[right] at (4.8,-1.4) {\small $\mathcal O_{R}(x')$};
\node[left] at (-4.8,-2.7) {\small $\mathcal O_L(y)$};
\node[below right, fill=white, inner sep=1pt] at (-3.15,-1.45) {\small $\tO_R(\by')$};
\end{tikzpicture}
\caption{}
\end{subfigure}
\begin{subfigure}{0.54\textwidth}
\centering
\begin{tikzpicture}[scale=0.62]
\draw[mgray] (-1.6,0) -- (7.4,0); \node[right, mgray] at (7.4,0) {\footnotesize $0$};
\draw[mgray] (-1.6,-1.3) -- (7.4,-1.3); \node[right, mgray] at (7.4,-1.3) {\footnotesize $-\tfrac{\beta_R}{2}$};
\draw[mgray, dotted] (3.9,-2.3) -- (7.4,-2.3); \node[right, mgray] at (7.4,-2.3) {\footnotesize $-\tfrac{\beta_+}{2}$};
\draw[mgray] (-1.6,-3.0) -- (7.4,-3.0); \node[right, mgray] at (7.4,-3.05) {\footnotesize $-\tfrac{\beta_R+\beta_L}{2}$};
\draw[mgray] (-1.6,-4.6) -- (7.4,-4.6); \node[right, mgray] at (7.4,-4.6) {\footnotesize $-\tfrac{\beta_R}{2}-\beta_L$};
\draw[mgray] (-1.6,-5.9) -- (7.4,-5.9); \node[right, mgray] at (7.4,-5.9) {\footnotesize $-(\beta_R+\beta_L)$};
\draw[->] (-1.6,-6.6) -- (-0.4,-6.6); \node[right] at (-0.4,-6.6) {\footnotesize ${\rm Re}\,t$};
\draw[thick] (-1.2,0.08) -- (4.8,0.08) -- (4.8,-2.3) -- (4.5,-2.3) -- (4.5,-0.08) -- (-1.2,-0.08) -- (-1.2,-2.92) -- (6.1,-2.92) -- (6.1,-3.08) -- (-1.2,-3.08) -- (-1.2,-5.9);
\draw (-1.3,0.08)--(-1.1,0.08) (-1.3,0.16)--(-1.1,0.16);
\draw (-1.3,-5.9)--(-1.1,-5.9) (-1.3,-5.98)--(-1.1,-5.98);
\filldraw (0.4,0.08) circle (2pt); \node[above] at (0.4,0.08) {$\mathcal O_R(x)$};
\filldraw (0.6,-0.08) circle (2pt); \node[below] at (0.6,-0.08) {$\mathcal O_R(x')$};
\filldraw (4.65,-2.3) circle (2pt); \node[left] at (4.45,-1.95) {$\tO_R(\by')$};
\filldraw (5.45,-3.08) circle (2pt); \node[below right] at (5.6,-3.1) {$\mathcal O_L(y)$};
\draw[mgray, dotted] (4.65,-2.3) -- (4.65,-4.1); \draw[mgray, dotted] (5.45,-3.08) -- (5.45,-4.1);
\draw[|<->|] (4.65,-4.1) -- (5.45,-4.1); \node[below] at (5.05,-4.1) {\scriptsize $\delta t_\Sigma$};
\draw[|<->|] (3.9,-2.3) -- (3.9,-3.0); \node[left] at (3.9,-2.65) {\scriptsize $\delta\tau_\Sigma$};
\filldraw[mred] (-1.3,-1.4) rectangle (-1.1,-1.2); \node[left] at (-1.3,-1.3) {$\mathcal O_{\rm shell}$};
\filldraw[mred] (-1.3,-4.7) rectangle (-1.1,-4.5); \node[left] at (-1.3,-4.6) {$\mathcal O^\dagger_{\rm shell}$};
\node[right] at (4.8,-1.15) {\footnotesize $e^{-\beta_+H/2}$};
\node[left] at (4.5,-1.15) {\footnotesize $e^{+\beta_+H/2}$};
\node[left] at (-1.2,-0.65) {\scriptsize $e^{-\beta_RH/2}$};
\node[left] at (-1.2,-2.15) {\scriptsize $e^{-\beta_LH/2}$};
\node[left] at (-1.2,-3.8) {\scriptsize $e^{-\beta_LH/2}$};
\node[left] at (-1.2,-5.25) {\scriptsize $e^{-\beta_RH/2}$};
\end{tikzpicture}
\caption{}
\end{subfigure}
\caption{The camera in a shell state. (a) Two black holes glued along the shell $\Sigma$. The legs of $\mathcal O_L(y)$ and $\tO_R(\by')$ reach $X$ along the same ray, so the four legs meet at $X$. The leg of $\mathcal O_L(y)$ (solid) comes from the left boundary and crosses $\Sigma$ smoothly. The leg of $\tO_R(\by')$ (dashed) lives in the continuation of the right black hole, which has no shell, and comes from the left boundary of that continuation (dotted), at a time $\delta t_\Sigma$ after $\mathcal O_L(y)$. That boundary is cut out by $\Sigma$ in the shell geometry; it is the level $-\beta_+/2$ in (b). (b) The contour. The excursion of $\tO_R(\by')$ has the depth $\beta_+/2$ fixed by the right black hole; it is displaced from $\mathcal O_L(y)$ by a Lorentzian time $\delta t_\Sigma$, and it sits a Euclidean time $\delta\tau_\Sigma$, eq.~\eqref{eq:euclidean-gap}, above the left boundary.}
\label{fig:ampint4pt}
\end{figure}

The heavy operator insertion can be used to expand the black hole interior by introducing a shell behind the horizon which constitutes a deformation of the thermal half-circle by gluing two different geometries across the shell using Israel junction conditions \cite{Israel:1966rt,Goel:2018ubv} (see also \cite{Balasubramanian:2022gmo,Antonini:2023hdh,Anous:2016kss,Chandra:2022fwi, Banerjee:2023xni, Geng:2024jmm}). Very loosely, considering the thermofield doubled state as the time evolution $| \Psi_{\rm TFD} \rangle \sim e^{-{\beta H/2}}$, where $e^{-{\beta H/2}}$ is the time evolution from the left CFT to the right CFT, these states can be expressed as follows:
\beq
| \Psi_{\rm PETS} \rangle \sim e^{-{\beta_L H/2}} \ \mathcal O_{\rm shell} \ e^{-{\beta_R H/2}}
\eeq
The complex time contour consists of a Euclidean thermal circle with the heavy insertions on it, with a Lorentzian timefold at each of the two boundaries as in Fig \ref{fig:ampint4pt}(b). The operator $\mathcal O_L(y)$ sits on the Euclidean arc carrying $\mathcal O_{\rm shell}$, between the right boundary and the left boundary, while $\tO_R(\by')$ sits on an excursion attached to the right boundary which never reaches the shell insertion. This is the boundary version of the statement, drawn in Fig \ref{fig:ampint4pt}(a), that in the bulk the leg of $\mathcal O_L(y)$ is the one which crosses the shell on its way to $X$. 

The heavy insertions are part of the state rather than probes on the same footing as the camera operators, so the camera in the shell state is
\begin{equation}
  \cG_H[O_i]
  = \langle\Psi_{\rm PETS}|\ \mathcal O_L(y)\ \mathcal O_R(x')\ \tO_R(\by')\ \mathcal O_R(x)\ |\Psi_{\rm PETS}\rangle\
  ,
  \label{eq:heavy-corr-shell}
\end{equation}
with the smearings of \eqref{eq:camera-def} understood.

The operator $\tO_R(\by')$ is where the state enters. Its Euclidean shift has to be half of the period of the Killing time of the right black hole, $\beta_+/2$, so that its leg is the leg of the eternal black brane $(+)$ entering the future interior from the left. So $\beta_+$ has to be known before we can build the camera, and it can be read off from right-boundary correlators. In the shell state $\beta_+/2$ is not the Euclidean distance $(\beta_L+\beta_R)/2$ between the two boundaries. The shell cuts out of the Euclidean disk of the right black hole a boundary arc of length $\beta_+-\beta_R$, which is the Euclidean time elapsed along the shell \cite{Balasubramanian:2022gmo, Antonini:2023hdh}. For the symmetric dust shell in BTZ of \S\ref{sec:shell-delay}, which has $\beta_L=\beta_R$, the Euclidean junction condition $\sqrt{f_r(R)-\dot R^2}=2Gm$ gives this time as $\frac{2}{r_h}\arctan\frac{r_h}{2Gm}$, and so
\begin{equation}
  \delta\tau_\Sigma\equiv\frac{\beta_L+\beta_R}{2}-\frac{\beta_+}{2}=\frac{\beta_+}{\pi}\,\phi_0,\qquad \tan\phi_0=\frac{2Gm}{r_h},
  \label{eq:euclidean-gap}
\end{equation}
with $\phi_0$ the angle of \eqref{eq:geodesic-kruskal} for the shell. In terms of the preparation data, the shift of $\tO_R(\by')$ is
\begin{equation}
  \frac{\beta_+}{2}=\frac{\beta_L+\beta_R}{2}-\delta\tau_\Sigma=\frac{\beta_L+\beta_R}{2}\Big(1+\frac{2\phi_0}{\pi}\Big)^{-1},\qquad \tan\phi_0=\frac{Gm\,\beta_+}{\pi},
  \label{eq:shift-preparation}
\end{equation}
which fixes $\beta_+$ in terms of $\beta_L+\beta_R$ and $m$, and reduces to $(\beta_L+\beta_R)/2$ only as $m\to0$. A shift by $(\beta_L+\beta_R)/2$ would have been right for a probe which commutes with $\mathcal O_{\rm shell}$. For the actual probe it overshoots half of the period of the right black hole by $\delta\tau_\Sigma$. This multiplies the Kruskal label $U$ of the leg of $\tO_R(\by')$ by $e^{\ic k\delta\tau_\Sigma}$, with $k=2\pi/\beta_+$, which takes the leg off the real section, and the legs of $\mathcal O_L(y)$ and $\tO_R(\by')$ would then not meet at a real point. This is why $\tO_R(\by')$ sits a Euclidean time $\delta\tau_\Sigma$ above the left boundary in Fig \ref{fig:ampint4pt}(b). For a light shell $\delta\tau_\Sigma\simeq4Gm/r_h^2$ is small, while the Lorentzian offset $\delta t_\Sigma$ between $\mathcal O_L(y)$ and $\tO_R(\by')$, which we find in \S\ref{sec:shell-delay}, grows like $Gm\,e^{k|t_L|}$. For large $|t_L|$ the shell is therefore recorded mainly through the Lorentzian offset.

\subsubsection{\texorpdfstring{The offset between the legs of $\mathcal O_L(y)$ and $\tO_R(\by')$}{The offset between the two legs at y}}
\label{sec:shell-delay}
\label{sec:detecting-shells}

Consider a bulk point $X$ lying between the right future horizon and the shell. The two legs of $\mathcal O_R(x)$ and $\mathcal O_R(x')$ are right propagators continued through the right future horizon. The region they cross on their way to $X$ lies to the right of the shell, where the geometry is the black brane $(+)$, so they never meet the shell. The leg of $\tO_R(\by')$ is the continuation of a right propagator through half of the thermal circle of the right black hole. By Birkhoff's theorem the right exterior of the shell geometry is the black brane $(+)$, so this leg is just the leg of the eternal black brane $(+)$ without the shell. These three legs know nothing about the shell. The leg of $\mathcal O_L(y)$ is different. It is the propagator from the actual left boundary, it lives in the glued geometry, and it can only reach $X$ by crossing the shell, as in Fig.~\ref{fig:ampint4pt}(a), while the leg of $\tO_R(\by')$ lives in the continued geometry, which is the eternal black brane $(+)$ without the shell. So only one leg of the fold of $\mathcal O_L(y),\tO_R(\by')$ crosses the shell.

Near $X$ the glued geometry and the eternal $(+)$ geometry coincide, so all four legs are local plane waves in one and the same geometry and the factorization of \S\ref{sec:wkb} goes through unchanged. The signal differs from \eqref{eq:camera-signal-kruskal} only through the leg of $\mathcal O_L(y)$, and two things happen to this leg at the shell. Its energy conjugate to the Killing time changes according to the refraction law \eqref{eq:shell-refraction-z}, and the null ray it follows after the crossing is not the ray it would have followed in the eternal geometry. Let us compute both for the symmetric dust shell in BTZ, where everything is explicit, and then state the general case.

The symmetric shell of dust is obtained by gluing two BTZ regions with the same horizon radius. The energy equation \eqref{eq:junction-energy} of \S\ref{sec:shell-general-d} then gives $2\sqrt{f_r(R)+\dot R^2}=8\pi G\sigma R$, with $\sigma R$ constant by the conservation law. Writing $m\equiv 2\pi\sigma R$ for the rest mass of the shell per $2\pi$ of the coordinate $X$ (its total mass when $X$ is an angle), the junction condition reads $\sqrt{f_r(R)+\dot R^2}=2Gm$. So the shell follows the radial geodesic \eqref{eq:radial-geodesic} with
\begin{equation}
  E=2Gm,\qquad r_{\max}=\sqrt{r_h^2+4G^2m^2},\qquad \cos\phi_0=\frac{r_h}{r_{\max}},
  \label{eq:shell-radial-eq-corrected}
\end{equation}
and in the Kruskal coordinates of the right patch its worldline is \eqref{eq:geodesic-kruskal} with $r_0=r_{\max}$ and $\sin\phi_0=2Gm/r_{\max}$. As for the particle of \S\ref{sec:dropped-objects}, the shooting data at which the bulk point reaches the shell follow from this curve, now drawn in the right patch. What the shell adds is that the size of the effect can be computed exactly, through the gluing of the two sides. The shell comes out of the past singularity, turns around at $r_{\max}>r_h$ in the region between the two patches, which lies behind both horizons, and falls into the future singularity. The glued geometry is symmetric under the exchange of the two boundaries, so in the Kruskal coordinates of the left patch the shell is the mirror image of \eqref{eq:geodesic-kruskal},
\begin{equation}
  V-U=\sin\phi_0\,(1-UV)\qquad(\text{left patch}),
  \label{eq:shell-kruskal-left}
\end{equation}
and a point $(U,V)$ of the shell in the left patch is identified with the point $(V,U)$ in the right patch.

The leg of $\mathcal O_L(y)$ is the ray $U=U_y$ in the left patch, with $U_y=e^{kt_L}$ as in \eqref{eq:kruskal-labels}. By \eqref{eq:shell-kruskal-left} it meets the shell at $V=(U_y+\sin\phi_0)/(1+U_y\sin\phi_0)$, which is the point
\begin{equation}
  (U,V)=\Big(\frac{U_y+\sin\phi_0}{1+U_y\sin\phi_0},\ U_y\Big)\qquad(\text{right patch}).
  \label{eq:shell-crossing}
\end{equation}
For $0<U_y<1$ this point has $V>0$ and $UV<1$, so the crossing happens after the shell has entered the future interior and before it reaches the singularity. The leg then continues into the right patch along the radial null ray through this point which keeps moving to the right, i.e. along constant $U$. So to the right of the shell the leg of $\mathcal O_L(y)$ is the ray
\begin{equation}
  U_{\rm out}=\frac{U_y+\sin\phi_0}{1+U_y\sin\phi_0}\equiv U_y\,e^{k\,\delta t_\Sigma},
  \qquad
  e^{k\,\delta t_\Sigma}=1+\frac{\sin\phi_0\,(1-U_y^2)}{U_y\,(1+U_y\sin\phi_0)}>1 ,
  \label{eq:shell-delay}
\end{equation}
while a left operator at the same time in the eternal geometry would have given the ray $U=U_y$. Note that nothing happens to the leg at $\Sigma$ itself. The induced metric is continuous and the leg crosses $\Sigma$ smoothly, without any jump of the kind a null shock produces. The label changes because $\Sigma$ does not sit on the midline $U=V$ of the right patch, where it would be for $m=0$. A point of $\Sigma$ is labelled $(U,V)$ from one side and $(V,U)$ from the other, and the distance of $\Sigma$ from the midline is the extra length of the wormhole. With $s\equiv\sin\phi_0$ the map \eqref{eq:shell-delay} is simply
\begin{equation}
  \frac{1+U_{\rm out}}{1-U_{\rm out}}=\frac{1+s}{1-s}\,\frac{1+U_y}{1-U_y},
  \label{eq:shell-map}
\end{equation}
a constant rescaling of $(1+U)/(1-U)$. With $\mathcal O_R(x),\mathcal O_R(x')$ at $t_R=0$, i.e. $V_x=1$, the ratio $(1-U)/(1+U)$ is $r/r_h$ at the collision point, so at fixed shooting times the shell moves the collision point inward by the constant factor $(1-s)/(1+s)$. For rays crossing $\Sigma$ near the horizon, $U_y\ll1$, the map becomes the constant shift $U\to U+s$, which is the form a null shock takes. To the right of $\Sigma$ the leg of $\mathcal O_L(y)$ is the leg that the continued right black hole would emit from its own left boundary at the later time $t_L+\delta t_\Sigma$, and this is what the leg of $\tO_R(\by')$, which lives in that continuation, has to match. For a light shell, $Gm\ll r_h$ and $\sin\phi_0\simeq 2Gm/r_h$, and \eqref{eq:shell-delay} reduces to
\begin{equation}
  \delta t_\Sigma\simeq\frac{2Gm}{r_h^2}\,\frac{1-U_y^2}{U_y}=\frac{4Gm}{r_h^2}\,\sinh\big(k|t_L|\big),
  \qquad t_L<0 ,
  \label{eq:shell-delay-light}
\end{equation}
which grows exponentially with the time separation between the two pairs, just like the Shenker-Stanford time delay of \S\ref{sec:interior-null-shocks}: the earlier $\mathcal O_L(y)$ is sent, the more blueshifted it is when it hits the shell. The offset vanishes for rays which cross the shell at the singularity, $U_y\to1$.

The same map fixes the energy of the leg of $\mathcal O_L(y)$. A radial null leg carries a WKB phase which is constant along each ray and continuous across $\Sigma$. To the left of $\Sigma$ a leg of frequency $E_-$ carries the phase $-E_-t_L$ on the ray labelled by $t_L$, and to the right of $\Sigma$ the same phase sits on the ray labelled by $t'=t_L+\delta t_\Sigma(t_L)$. The energy conjugate to the right Killing time is therefore
\begin{equation}
  \frac{E_+}{E_-}=\frac{\dd t_L}{\dd t'}=\frac{(U_y+s)(1+sU_y)}{U_y\cos^2\phi_0}\ \simeq\ 1+\frac{4Gm}{r_h}\cosh(kt_L),
  \label{eq:radial-refraction}
\end{equation}
with the last form valid for a light shell. The leg of $\mathcal O_L(y)$ is blueshifted, because earlier rays are offset more. This is the radial form of the general refraction law \eqref{eq:shell-refraction-z} of \S\ref{sec:shell-general-d}, obtained here without parametrizing the shell. It agrees with the continuity of the shell-frame energy $\omega_\Sigma$ once we take into account that the leg is ingoing with respect to one patch and outgoing with respect to the other, so that its normal momentum changes sign across $\Sigma$ in the orientation of the junction condition \eqref{eq:junction-energy}. The general form also covers asymmetric shells, $z_{h,+}\neq z_{h,-}$, where the two patches have different temperatures $\beta_\pm=2\pi z_{h,\pm}$, so that the left time has to be measured in units of $\beta_-$ and the continued right time in units of $\beta_+$ when comparing the two legs.

\subsubsection{The camera signal}

Let us now put this together. The fold of $\mathcal O_L(y),\tO_R(\by')$ consists of the leg of $\mathcal O_L(y)$, which to the right of the shell is the ray $U_{\rm out}$ of \eqref{eq:shell-delay}, and of the leg of $\tO_R(\by')$, which is the ray $U=e^{kt_{\rm in}}$ with $t_{\rm in}$ the left time assigned to $\tO_R(\by')$ by \eqref{eq:tfd-left-right-kms-bra}. In the forward camera \eqref{eq:camera-def}, with $t_L$ the center of the pair $\mathcal O_L(y),\tO_R(\by')$, the operator $\mathcal O_L(y)$ sits at $t_y=t_L+\delta y^0/2$ and $\tO_R(\by')$ at $t_{y'}=t_L-\delta y^0/2$. The two legs coincide when $t_{y'}=t_y+\delta t_\Sigma(t_y)$, that is when
\begin{equation}
  \delta y^0=-\delta t_\Sigma ,
  \label{eq:fold-closure-shell}
\end{equation}
with $\delta t_\Sigma$ evaluated at the time $t_L-\delta t_\Sigma/2$ of $\mathcal O_L(y)$; to leading order in $\dd\delta t_\Sigma/\dd t_L$ this is just $\delta t_\Sigma(t_L)$. This is the separation between $\mathcal O_L(y)$ and $\tO_R(\by')$ drawn in Fig \ref{fig:ampint4pt}(b). Their Lorentzian times differ by $\delta t_\Sigma$, while the Euclidean shift $\beta_+/2$ of $\tO_R(\by')$, which is fixed by the right black hole, stays put. At the closure the two operators sit at $t_L\mp\delta t_\Sigma/2$, and their common ray is $U=U_y\,e^{k\delta t_\Sigma/2}$, with $U_y=e^{kt_L}$. The folded singularity of the pair $\mathcal O_L(y),\tO_R(\by')$, which in the eternal geometry sits at $\delta y\to0$, has therefore moved to \eqref{eq:fold-closure-shell}. Its transform with the kernel \eqref{eq:psi}, which carries $e^{-\ic q\cdot\delta y}=e^{\ic\omega_q\delta y^0+\ldots}$ together with the Gaussian envelope, produces the phase and the suppression described at the start of this section, and the meeting point moves to $(U_y e^{k\delta t_\Sigma/2},V_x)$. Putting everything together, for a transparent symmetric shell,
\begin{equation}
  \mathcal G^{\rm con}_{\rm shell}(t_L,t_R)
  \simeq
  \mathcal G^{\rm con}_0\big(U_y e^{k\delta t_\Sigma/2},V_x\big)\;
  e^{-\ic\,\omega_q\,\delta t_\Sigma}\;
  e^{-\delta t_\Sigma^2/(2\sigma_t^2)}
  \label{eq:shell-modified-camera-signal}
\end{equation}
with $\mathcal G^{\rm con}_0$ the function \eqref{eq:camera-signal-kruskal}, $\delta t_\Sigma$ the offset \eqref{eq:shell-delay} evaluated as above, and $\sigma_t^2\equiv1/(\sigma^{-1})_{00}$ the temporal width of the wavepacket $\psi_q$. The three factors are the three things the shell does. The collision now takes place at the displaced point, which changes the local center of mass energy; the phase is the offset times the frequency; and the Gaussian is the loss of overlap between the two legs of the fold. For a light shell, $\delta t_\Sigma\ll\sigma_t$, the shell is recorded as the phase \eqref{eq:shell-delay-light}. For a heavy shell, $\delta t_\Sigma\gg\sigma_t$, the forward signal is lost altogether, which is itself a detection. If the shell has absorptive degrees of freedom the leg of $\mathcal O_L(y)$ is further attenuated by $e^{-\mathcal A_q}$, with $\mathcal A_q=S^{\rm I}_H$ of \eqref{eq:imaginary-action-local} evaluated along that leg. The energy shift \eqref{eq:radial-refraction} does not open the fold of $\mathcal O_L(y),\tO_R(\by')$. Its two legs are tied together by the single relative coordinate $\delta y$ and reach $X$ with a common momentum, which differs from the eternal value at relative order $\dd\delta t_\Sigma/\dd t_L$. We neglect this next to the phase.

The non-forward observable of \S\ref{sec:contact}, in which $\mathcal O_L(y_3)$ and $\tO_R(\by'_4)$ are smeared independently, is not suppressed. Its signal peaks when the leg of $\mathcal O_L(y_3)$ matches the leg of $\tO_R(\by'_4)$, i.e. at
\begin{equation}
  t_{y_3}=t_{{\rm in},4}-\delta t_\Sigma(t_{y_3}),\qquad E_+(q_3)=\omega_{q_4},
  \label{eq:nonforward-shell-peak}
\end{equation}
and the position of the peak directly measures the offset \eqref{eq:shell-delay} and the energy shift \eqref{eq:radial-refraction}. Scanning the shooting data then maps out $\delta t_\Sigma(t_L)$, from which $\sin\phi_0$, and hence the mass $m$, follows through \eqref{eq:shell-delay}.

Suppose instead that the shell lies between $X$ and the right boundary. Then the two legs of $\mathcal O_R(x)$ and $\mathcal O_R(x')$ cross it, and both are ordinary propagators. Since the shell is part of the state, on the bra as well as on the ket, both legs are shifted by the same amount and their fold is displaced as a whole, with no net phase. The displaced fold now meets the legs of $\mathcal O_L(y)$ and $\tO_R(\by')$, which no longer share a geometry near $X$, at a point to the left of the shell. So the configuration to use is the one with $X$ between the right horizon and the shell, where the shell is crossed by one leg only. The same cancellation is also why the real-time correlator \eqref{eq1.1}, in which every fold is made of two ordinary propagators seeing the same state, could never record a phase.

\subsubsection{How much of the shell is accessible}

As $U_y$ runs from $0$ to $1$, the crossing point \eqref{eq:shell-crossing} moves along the shell from its horizon crossing, $(U,V)=(\sin\phi_0,0)$, to the singularity, $(1,1)$. So the leg of $\mathcal O_L(y)$ can reach every point of the collapsing branch inside the horizon. The bulk point itself sits at the radius $r_X=r_h\tanh(r_h|t_R|/2)$ of \eqref{eq:interior-radial-intersection} when the left detector is fixed at $t_L=0$. As $|t_R|$ runs over the allowed window $|t_R|<t_{\rm cut}$ of \S\ref{sec:extensions}, this sweeps the whole interval $0<r_X<r_h\tanh(r_ht_{\rm cut}/2)$, and since $r_ht_{\rm cut}\gg1$ the upper end is essentially $r_h$. The crossing happens before the leg reaches $X$, so by choosing $X$ deep enough we can access the whole interior part of the collapsing branch. By the junction condition, $\dot R^2=r_{\max}^2-R^2$ along the shell, with the dot the derivative with respect to its proper time, so this part of the shell lasts
\begin{equation}
  \delta\tau_{\rm in}
  =\int_0^{r_h}\frac{\dd R}{\sqrt{r_{\max}^2-R^2}}
  =\sin^{-1}\frac{r_h}{r_{\max}}
  \ \simeq\ \frac{r_h}{2Gm}\quad (Gm\gg r_h),
  \qquad
  \frac{\delta\tau_{\rm in}}{\Delta\tau_{\rm shell}}
  \sim
  \frac{r_h}{Gm},
  \label{eq:accessible-fraction-scaling}
\end{equation}
where $\Delta\tau_{\rm shell}=\pi$ is the proper time from the past to the future singularity. So a low-point camera experiment gives, at best, a late time snapshot of a very massive shell as it approaches the future singularity. This is enough to detect coarse features, such as where the shell is, its mass through \eqref{eq:shell-delay}, and its absorption. To understand the full microstate of the shell however, instead of this simple bulk effect, we would need to need to perform measurements in a way to keep track of extremely precise contributions to the four-point correlator. 

Finally for a nice discussion complementing ours but attacking a similar problem, we urge readers to see the analysis of \cite{Balasubramanian:2025gtv}.

\subsubsection{The shell in general dimension}
\label{sec:shell-general-d}

The BTZ computation used two ingredients, the junction condition for the shell and the matching of a probe leg across it. Both hold for planar black branes in any dimension, and we collect them here, leaving the derivations to Appendix~\ref{app:shell}. The two sides of the shell are the black branes
\begin{equation}
  \dd s_\pm^2
  =
  \frac{1}{z^2}
  \left(
    -f_\pm(z)\,\dd T_\pm^2
    +
    \frac{\dd z^2}{f_\pm(z)}
    +
    \dd X^2
  \right),
  \label{eq:shell-metrics-z}
\end{equation}
with $f_\pm(z)=1-z^d/z_{h,\pm}^d$ and $\dd X^2=\delta_{ab}\dd X^a\dd X^b$. The $(+)$ region is attached to the right boundary and the $(-)$ region to the left, with temperatures $\beta_\pm=4\pi z_{h,\pm}/d$. The shell sits at $(T_\pm(\tau),Z(\tau),X)$, with $\tau$ its proper time, and we write $A_\pm\equiv\sqrt{\dot Z^2+Z^2f_\pm(Z)}$, so that $\dot T_\pm=A_\pm/f_\pm(Z)$. The trajectory is fixed by the Israel condition
\begin{equation}
  \left(
    K_{ab}-K\gamma_{ab}
  \right)_+^-
  =
  -8\pi G\,S_{ab}^{\Sigma}.
  \label{eq:israel-shell-z}
\end{equation}
For a shell of dust, $S^\Sigma_{ab}=\sigma\,u_au_b$, with the normals on the two sides pointing towards the respective boundaries, which is the orientation of the two-sided gluing, it reduces to
\begin{equation}
  A_+ + A_- = \frac{8\pi G\,\sigma\,Z}{d-1},
  \qquad
  \sigma\,Z^{1-d}=\text{const},
  \label{eq:junction-energy}
\end{equation}
where the second relation is the conservation of the dust along the shell. In terms of $r=1/z$ and $f_r(r)=r^2f(1/r)$ one has $A_\pm=\sqrt{f_{r,\pm}(R)+\dot R^2}/R^2$, so for $d=2$ the energy equation reads $\sqrt{f_{r,+}(R)+\dot R^2}+\sqrt{f_{r,-}(R)+\dot R^2}=8\pi G\sigma R$ with $\sigma R$ constant, which is what we used in \S\ref{sec:shell-delay}.

A probe leg crosses a transparent shell with its phase continuous, and with its energy $\omega_\Sigma=-u^Mp_M$ in the shell frame and its transverse momentum $p_X$ continuous. The normal momentum $p_n=n^Mp_M$ then keeps its magnitude, so a thin transparent shell does not bend the ray in its own frame. What changes are the energies conjugate to the two Killing times,
\begin{equation}
  E_\pm
  =
  \frac{
    \omega_\Sigma A_\pm+\dot Z\,p_n
  }{Z^2}.
  \label{eq:shell-refraction-z}
\end{equation}
which is the refraction law. Outside the horizon a static shell gives the gravitational redshift $E_\pm=\omega_\Sigma\sqrt{f_\pm(Z)}/Z$, and a slowly moving one adds a Doppler shift,
\begin{equation}
  E_\pm
  =
  \frac{\omega_\Sigma\sqrt{f_\pm(Z)}}{Z}
  +
  \frac{\dot Z\,p_n}{Z^2}
  +
  O(\dot Z^2).
  \label{eq:slow-shell-energy-z}
\end{equation}
Behind the horizon $f_\pm<0$, the shell is necessarily moving, and the full formula is needed. With the orientation of the junction condition a leg that crosses the shell is ingoing with respect to one side and outgoing with respect to the other, so $p_n$ changes sign across $\Sigma$, and for radial legs in BTZ the ratio $|E_+/E_-|$ reproduces \eqref{eq:radial-refraction}.

\label{subsec:folded-contours-shell-phases}

An interior
object can also change the phase-shift of the camera signal in other non-trivial ways, which depends upon the microscopic constituents of the object (or boundary inserted operator). These effects
should be expressed through the WKB action and transmission coefficient of the folded
bulk geodesics.  If the object has absorptive degrees of freedom, the transmitted WKB action
can acquire an imaginary part,
\begin{equation}
  S_H
  =
  S_H^{\rm R}
  +
  \ic S_H^{\rm I},
  \qquad
  S_H^{\rm I}\geq 0,
  \label{eq:complex-shell-action}
\end{equation}
so that a single transmitted leg contributes an exponentially suppressed factor
\begin{equation}
  K_H
  \sim
  \cA_H
  \exp\!\left(\ic S_H^{\rm R}\right)
  \exp\!\left(-S_H^{\rm I}\right).
  \label{eq:attenuated-wkb-leg}
\end{equation}
In a local optical approximation, one may write along the geodesic path:
\begin{equation}
  S_H^{\rm I}
  =
  \int_{\Gamma\cap{\rm object}}
  \dd\lambda\,
  \Gamma_{\rm abs}(\omega_{\rm loc}(\lambda),X(\lambda)),
  \label{eq:imaginary-action-local}
\end{equation}
where $\Gamma_{\rm abs}$ is a local absorption rate along the geodesic and
$\omega_{\rm loc}$ is the frequency measured in the local rest frame of the object.

\section{Generalizations and limitations}
\label{sec:extensions}
\subsection{The past interior}
We study how the signal behaves if we change the operator times accordingly. Till now our signal probes the future interior. The bulk dynamics has an interesting interpretation if operator insertions are made at late time using past directed geodesics. This configuration can arise since we are essentially decomposing the four point function into products of Wightman bulk-to-boundary propagators, which have support both when the bulk point is to the future and when it is to the past of the boundary insertion (see discussion in \cite{Caron-Huot:2025hmk}).

\begin{figure}[t]
\centering
\begin{subfigure}{0.45\textwidth}
\centering
\begin{tikzpicture}[scale=0.5]
\node (I)    at ( 4,0)   {};
\node (II)   at (-4,0)   {};
\path  (II) +(90:4)  coordinate (IItop) +(-90:4) coordinate (IIbot) +(0:4) coordinate (IIright);
\draw (IItop) -- (IIright) -- (IIbot)--cycle;
\draw[dashed, mred, thick]  (-4,4) ..controls (-1.0,3) and (1.0,3).. (4,4);
\path (I) +(90:4)  coordinate (Itop) +(-90:4) coordinate (Ibot) +(180:4) coordinate (Ileft);
\draw  (Ileft) -- (Itop) -- (Ibot) -- (Ileft) -- cycle;
\draw[dashed, mred, thick] (IIbot) to[out=+20,in=-200] (Ibot) ;
\draw[mblue, thick] (4,1.5) -- (0.05,-2.6) to[out=-200,in=150] (0.05, -2.55) -- (4,1.8);
\draw[mblue, thick] (-4,1.5) -- (-0.05,-2.6) to[out=350,in=-50] (-0.05, -2.55);
\draw[mblue, thick, dashed] (-0.2, -2.35) -- (-4,1.8); 
\node[above] at (0,-2.2) {$X$};
\filldraw (4,1.5) circle (2pt); \filldraw (4,1.8) circle (2pt); \filldraw (-4,1.5) circle (2pt); \filldraw (-4,1.8) circle (2pt); 
\node[right] at (4,2.05) {$\mathcal O_{R}(x)$};
\node[right] at (4,1.15) {$\mathcal O_{R}(x')$};
\node[left] at (-4,1.15) {$\mathcal O_L(y)$};
\node[left] at (-4,2.05) {$\tO_R(\by')$};
\end{tikzpicture}
\caption{}
\end{subfigure}
\begin{subfigure}{0.45\textwidth}
\centering
\begin{tikzpicture}[scale=0.72]
\draw[mgray] (-1.60,0.00) -- (6.10,0.00); \node[right, mgray] at (6.10,0.00) {\footnotesize $0$};
\draw[mgray] (-1.60,2.00) -- (6.10,2.00); \node[right, mgray] at (6.10,2.00) {\footnotesize $+\tfrac{\beta}{2}$};
\draw[mgray] (-1.60,4.00) -- (6.10,4.00); \node[right, mgray] at (6.10,4.00) {\footnotesize $+\beta$};
\draw[->] (-1.6,-1.2) -- (-0.4,-1.2); \node[right] at (-0.4,-1.2) {\footnotesize ${\rm Re}\,t$};
\draw[thick] (-1.2,4) -- (-1.2,2.24) -- (5.3,2.24) -- (5.3,2.08) -- (-1.2,2.08) -- (-1.2,0.08) -- (4.5,0.08) -- (4.5,1.92) -- (4.8,1.92) -- (4.8,-0.08) -- (-1.2,-0.08);
\draw (-1.3,4)--(-1.1,4) (-1.3,4.08)--(-1.1,4.08);
\draw (-1.3,-0.08)--(-1.1,-0.08) (-1.3,-0.16)--(-1.1,-0.16);
\filldraw (0.4,-0.08) circle (2pt); \node[below] at (0.4,-0.08) {$\mathcal O_R(x)$};
\filldraw (0.6,0.08) circle (2pt); \node[above] at (0.6,0.08) {$\mathcal O_R(x')$};
\filldraw (4.65,1.92) circle (2pt); \node[left] at (4.4,1.58) {$\tO_R(\by')$};
\filldraw (4.5,2.24) circle (2pt); \node[above] at (4.5,2.24) {$\mathcal O_L(y)$};
\node[right] at (4.8,0.7) {\footnotesize $e^{-\beta H/2}$};
\node[left] at (4.5,0.7) {\footnotesize $e^{+\beta H/2}$};
\node[left] at (-1.2,1.1) {\footnotesize $e^{-\beta H/2}$};
\node[left] at (-1.2,3.1) {\footnotesize $e^{-\beta H/2}$};
\end{tikzpicture}
\caption{}
\end{subfigure}
\caption{The past-interior camera \eqref{eq:past-camera}. (a) Both pairs are at late times and the legs meet at $X$ in the past interior; the dashed leg carries the shift $+\ic\beta/2$ of \eqref{eq:tfd-left-right-kms}. (b) Its contour, the mirror image of Fig \ref{fig: 1.2}(b), with the excursion going up to ${\rm Im}\,t=+\beta/2$.}
\label{jurassic2}
\end{figure}
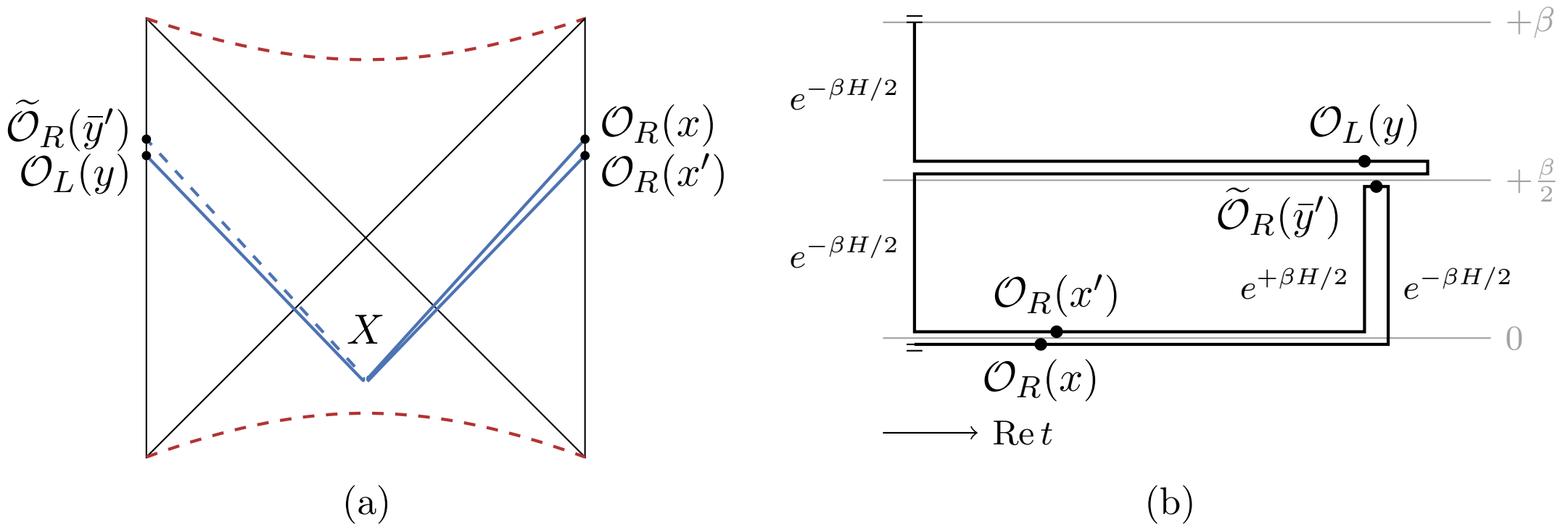

We consider the configuration where the operators on the first timefold swap places with each other (and similarly on the second timefold). Geometrically this is shown in Fig \ref{jurassic2} (a). The construction is mapped to the past interior by the time-reversal symmetry of the eternal geometry, which exchanges the roles of emission and absorption on each pair. Concretely, the past-interior camera is the Hermitian conjugate of \eqref{eq:camera-def} with all boundary times reversed,
\beq \label{eq:past-camera}
\begin{aligned}
\mathcal G^{\rm past}(x,p;y,q) &= \int {\rm d}\delta x\,{\rm d}\delta y\ \psi_p(\delta x)\,\psi_q(\delta y)\ \langle\Psi|\ \mathcal O_R(x)\ \tO_R(\by'_+)\ \mathcal O_R(x')\ \mathcal O_L(y)\ |\Psi\rangle \\
&= \langle\Psi|\ \mathcal O_{x,p,\sigma_x}\ \tO^\dagger_{\by,q,\sigma_y}\ \mathcal O^\dagger_{x,p,\sigma_x}\ \mathcal O_{y,q,\sigma_y}\ |\Psi\rangle,
\qquad \by'_+ \equiv \big(-t_{y'}+\ic\tfrac{\beta}{2},\ \mathbf y'\big),
\end{aligned}
\eeq
where now both pairs are at late times, the ordinary left operator $\mathcal O_L(y)$ is adjacent to the ket, and the right operator $\tO_R(\by'_+)=e^{-\beta H_R/2}\tO_R(-t_{y'},\mathbf y')e^{+\beta H_R/2}$ is the other member of the pair at $y$, with the shift $+\ic\beta/2$ of \eqref{eq:tfd-left-right-kms} instead of the $-\ic\beta/2$ of \eqref{eq:tfd-left-right-kms-bra}. In the trace form the backward Euclidean segment now sits between $\tO_R(-t_{y'})$ and $\mathcal O_R(x')$, and the contour is the mirror image of the one in Fig \ref{fig: 1.2}(b), with the excursion from the right boundary going up to ${\rm Im}\,t=+\beta/2$ instead of down to $-\beta/2$, as drawn in Fig \ref{jurassic2}(b). Smearing the operators appropriately, we can again obtain the bulk point. This gives rise to the following bulk amplitude in the factorization limit:
\beq 
\begin{split}
\langle\Psi|\ \mathcal O_{x,p,\sigma_x}\ \tO^\dagger_{\by,q,\sigma_y}\ \mathcal O^\dagger_{x,p,\sigma_x}\ \mathcal O_{y,q,\sigma_y}\ |\Psi\rangle  \quad \mapsto \quad  \langle  \, b_{X,P_1}\, \tilde b_{X,Q_4} \,  b^\dagger_{X,P_2}  \, \tilde b^\dagger_{X,Q_3}  \,   \rangle
\end{split}
\eeq
where we have now associated the modes $b$ with both pairs, since both pairs are to the future of the bulk point, and where the tilde marks the modes supplied by the pair containing $\tO_R(\by'_+)$, whose momenta enter with the opposite sign as before. The factorization formula for the correlator again takes the following form:
\beq 
\begin{split}
\langle\Psi|\ & \mathcal O_{x,p,\sigma_x}\ \tO^\dagger_{\by,q,\sigma_y}\ \mathcal O^\dagger_{x,p,\sigma_x}\ \mathcal O_{y,q,\sigma_y}\ |\Psi\rangle \\
 & \approx \int \qty[\prod \diff P_i\,\tilde\psi_i] \int \qty[\prod \diff Q_j\,\tilde\psi_j]  \sqrt{-g} \, (2\pi)^{d+1}\ \delta^{d+1}\qty(\textstyle{P_1 - P_2 + Q_4 - Q_3})\, \ \ic \cM(\{P_i, Q_j\}), \\
\end{split}
\eeq

Let us fix the left boundary detector time at $t_L=0$. 
For the BTZ solution, with $t_L=0$, the future-interior bulk point reaches the future singularity as $t_R\to0^-$ and the past-interior bulk point reaches the past singularity as $t_R\to0^+$, so as we go from $t_R= -\epsilon$ to $t_R =\epsilon$, the bulk point jumps from the future singularity to the past singularity, with the operator ordering changed as in Fig \ref{jurassic2}. For the AdS$_5$ black brane, we do not get a bulk point in the interior within the range 
\beq
-\tfrac{\beta}{2}<t_R<\tfrac{\beta}{2}.
\eeq
While the signal exists within the above range, it no longer describes a bulk point within the future interior. The camera in the past interior operates in the following time domain:
\beq
0\leq t_R; \quad {\rm BTZ} \qquad \qquad \tfrac{\beta}{2}\leq t_R; \quad {\rm AdS}_5
\eeq

This is clear by simply looking at the geodesics in question.

\textbf{Saddles at intermediate time:} Intermediate time saddles can also dominate the four point correlator within the above time domain. In this intermediate time range the dominant contribution comes from the product of two point functions with lightcone singularities such as bouncing geodesics \cite{Fidkowski:2003nf,Festuccia:2005pi, Ceplak:2024bja, Afkhami-Jeddi:2025wra, Ceplak:2025dds, Chakravarty:2026fdm}. 

One such saddle can appear at intermediate times when one of the wavepackets' energy is much larger than the other, for instance when the pair $\mathcal O_R(x),\mathcal O_R(x')$ is much more energetic than the pair $\mathcal O_L(y),\tO_R(\by'_+)$, $\omega_p \gg \omega_q$. We arrive at this new saddle provided that the suppression factor coming from the Fourier transform dominates over our signal saddle, i.e.
\beq
\exp\lc -{ \beta \omega_q \over 2}-{{1 \over \sigma_y^2}\lc t_{x'} + \ic {\beta / 2} + t_y\rc^2}\rc \gg \exp \lc -{\beta \over 2} (\omega_q + \omega_p)\rc 
\eeq
where $t_{x'}$ and $t_y$ are the (real) right and left boundary times of the two operators which pair up. In this limit, the contribution is the factorized contribution of two two-point correlators, with $\mathcal O_L(y)$ localizing near $\mathcal O_R(x')$, which are adjacent in \eqref{eq:past-camera}, and similarly for the two other operators:
\beq
\langle \Psi | \ \mathcal O_R(x) \ \tO_R(\by'_+) \ \mathcal O_R(x') \ \mathcal O_L(y) \ |  \Psi \rangle \approx \langle \,\mathcal  O_R(x) \, \tO_R(\by'_+)  \, \rangle \  \langle \, \mathcal O_R(x') \, \mathcal O_L(y) \, \rangle .
\eeq
Within the previously missing time range, the two two-point correlators get contributions from geodesics between the boundaries reflected off the singularity \cite{Fidkowski:2003nf,Festuccia:2005pi, Ceplak:2024bja, Afkhami-Jeddi:2025wra, Ceplak:2025dds}.

With the cutoff $t_{\rm cut}$ set by backreaction, \S\ref{sec:finite-N}, the description of the past interior in the BTZ geometry is valid in the time range:
\beq
0 < \,  t_R  \, \lesssim t_{\rm cut}.
\eeq
Decreasing the time, we gradually reach the past singularity at $t_R \to 0$. For the AdS$_5$ black brane, the description of the interior works in the time range:
\beq
\frac{\beta}{2} < \, t_R  \, \lesssim t_{\rm cut}, \qquad t_L =0.
\eeq
Beyond these regions there is no meaningful signal from the past interior region. 

Finally at intermediate timescales, we expect the contribution from two-point function. Very close to the singularity we expect the two-point function to break down as well as the probe approximation.

\subsection{Limits of the description}
\label{sec:finite-N}

The graviton exchange which produces the camera signal is in addition to the disconnected contribution of the product of 2 pt correlators. What limits the description is the $G_{\rm N}$ combined with the growth of the local center of mass energy with the separation between the two pairs, which can invalidate the probe approximation if the coincident limit is not taken carefully. The main requirement here is to ensure that the geodesic approximation remains valid. 

Apart from the leading black hole geometry, we may also have contributions from other subleading saddles to the correlation functions. Going to finite $N$ needs a treatment of nonperturbatively small corrections, of order $e^{-S}$ with $S$ the black hole entropy, to correlators arising from subleading saddles (the existence of such a form is motivated from the gravitational path integral). See also the recent discussion of finite-$N$ effects in the context of the two point function in \cite{Simon_unpub, Komargodski:2026oob, Batra:2026gbq}.

In line with our present analysis, we remain sensitive only to the perturbative corrections in the scalar coupling $\lambda$ and in $G_{\rm N}$, and are not sensitive to nonperturbative effects. 

As discussed, we can think of the geodesic description as introducing a pair of detectors, one on the left boundary and another on the right boundary. Using the boost isometry, we fix the left boundary detector at $t_L=0$. 

Now the time difference on each side in the initial Regge configuration cannot be too large. Timescales larger than this difference for the starting point can induce backreaction. This is essentially since the past geodesic is at very early time and stays close to the horizon where it is acted upon by the boost operator and its energy is highly boosted leading to strong backreaction which takes it outside the test geodesic approximation  \cite{Shenker:2013pqa, Shenker:2013yza}. In the language of \S\ref{sec:wkb}, the center of mass energy grows like $s_{\rm phys}\propto e^{2\pi|t_R|/\beta}$, cf. \eqref{eq:thermal-regge-camera}, and the eikonal phase \eqref{eq:eikonal-exterior} becomes of order one when $G_{\rm N}s_{\rm phys}\sim1$. Due to this, in the case of the BTZ geometry, we have a valid description of the future interior in the time range:
\beq
-t_{\rm cut} \, < \,  t_R  \, < \, 0
\eeq
Here $t_{\rm cut}$ is of the order of the scrambling time scale, $t_{\rm cut}\simeq\frac{\beta}{2\pi}\log\frac{1}{G_{\rm N}\omega_p\omega_q}$, since at that scale the energy becomes very large. Increasing the time $t_R$ we gradually reach the future singularity at $t_R\to 0$. In contrast to BTZ, for the AdS$_5$ black brane, since the singularities are pulled inwards, the description of the interior works in the  time range:
\beq
-t_{\rm cut} \, < \,  t_R  \, < -\tfrac{\beta}{2}.
\eeq
where we have again fixed the left detector at $t_L =0$, while the finite upper-bound results since the singularity in higher dimensions is pulled inwards \cite{Fidkowski:2003nf}. The upper bound is saturated when the intersection takes place on the future singularity, while the lower bound is saturated when the signal becomes very large. Beyond this, there is no meaningful bulk signal from the future interior region.

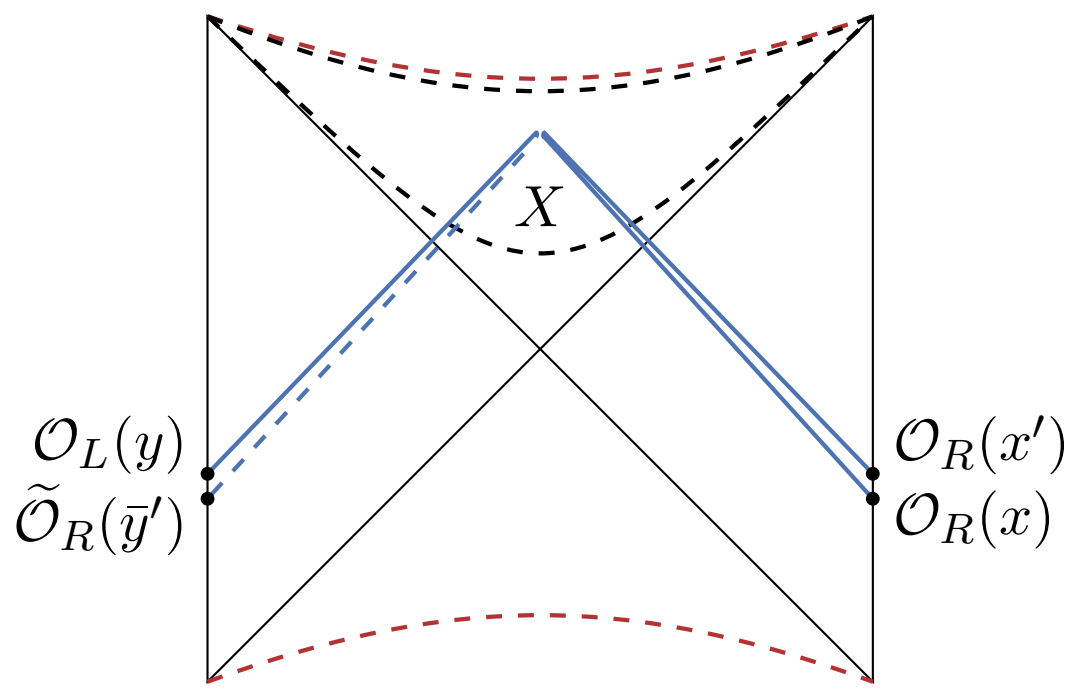
\begin{figure}[t!]
\centering
\begin{tikzpicture}[scale=0.55]
\node (I)    at ( 4,0)   {};
\node (II)   at (-4,0)   {};
\path  (II) +(90:4)  coordinate (IItop) +(-90:4) coordinate (IIbot) +(0:4) coordinate (IIright);
\draw (IItop) -- (IIright) -- (IIbot)--cycle;
\draw[dashed, mred, thick]  (-4,4) ..controls (-1.0,3) and (1.0,3).. (4,4);
\path (I) +(90:4)  coordinate (Itop) +(-90:4) coordinate (Ibot) +(180:4) coordinate (Ileft);
\draw  (Ileft) -- (Itop) -- (Ibot) -- (Ileft) -- cycle;
\draw[dashed, mred, thick] (IIbot) to[out=+20,in=-200] (Ibot) ;
\draw[dashed, thick] (-4,4) ..controls (-0.0,0.2) and (0.0,0.2).. (4,4);
\draw[dashed, thick] (-4,4) ..controls (-1,2.8) and (1,2.8).. (4,4);
\draw[mblue, thick] (4,-1.5) -- (0.05,2.6) to[out=200,in=-150] (0.05, 2.55) -- (4,-1.8);
\draw[mblue, thick] (-4,-1.5) -- (-0.05,2.6) to[out=-350,in=50] (-0.05, 2.55);
\draw[mblue, thick, dashed] (-0.2, 2.35) -- (-4,-1.8); 
\node[below] at (0,2.2) {$X$};
\filldraw (4,-1.5) circle (2pt); \filldraw (4,-1.8) circle (2pt); \filldraw (-4,-1.5) circle (2pt); \filldraw (-4,-1.8) circle (2pt); 
\node[right] at (4,-2.05) {$\mathcal O_{R}(x)$};
\node[right] at (4,-1.15) {$\mathcal O_{R}(x')$};
\node[left] at (-4,-1.15) {$\mathcal O_L(y)$};
\node[left] at (-4,-2.05) {$\tO_R(\by')$};
\end{tikzpicture}
    \caption{Backreaction limits the camera to the region between the dashed curves in the interior, away from the singularity and the horizon.}
    \label{finite-N}
\end{figure}

\section{Discussion}
\label{sec:discussion}

Let us summarize what we have found. The observable \eqref{eq:camera-def} is a smeared four-point function with three ordinary boundary insertions and one right operator moved by half of the thermal circle corresponding to the right black hole. At high frequencies we can tune the wavepackets in a way such that it focuses onto a scattering event at a point $X$ behind the horizon and factorizes into a local flat-space amplitude for the collision of two infalling wavepackets, with the momenta of the legs of $\mathcal O_L(y)$ and $\tO_R(\by')$ reversed. In the thermofield double state the observable is the analytic continuation of an exterior camera: the signal for planar BTZ is \eqref{eq:continued-angular-image}, with the opposite sign to the exterior camera and with the cusp $e^{-|\phi_2|}$ marking the bulk point. Away from the Einstein gravity limit, the cusp broadens. The sharpness of the image is also a measure of how local the bulk is. 

The camera sees whatever the two routes to the bulk point in the interior see differently. The leg of $\mathcal O_L(y)$ comes from the actual left boundary, the leg of $\tO_R(\by')$ is the analytic continuation of the right exterior, and the fact that an object behind the horizon which is crossed by one route and not by the other shows up as a jump in the signal as the shooting data are scanned. For a heavy worldline the jump sits on the curve \eqref{eq:worldline-image}, which gives the energy of the particle and the time of its turning point. For a null shock the image is displaced by the Shenker-Stanford time delay, \eqref{eq:shock-camera-first-order}, and the non-forward observable reads the shock profile off the phase per unit momentum transfer. For a shell the leg of $\mathcal O_L(y)$ is offset by $\delta t_\Sigma$, given in closed form in \eqref{eq:shell-delay}, and the camera records a phase $e^{-\ic\omega_q\delta t_\Sigma}$, a displaced collision point and a Gaussian loss of overlap between the legs of $\mathcal O_L(y)$ and $\tO_R(\by')$, \eqref{eq:shell-modified-camera-signal}. The mass of the shell follows from the offset, and absorption from the attenuation of the leg of $\mathcal O_L(y)$. In every case the data are coarse: specifically positions, time delays, phase shifts and absorption factors are all we can extract with a coarse grained signal. Note also that a similar gravitational effect which drives scrambling, i.e. the shift in time produced by a highly boosted perturbation as in \cite{Shenker:2013pqa,Shenker:2013yza,Maldacena:2015waa,Hernandez:2025seq}, is closely related to the imaging mechanism using smeared wavepackets.

A low-point correlator may not be tell apart the exponentially many microstates of the black hole unless we measure it to arbitrary precision\footnote{We can also try to measure the microstate by using very high-point correlators which may as well lead to issues with loss of bulk locality \cite{Ghosh:2017pel}.}, which fits with the picture in which many semiclassical interior excitations have exponentially small overlaps which account for the BH entropy \cite{Chakravarty:2020wdm}. 

We now discuss two features of the observable. The first is the backward Euclidean segment in \eqref{eq:camera-trace}. From the point of view of the complexified geometry we evolve against the direction of Euclidean time, which violates diagnostics such as the Kontsevich-Segal criterion \cite{Kontsevich:2021dmb, Witten:2021nzp}\footnote{Even with the correct sign of the Euclidean evolution one may have to work harder to make sense of the complexified geometry beyond the horizon using KSW, see the two similarly-posed questions in two different settings which require completely different analysis in \cite{Chakravarty:2024bna, Chakravarty:2025sbg}.}. We think of this as the price for seeing behind the horizon. This fits with the fact that the real-time algebras of the two boundaries describe the two exteriors \cite{Leutheusser:2021qhd,Leutheusser:2021frk}; reaching the interior needs an additional operation which here is given by our shifted operator in the correlator. 

On a similar note, a traversable wormhole \cite{Gao:2016bin, Maldacena:2017axo} is the physical counterpart of the insertion $\tO_R(\by')$. The coupling between the two boundaries places a genuine detector to the future, where a small window of the former interior opens up next to the horizon, and an exterior camera in that geometry would image the window with real-time operators. The insertion $\tO_R(\by')$ reaches deeper, at the cost of being a complex-time operation. 

On a slightly different note, the number of boundary timefolds does not by itself fix the number of bulk timefolds; it is the operator insertions which select the bulk fold \cite{Caron-Huot:2025hmk, Chakravarty:2025ncy}, and one can think of this as a choice of dressing.

The second feature is state dependence, in the sense of \cite{Papadodimas:2012aq,Papadodimas:2013jku,Papadodimas:2015jra,Harlow:2014yoa}. The shift of $\tO_R(\by')$ is half of the period of the Killing time of the right black hole, and it has to be chosen correctly for the leg of $\tO_R(\by')$ to meet the leg of $\mathcal O_L(y)$. The dependence is mild in one respect: the only data which enters is this period, which is also the KMS period of right two-point functions and can be measured by a right observer. In a state whose interior geometry is not known in advance, one would need to scan the Euclidean shift and look for the value at which the signal \eqref{eq:continued-einstein-camera} appears. It would be interesting to try this in a typical microstate, where the right exterior is unchanged but the interior is not the eternal one \cite{deBoer:2018ibj,deBoer:2019kyr}, and to see whether the difference is again recorded through the leg of $\mathcal O_L(y)$ alone, as for the shell of \S\ref{sec:timelike-shells}. On a similar note, doing this on a bags of gold geometry (which are atypical states) would also be useful.

Let us end with a few directions worth pursuing. In our present work, on the boundary side, the Regge analysis is presently restricted to two dimensions, where the cylinder-to-plane map lets us use the vacuum conformal Regge theory \cite{Costa:2012cb}. However in higher dimensions the thermal correlator is not a vacuum correlator in disguise, and it would be worth doing the Regge analysis from the boundary side, it might contain additionally insighful features. While we do not forsee a clear way, perhaps tools related to recent important developments in \cite{Benjamin:2023qsc, Buric:2025anb,Buric:2025fye,Barrat:2025twb, Barrat:2025nvu, Barrat:2026jfg, Niarchos:2025cdg, Niarchos:2026vow,Alkalaev:2026sha,Buric:2026cmn,Araya:2026shz,Correia:2026fnr,Sivaramakrishnan:2024ydy,Sivaramakrishnan:2025srr,Bohra:2025mhb, Kakkar:2026czr} could eventually prove useful to do such calculation. 

In our present work, we have looked at large but finite $N$. At finitely smaller $N$, subleading saddles and Stokes phenomena may become important in the computation of teh four point correlator. In principle, understanding which saddles dominate after the folding is what would promote the camera from a semiclassical probe to a finite-$N$ observable. While we do not directly encounter bouncing geodesics in our four-point description here, the finite-$N$ understanding of the two-point function motivated by \cite{Simon_unpub,Komargodski:2026oob,Batra:2026gbq} could be the natural place to start. 

On the bulk side, the shell computation is nice for the symmetric dust shell in BTZ, where everything follows from the map \eqref{eq:shell-map}. We can also try to probe the internal degrees of freedom, whose absorption should be derived from a microscopic influence functional rather than modelled by an imaginary WKB action, for instance an extremely high frequency laser will simply burn glass rather than refract through it. A very careful reconstruction of such a scattering process should in principle give us the full microstate, however we do not yet see a clean picture of how to extract such contributions. Finally, it would be useful to see what the observable has to say about the geometry near the singularity \cite{Blacker:2026poc, Hartnoll:2026vhu, Coviello:2026hdc, Coviello:2026tol, Gao:2026ieh, Oh:2026dky, Ceplak:2026ard,  Caceres:2023zft, Caceres:2026mug}, especially with respect to the Kasner like features, and how measurements such as these are connected to holography of information \cite{ Laddha:2020kvp,  Chowdhury:2021nxw, Chowdhury:2020hse, Chakraborty:2023los, Chakraborty:2023yed, Chakravarty:2023cll, Gaddam:2024mqm}.

\section*{Acknowledgements}
We thank Simon Caron-Huot for collaboration in the initial stages of this project, and for clarifying many physical aspects over the course of this and related projects. We thank Nima Afkhami-Jeddi, Keshav Dasgupta, Hao Geng, Matthew Headrick, Diksha Jain, Daniel Jafferis, Alex Maloney, Viraj Meruliya, Gautam Mandal, Rob Myers, Keivan Namjou, Ashoke Sen, Shivam Kumar Sharma, Brian Swingle, Spenta Wadia, Aron Wall and Yoav Zigdon for various helpful discussions. We are grateful to Claude Fable 5.1 for spotting an important physics error, spotted while wrapping up the project. JC is presently supported by J. Cotler’s Alfred P. Sloan Research Fellowship, and was earlier supported by the Natural Sciences and Engineering Research Council of Canada (NSERC) and the Canada Research Chair program, reference number CRC-2022-00421.

\appendix

\section{Suppression of smooth contributions to the correlator}
\label{app:fourier-asymptotics}

We will show now that within the object in \eqref{eq:two-sided-fourier-rescaling}, i.e. $ \int \diff^d\delta\, \psi_p(\delta)\,F(\delta)$, certain analytic contributions are exponentially suppressed compared to contributions non-analytic in $\delta$ at the separation $\delta =0$. Further even after Fourier transformation, if the singular contribution in \eqref{eq:two-sided-fourier-rescaling} has $\alpha < d$, it is only power law suppressed, which is much weaker than the exponential suppression of analytic contributions. Thus the Fourier transform filters out smooth dependence on the relative position and enhances any nonanalytic behaviour near $\delta=0$.

For comparison, consider a contribution $F(\delta)$ which is
analytic in a finite neighborhood (in the complex plane) of the real $\delta$ contour.  To see the large-momentum behavior of this object, it is sufficient to consider the relative displacement as one-dimensional, i.e., just one
component of the displacement $\delta$.  Upon Fourier transforming, we obtain:
\begin{equation}
  I(p) \equiv \int_{-\infty}^{\infty}
  \diff\delta\, \ee^{-\ic p\delta} \ee^{-\delta^2/(2\sigma)} \, F(\delta),
  \qquad
  p>0.
  \label{eq:analytic-wavepacket-transform}
\end{equation}
Now suppose that $F(\delta)$ is analytic in the strip $  -a<\Im\delta<0,$ and that the integrand (which has Gaussian weight) remains integrable throughout this strip.  For any $0<\eta<a$, the contour in
\eqref{eq:analytic-wavepacket-transform} may then be deformed to
$\Im\delta=-\eta$, thereby giving us:
\begin{align}
  I(p)
  &=
  \int_{-\infty}^{\infty}
  \diff x\,
  \ee^{-\ic p(x-\ic\eta)}
  \ee^{-(x-\ic\eta)^2/(2\sigma)}
  F(x-\ic\eta)=
  \ee^{-p\eta}
  \int_{-\infty}^{\infty}
  \diff x\,
  \ee^{-\ic px}
  \ee^{-(x-\ic\eta)^2/(2\sigma)}
  F(x-\ic\eta).
  \label{eq:analytic-contour-deformation}
\end{align}
We can use this to put some simple bounds as follows:
\begin{equation}
  |I(p)|
  \leq
  C_\eta\,\ee^{-p\eta}, \qquad {\rm where} \qquad
  C_\eta
  \equiv
  \int_{-\infty}^{\infty}
  \diff x\,
  \left|
    \ee^{-(x-\ic\eta)^2/(2\sigma)}
    F(x-\ic\eta)
  \right|.
  \label{eq:analytic-exponential-suppression}
\end{equation}
Thus an analytic contribution is exponentially suppressed at large positive
$p$, with the suppression scale set by the distance from the real contour to
the nearest singularity in the lower half-plane.  For  a negative $p$, one
instead deforms the contour into the upper half-plane.

As an example, if $F(\delta)$ has its nearest singularity at
$\delta=-\ic a$, its transform behaves as follows, where we work in the limit $ p\to+\infty.$:
\begin{equation}
  I(p)
  \sim
  \ee^{-ap}
  \times
  \text{powers of $p$},
  \label{eq:analytic-asymptotics}
\end{equation}
In contrast, a singularity on the Lorentzian contour, such as
\eqref{eq:two-sided-small-delta-singularity}, obstructs the contour
deformation.  Its Fourier transform therefore gives the power-law behavior
in \eqref{eq:two-sided-fourier-rescaling}, rather than the exponential
suppression in \eqref{eq:analytic-exponential-suppression}. Note that the physics we obtain in this way is not very different from what we expect, for instance from Breit-Wigner resonances (see \S 4 of \cite{Caron-Huot:2025hmk} for some discussion).

As in \S\ref{sec:folded-oscillators}, for the exterior camera, a similar nonanalyticity  gives the folded OPE singularity which gives the light transform in
\eqref{eq:light-transform}.  In the two-sided case this is also why a pair made of one left and one right operator, which is separated by $\ic\beta/2$ for any real displacement, does not produce a fold: its Fourier transform is of the type \eqref{eq:analytic-asymptotics} with $a=\beta/2$. The pair $\mathcal O_L(y),\tO_R(\by')$ of the camera \eqref{eq:camera-def} avoids this because, by \eqref{eq:tfd-left-right-kms-bra}, both of its members are right operators at the same imaginary time, and their light-cone singularity at real $\delta y\to0$ is of the type \eqref{eq:two-sided-small-delta-singularity}.  As we show in
\S\ref{sec:rindler-cross-ratios}, the cross-ratios can then reach the interior
bulk-point configuration \eqref{eq:folded-bulkpoint-edge} on the scattering sheet.  In the WKB
description of \S\ref{sec:wkb}, this configuration is represented by two folded bulk
geodesics which meet at an interior bulk point.

\section{Cross-ratios along the analytically continued path}
\label{app:cross-ratios}

This appendix gives the computation summarized in \S\ref{sec:rindler-cross-ratios}. We keep track of the $\ic\epsilon$ prescription along the complex path \eqref{eq:rindler-time-continuation} and check that its endpoint stays on the sheet where the bulk-point singularity sits, using the known form of the contact Witten diagram. This is also where the difference with the real-time two-sided correlator \eqref{eq1.1} is seen most directly.

\subsection*{From Euclidean to exterior kinematics}
Let us place the four operator insertions $x,x',y,y'$ of \eqref{fig1.2}, numbered $1,2,3,4$ as in \S\ref{sec:cam_ext}, at the spatial positions
\begin{equation}
    (x_1,x_2,x_3,x_4)=(-a,a,a,-a),
    \qquad a>0,
    \label{eq:spatial-camera-separations}
\end{equation}
so that, as before, the pair $x,x'$ is centered at $(0,0)$ and the pair $y,y'$ at $(T,0)$.

The operator ordering in \eqref{eq:ext-otoc} is $3241$, reading the operators from left to right. We implement it with the $\ic\epsilon$ prescription
\begin{equation}
    (\epsilon_1,\epsilon_2,\epsilon_3,\epsilon_4)
    =
    (\epsilon,3\epsilon,4\epsilon,2\epsilon),
    \qquad
    0<\epsilon\ll a,
    \label{eq:explicit-regulators}
\end{equation}
which fixes the Euclidean times of the starting configuration. We then bring the pair $y,y'$ to real time by dialling a parameter $s$ from $0$ to $1$,
\begin{equation}
    (t_1,t_2,t_3,t_4)
    =
    (-\ic\epsilon,-3\ic\epsilon,
       sT-4\ic\epsilon,sT-2\ic\epsilon),
    \qquad 0\leq s\leq1,
    \label{eq:euclidean-to-exterior}
\end{equation}
keeping $T>2a$ and $\epsilon$ nonzero throughout, as in the causally scattering configurations of \cite{Chandorkar:2021viw}.

With the exponential map of \S~\ref{sec:rindler-continuation}, the cross-ratios\footnote{It is a very unfortunate clash of notation, but the notation $z$ and $\bar z$ denote cross-ratios here and should not be confused with the radial coordinate which is also denoted as $z$. It is clear from the context but readers are advised to exercise caution.} are
\begin{equation}
    z=
    \frac{
        \sinh\frac{k u_{12}}2\,
        \sinh\frac{k u_{34}}2
    }{
        \sinh\frac{k u_{13}}2\,
        \sinh\frac{k u_{24}}2
    },
    \qquad
    \bar z=
    \frac{
        \sinh\frac{k v_{12}}2\,
        \sinh\frac{k v_{34}}2
    }{
        \sinh\frac{k v_{13}}2\,
        \sinh\frac{k v_{24}}2
    },
    \label{eq:exact-thermal-cross-ratios}
\end{equation}
where the exponential prefactors have cancelled. For the configuration above this gives
\begin{equation}
\begin{aligned}
    z(s)
    &=
    -\frac{\sinh^2[k(a+\ic\epsilon)]}{
        \sinh\frac{k(sT-2a-3\ic\epsilon)}2\,
        \sinh\frac{k(sT+2a+\ic\epsilon)}2
    }, \quad 
    \bar z(s)
    =
    -\frac{\sinh^2[k(a-\ic\epsilon)]}{
        \sinh\frac{k(sT+2a-3\ic\epsilon)}2\,
        \sinh\frac{k(sT-2a+\ic\epsilon)}2
    },
    \label{eq:regulated-exterior-cross-ratios}
\end{aligned}
\end{equation}
and for $1-z$ and $1-\bar z$
\begin{equation}
\begin{aligned}
    1-z(s)
    =
    \frac{\sinh^2\frac{k(sT-\ic\epsilon)}2}{
        \sinh\frac{k(sT-2a-3\ic\epsilon)}2\,
        \sinh\frac{k(sT+2a+\ic\epsilon)}2
    }, \qquad
    1-\bar z(s)
    =
    \frac{\sinh^2\frac{k(sT-\ic\epsilon)}2}{
        \sinh\frac{k(sT+2a-3\ic\epsilon)}2\,
        \sinh\frac{k(sT-2a+\ic\epsilon)}2
    }.
    \label{eq:regulated-one-minus-cross-ratios}
\end{aligned}
\end{equation}
At $s=0$ we start from a Euclidean configuration, where $\bar z$ is the complex conjugate of $z$. As $sT$ passes through $2a$, the factor $\sinh\frac{k(sT-2a-3\ic\epsilon)}2$ in \eqref{eq:regulated-one-minus-cross-ratios} passes below zero while the factor $\sinh\frac{k(sT-2a+\ic\epsilon)}2$ passes above zero. This is what decides the relative winding of the two cross-ratios.

Following all factors continuously from $s=0$ to $s=1$, we find to leading order in $\epsilon$
\begin{equation}
    z=-c_0-\ic c_1\epsilon,\qquad \bar z=-c_0+\ic c_2\epsilon,\qquad
    c_0=\frac{\sinh^2(ka)}{\sinh\frac{k(T-2a)}2\,\sinh\frac{k(T+2a)}2},
\label{eq:exterior-bulk-point-values}
\end{equation}
with $c_1,c_2>0$; for the parameters of Fig \ref{fig:crpath}, $c_1\simeq0.37$ and $c_2\simeq0.22$. So the two cross-ratios approach the same negative value from opposite sides of the real axis. Their logarithms, followed from the Euclidean configuration, satisfy
\begin{equation}
    \log\bar z(0,\epsilon)-\log z(0,\epsilon)
    =2\pi\ic+O(\epsilon),
\label{eq:exterior-branch-differences}
\end{equation}
where we write the cross-ratios as functions of the path parameter $\theta$ of the next subsection and of the regulator. Similarly,
\beq \label{eq:exterior-one-minus-branch-differences}
    \log(1-\bar z(0,\epsilon))
    -\log(1-z(0,\epsilon))
    =2\pi\ic+O(\epsilon),
\eeq
which can also be read off from \eqref{eq:regulated-one-minus-cross-ratios}: the numerators of $1-z$ and $1-\bar z$ are the same, while their denominators pass on opposite sides of zero, so $1-\bar z$ picks up one extra counterclockwise winding relative to $1-z$. The exterior configuration has therefore reached the scattering sheet.

\subsection*{Continuation to the opposite wedge}
\label{sec:continuation-opposite-wedge}

Starting from the exterior configuration on the second sheet, we now continue the insertion times $t_3,t_4$ of the pair $y,y'$ along the complex path \eqref{eq:rindler-time-continuation}. We want to show that the endpoint still has the bulk-point discontinuity, which is the boundary counterpart of the interior bulk point of \S\ref{sec:btz-bulk-point}.

The cross-ratios along the path follow from \eqref{eq:regulated-exterior-cross-ratios} by setting $sT\longrightarrow T-\ic\theta$, with $0\leq\theta\leq\frac{\beta}{2}$ as usual. We write them as $z(\theta,\epsilon)$ and $\bar z(\theta,\epsilon)$.

Let us first see that the offset \eqref{eq:exterior-branch-differences} is preserved. Choose $a$ small enough at fixed $T$ that the value $c_0$ in \eqref{eq:exterior-bulk-point-values} is less than one. Since $|\sinh(x-\ic y)|\geq|\sinh x|$, we have along the whole path
\begin{equation}
    |z(\theta,\epsilon)|,\,
    |\bar z(\theta,\epsilon)|
    \leq
    \frac{\sinh^2(ka)}{
        \sinh\frac{k(T-2a)}2\,
        \sinh\frac{k(T+2a)}2
    }
    +O(\epsilon),
\end{equation}
so for small $\epsilon$ both cross-ratios stay inside the unit disk. Both numerators and denominators are nonzero along the way, so neither cross-ratio meets $0$, $1$ or $\infty$.

Inside the unit disk the principal branch of $\log(1-z)$ is single valued. The offset of $2\pi\ic$ between the two logarithms can only change if the path winds around $1$, and a path inside the unit disk cannot do that. So throughout the continuation
\begin{equation}
    \log(1-z)=
    \log(1-z)_{\mathrm{principal}},
    \qquad
    \log(1-\bar z)
    =
    \log(1-\bar z)_{\mathrm{principal}}+2\pi\ic.
    \label{eq:retained-one-minus-branches}
\end{equation}
These are exact statements at nonzero $\epsilon$, with the branches fixed by the Euclidean preparation.

We also have to check the approach to $z=\bar z$. Expanding in $\epsilon$,
\begin{equation}
    z(\theta,\epsilon)-\bar z(\theta,\epsilon)
    =
    -4\ic k\epsilon\,
    \frac{
        \sinh(ka)\cosh(ka)\,
        \sinh^2[k(T-\ic\theta)/2]
    }{
        \left(
            \sinh^2[k(T-\ic\theta)/2]
            -\sinh^2(ka)
        \right)^2
    }
    +O(\epsilon^2).
    \label{eq:regulated-cross-ratio-difference}
\end{equation}
For $T>2a>0$ the coefficient of $\epsilon$ is finite and nonzero on the whole closed interval. So the path reaches $z=\bar z$ only when we actually set $\epsilon=0$; with the regulator in place it never does. The same equation and its derivative also give
\begin{equation}
    \frac{\partial_\theta\bar z}{\bar z}
    -
    \frac{\partial_\theta z}{z}
    =
    O(\epsilon),
\end{equation}
and integrating this with the initial value \eqref{eq:exterior-branch-differences} gives, along the whole path,
\begin{equation}
    \log\bar z(\theta,\epsilon)
    -
    \log z(\theta,\epsilon)
    =
    2\pi\ic+O(\epsilon).
    \label{eq:retained-logarithm-difference}
\end{equation}
At the endpoint, using $k\beta=2\pi$, the difference between the two cross-ratios is given by \eqref{eq:regulated-cross-ratio-difference} and their common value is
\begin{equation}
    z\!\left(\tfrac{\beta}{2},0\right)=\bar z\!\left(\tfrac{\beta}{2},0\right) = \frac{\sinh^2(ka)}{\cosh^2\frac{kT}{2}+\sinh^2(ka)}\in(0,1).
    \label{eq:folded-bulkpoint-edge}
\end{equation}
So \eqref{eq:retained-one-minus-branches} and \eqref{eq:retained-logarithm-difference} carry the same relative offsets as the exterior experiment, which are the ones that put the cross-ratios on the second sheet, and the pinch at $z=\bar z$ is the bulk-point singularity. Fig \ref{fig:crpath} shows the path numerically, with the branches of the logarithms followed from the Euclidean configuration: the two cross-ratios travel together from the exterior value to the interior value inside the unit disk, and the offset of $2\pi\ic$ between their logarithms is carried along unchanged.

Let us be clear about what this does and does not show. The bulk-point singularity of a holographic correlator sits at $z=\bar z$ on the sheet reached when $\bar z$ winds once around $1$ relative to $z$ \cite{Maldacena:2015iua,Caron-Huot:2022lff}. The exterior camera is on that sheet, which is the content of \eqref{eq:exterior-one-minus-branch-differences}. Along the path neither cross-ratio passes through $0$, $1$ or $\infty$, so the sheet cannot change, and $z-\bar z$ stays of order $\epsilon$, so the endpoint pinches the same singularity. This settles the analytic structure in cross-ratio space. It does not by itself say that the coefficient of the singularity, i.e. the signal, is nonzero at the endpoint; that is what the Regge computation of \S\ref{sec:rindler-camera-signal} and the contact diagram below do.

It is instructive to compare with the real-time two-sided correlator \eqref{eq1.1}, which is reached from the Euclidean region by a different route: the pair $y,y'$ is placed at imaginary time $-\ic\beta/2$ from the start, which is still a Euclidean configuration, and its real time is then dialled from $0$ to $T$. Along that route $\sinh^2\frac{k(sT-\ic\beta/2)}{2} = -\cosh^2\frac{ksT}{2}$ never vanishes, so $z(s)$ and $\bar z(s)$ stay real and inside $(0,1)$ for all $s$, which is the statement of \S\ref{sec:interior-camera-def} that a pair of operators on each boundary never winds around $z=1$, and both logarithms in \eqref{eq:retained-one-minus-branches} stay principal. The two routes end at the same point \eqref{eq:folded-bulkpoint-edge} of cross-ratio space but on different sheets, and only the sheet reached along \eqref{eq:rindler-time-continuation} carries the bulk-point singularity.

\subsection*{A check using a contact diagram}

We now follow a bulk contact Witten diagram from the exterior scattering configuration to the interior along \eqref{eq:rindler-time-continuation}. We take external dimensions $\Delta_i=1$, for which the contact diagram is explicit: after stripping off the coupling and an analytic nonzero prefactor, its dependence on the cross-ratios is $N(z,\bar z)/(z-\bar z)$ with $N$ given in \eqref{eq:contact-numerator}. The numerator is regular along the regulated path. We first find its value in the exterior scattering configuration and then show that it does not change along the continuation.

Consider first the Euclidean preparation \eqref{eq:euclidean-to-exterior} of the exterior experiment. Following the dilogarithms through it gives, near the exterior configuration,
\begin{equation}
\begin{aligned}
    \operatorname{Li}_2(z)
    =
    \operatorname{Li}_2(z)_{\mathrm{principal}},
    \qquad 
    \operatorname{Li}_2(\bar z)
    =
    \operatorname{Li}_2(\bar z)_{\mathrm{principal}}
    -2\pi\ic\log\bar z,
\end{aligned}
    \label{eq:exterior-dilogarithm-branches}
\end{equation}
where the logarithm in the second expression is the continued one fixed by the Euclidean preparation, not a new principal value. The additive term comes from the crossing of the cut of the principal dilogarithm during \eqref{eq:euclidean-to-exterior}: at the crossing the continued $\log(1-\bar z)$ exceeds its principal value by $2\pi\ic$, and the identity
\begin{equation}
    \diff\operatorname{Li}_2(\bar z)
    =
    -\log(1-\bar z)\,\diff\log\bar z
\end{equation}
together with continuity across the cut gives \eqref{eq:exterior-dilogarithm-branches}. In the exterior scattering configuration the two principal dilogarithms differ by $O(\epsilon)$, so substituting \eqref{eq:exterior-dilogarithm-branches} into \eqref{eq:contact-numerator} and using \eqref{eq:exterior-branch-differences} we find
\begin{equation}
    N\bigl(z(0,\epsilon),\bar z(0,\epsilon)\bigr)
    =
    -4\pi^2+O(\epsilon),
    \label{eq:exterior-contact-residue}
\end{equation}
where $\theta=0$ is the exterior scattering configuration at the start of \eqref{eq:rindler-time-continuation}, with the regulator still nonzero.

Now follow \eqref{eq:rindler-time-continuation}, continuing the insertion times of the pair $y,y'$ by $-\ic\theta$ at fixed $x,x'$, keeping $\epsilon>0$ and all the branches inherited from the Euclidean preparation. Differentiating the numerator along the path gives the exact identity
\begin{equation}
\begin{aligned}
    \diff N
    =
    \bigl[\log(1-z)+\log(1-\bar z)\bigr]
    \bigl[\diff\log\bar z-\diff\log z\bigr]
    +
    \bigl[\log z+\log\bar z\bigr]
    \bigl[\diff\log(1-z)-\diff\log(1-\bar z)\bigr],
\end{aligned}
    \label{eq:contact-numerator-derivative}
\end{equation}
with all functions on their continuously followed branches. We saw above that the regulated cross-ratios stay away from zero and one along the path, so at fixed $T$ and $a$ their continued logarithms stay bounded, and the cross-ratios and their $\theta$ derivatives differ by $O(\epsilon)$ uniformly along the path. Hence
\begin{equation}
\begin{aligned}
    \partial_\theta\log\bar z
    -
    \partial_\theta\log z
    =O(\epsilon), \qquad    \partial_\theta\log(1-z)
    -
    \partial_\theta\log(1-\bar z)
    =O(\epsilon),
\end{aligned}
\end{equation}
and \eqref{eq:contact-numerator-derivative} gives $\partial_\theta N=O(\epsilon)$. Integrating from the exterior value \eqref{eq:exterior-contact-residue},
\begin{equation}
    N\bigl(z(\theta,\epsilon),
           \bar z(\theta,\epsilon)\bigr)
    =
    -4\pi^2+O(\epsilon),
    \qquad
    0\leq\theta\leq\frac{\beta}{2}.
    \label{eq:continued-contact-numerator}
\end{equation}
So the continuation from the exterior to the interior cannot kill the pole by making its numerator vanish. The numerator starts at $-4\pi^2+O(\epsilon)$ and changes only by $O(\epsilon)$ along the whole path; the cancellation which makes the principal-sheet contact function regular on the diagonal simply does not happen here.

At the interior endpoint $\theta=\beta/2$, \eqref{eq:regulated-cross-ratio-difference} gives a denominator of order $\epsilon$ and \eqref{eq:continued-contact-numerator} a numerator of order one, so
\begin{equation}
    \left.
    \frac{N(z,\bar z)}{z-\bar z}
    \right|_{\theta=\beta/2}
    =
    -\frac{4\pi^2}{z-\bar z}+O(1),
\end{equation}
and restoring the coupling and the analytic prefactor, the singular part of the correlator is
\begin{equation}
    G_{\mathrm{con}}
    \ \propto\
    -\frac{4\pi^2\lambda}{z-\bar z},
    \qquad \lambda\neq0,
    \label{eq:continued-contact-bulk-point}
\end{equation}
which diverges as $1/\epsilon$ when the regulator is removed at the interior endpoint. Note that both the Euclidean preparation and the complex path were followed at nonzero $\epsilon$. We never used the divergent exterior correlator as an initial value; we continued the regulated correlator and followed its nonzero pole coefficient. The same estimates show that removing the regulator at an intermediate $\theta$ also leads to the continued pole, but only the initial and final configurations correspond to real exterior and future-interior bulk points. In between, the configurations are complex.

The principal-sheet contact function is regular at the positive diagonal point \eqref{eq:folded-bulkpoint-edge} reached at $\theta=\beta/2$, so its difference from the continued contact function has a nonzero singular part. This establishes the bulk-point discontinuity for this contact diagram on the sheet of the interior endpoint. Since the real-time two-sided correlator \eqref{eq1.1} lives on the principal sheet, it also confirms, at the level of a contact diagram, that it has no interior bulk-point singularity. This is a perturbative contact-interaction check; the graviton contribution is the subject of \S\ref{sec:rindler-camera-signal}.

\subsection*{Relation to the camera}
The calculation above concerns the spacelike separations within the Fourier integral. It does not replace the radial shooting momenta by spacelike momenta: the pair centers and the radial shooting configuration are unchanged. In the coincidence limit the continued pair centers have the future-interior bulk point of \S~\ref{sec:rindler-continuation}, the contact-diagram calculation shows that the continued boundary configuration keeps its bulk-point singularity, and \S~\ref{sec:rindler-camera-signal} shows independently that the continued Einstein-gravity camera signal stays nonzero and sharply focused at the radial angle. These are two complementary checks of the continued exterior experiment: its bulk point reaches the future interior, and its boundary continuation reaches a sheet on which a contact interaction has a nonzero bulk-point singularity. Neither of them requires the principal cross-ratios to stay negative. The singularity statement is perturbative, and the finite wavepacket widths regulate the camera signal.

\section{The shell in general dimension}
\label{app:shell}

This appendix collects the derivations behind \S\ref{sec:shell-general-d}. With the metrics \eqref{eq:shell-metrics-z} and the embedding $X_\pm^M(\tau,X)=(T_\pm(\tau),Z(\tau),X)$, the induced metric on the shell is
\begin{equation}
  \dd s_\Sigma^2
  =
  -\dd\tau^2
  +
  \frac{\dd X^2}{Z(\tau)^2}.
  \label{eq:induced-shell-z}
\end{equation}
and its velocity is
\begin{equation}
  u_\pm^M
  =
  \left(\dot T_\pm,\dot Z,0\right),
  \label{eq:shell-velocity-z}
\end{equation}
with a dot denoting $\dd/\dd\tau$. The normalization $g^{(\pm)}_{MN}u_\pm^Mu_\pm^N=-1$ reads
\begin{equation}
  \frac{1}{Z^2}
  \left(
    -f_\pm(Z)\dot T_\pm^2
    +
    \frac{\dot Z^2}{f_\pm(Z)}
  \right)
  =
  -1.
  \label{eq:shell-normalization-z}
\end{equation}
which gives $\dot T_\pm=A_\pm/f_\pm(Z)$. The unit normal, a spacelike vector orthogonal to the velocity, is
\begin{equation}
  n_\pm^M
  =
  \left(
    \frac{\dot Z}{f_\pm(Z)},
    A_\pm,
    0
  \right).
  \label{eq:shell-normal-z}
\end{equation}
and the corresponding covectors are
\begin{equation}
  u^{(\pm)}_M
  =
  \left(
    -\frac{A_\pm}{Z^2},
    \frac{\dot Z}{Z^2 f_\pm(Z)},
    0
  \right),
  \qquad
  n^{(\pm)}_M
  =
  \left(
    -\frac{\dot Z}{Z^2},
    \frac{A_\pm}{Z^2 f_\pm(Z)},
    0
  \right).
  \label{eq:shell-covectors-z}
\end{equation}
The unit vector tangent to the shell in the $X$-direction is
\begin{equation}
   e_{\parallel}^M
   =
   Z\,\delta_X^M,
   \qquad
   e^{\parallel}_M
   =
   \frac{1}{Z}\,\delta_M^X.
   \label{eq:shell-spatial-unit}
\end{equation}

For the junction condition, the extrinsic curvature of the transverse directions is $K_{XX}=\tfrac12 n^z\partial_z(z^{-2})$, and with the normal \eqref{eq:shell-normal-z} this gives $K^X{}_X=-A_\pm/Z$ on the two sides, for normals pointing towards the respective boundaries. The $XX$ components of \eqref{eq:israel-shell-z} fix $[K_{\tau\tau}]=(d-2)[K^X{}_X]$, and the $\tau\tau$ component is the energy equation \eqref{eq:junction-energy}. The conservation law follows from $\nabla_aS^{ab}_\Sigma=0$ with $\sqrt\gamma\propto Z^{1-d}$.

For the refraction law, consider a minimally coupled scalar in the three-dimensional bulk,
\begin{equation}
  S_\Phi
  =
  -\frac12
  \int \dd^3X\,\sqrt{-g}\,
  \left(
    g^{MN}\partial_M\Phi\partial_N\Phi
    +
    m^2\Phi^2
  \right).
  \label{eq:scalar-shell-action}
\end{equation}
Varying the action in the two regions gives a boundary term on the shell,
\begin{equation}
  \delta S_\Phi\big|_{\Sigma}
  =
  -
  \int_{\Sigma}
  \dd^2\sigma\sqrt{|\gamma|}\,
  \delta\Phi
  \left(
    n_+^M\partial_M\Phi_+
    -
    n_-^M\partial_M\Phi_-
  \right).
  \label{eq:flux-variation-shell}
\end{equation}
For a transparent shell with no scalar charge, the field and the normal flux are continuous,
\begin{equation}
  \Phi_+\big|_\Sigma=\Phi_-\big|_\Sigma,
  \qquad
  n_+^M\partial_M\Phi_+
  =
  n_-^M\partial_M\Phi_- .
  \label{eq:field-flux-continuity-shell}
\end{equation}
In the WKB approximation $\Phi_\pm\sim\cA_\pm e^{\ic S_\pm}$ the phase is then continuous along the shell, and the crossing point is stationary under tangential variations of the shell coordinates. Decomposing the covariant momentum $p^{(\pm)}_M=\partial_MS_\pm$ in the orthonormal shell frame,
\begin{equation}
  p_M
  =
  \omega_\Sigma u_M
  +
  p_n n_M
  +
  k_\Sigma e^\parallel_M ,
  \label{eq:momentum-decomposition-shell}
\end{equation}
with $\omega_\Sigma=-u^Mp_M$, $p_n=n^Mp_M$ and $k_\Sigma=e_\parallel^Mp_M=Zp_X$, the tangential stationarity conditions are
\begin{equation}
  \omega_\Sigma^{(+)}
  =
  \omega_\Sigma^{(-)},
  \qquad
  p_X^{(+)}
  =
  p_X^{(-)}.
  \label{eq:tangential-continuity-shell}
\end{equation}
Since the crossing happens at a single value of $Z$, the shell-frame transverse momentum $k_\Sigma$ is continuous as well. The mass-shell condition $g^{MN}p_Mp_N=-m^2$ becomes
\begin{equation}
  -\omega_\Sigma^2+p_n^2+k_\Sigma^2=-m^2.
  \label{eq:mass-shell-frame-shell}
\end{equation}
so $|p_n|$ is continuous too. The energy conjugate to $T_\pm$ is $E_\pm=-p^{(\pm)}_{T_\pm}$, and \eqref{eq:shell-covectors-z} with \eqref{eq:momentum-decomposition-shell} gives the refraction law \eqref{eq:shell-refraction-z}. Solving the junction condition supplies $Z(\tau)$, $\dot Z(\tau)$ and the relation between $T_+$ and $T_-$ along the shell, and once these are known the refraction law follows locally. The shell dynamics and the WKB matching are therefore separate problems: the first fixes the background, the second how each probe leg crosses it.

\bibliographystyle{JHEP}
\bibliography{citation}

\end{document}